\documentclass[prd,amsmath,aps,floats,amssymb, floatfix, superscriptaddress,nofootinbib,twocolumn,preprintnumbers]{revtex4-1}  
\usepackage{amsmath,amssymb,natbib,latexsym,times,subfigure}

\usepackage[switch,columnwise]{lineno}

\usepackage[T1]{fontenc}
\usepackage{aecompl}

\newcommand{\aap}{A\&A}
\newcommand{\apjs}{ApJS}
\newcommand{\aj}{A.J.}
\newcommand{\mnras}{MNRAS}

\newcommand{\jcap}{JCAP}
\newcommand{\ssr}{SSRv}
\newcommand{\apjl}{ApJ}
\newcommand{\pasp}{PASP}

\newcommand{\Msol}{\mbox{$M_{\odot}$}}

\newcommand{\Var}{\mbox{$\mathrm{Var}$}}

\newcommand{\Cov}{\mbox{$\mathrm{Cov}$}}

\newcommand{\Mpch}{\mbox{$h^{-1}\mathrm{Mpc}$}}

\usepackage{verbatim}
\usepackage[usenames,dvipsnames]{xcolor}
\usepackage{tabulary}
\usepackage{tabularx}
\usepackage{todonotes}
\usepackage{hyperref}

\numberwithin{equation}{section}

\newcommand{\imshape}{{\textsc{im3shape}}}

\newcommand\metacal{{\textsc{metacalibration}}} 

\newcommand\be{\begin{equation}}
\newcommand\ee{\end{equation}}
\def\bea{\begin{eqnarray}}
\def\eea{\end{eqnarray}}

\allowdisplaybreaks
\begin{document}

\title{Density split statistics: Cosmological constraints from counts and lensing in cells\\ in DES Y1 and SDSS data}


\author{D.~Gruen}
\email{Einstein Fellow; e-mail: dgruen@stanford.edu}
\affiliation{Kavli Institute for Particle Astrophysics \& Cosmology, P. O. Box 2450, Stanford University, Stanford, CA 94305, USA}
\affiliation{SLAC National Accelerator Laboratory, Menlo Park, CA 94025, USA}
\author{O.~Friedrich}
\affiliation{Universit\"ats-Sternwarte, Fakult\"at f\"ur Physik, Ludwig-Maximilians Universit\"at M\"unchen, Scheinerstr. 1, 81679 M\"unchen, Germany}
\affiliation{Max Planck Institute for Extraterrestrial Physics, Giessenbachstrasse, 85748 Garching, Germany}
\author{E.~Krause}
\affiliation{Kavli Institute for Particle Astrophysics \& Cosmology, P. O. Box 2450, Stanford University, Stanford, CA 94305, USA}
\author{J.~DeRose}
\affiliation{Department of Physics, Stanford University, 382 Via Pueblo Mall, Stanford, CA 94305, USA}
\affiliation{Kavli Institute for Particle Astrophysics \& Cosmology, P. O. Box 2450, Stanford University, Stanford, CA 94305, USA}
\author{R.~Cawthon}
\affiliation{Kavli Institute for Cosmological Physics, University of Chicago, Chicago, IL 60637, USA}
\author{C.~Davis}
\affiliation{Kavli Institute for Particle Astrophysics \& Cosmology, P. O. Box 2450, Stanford University, Stanford, CA 94305, USA}
\author{J.~Elvin-Poole}
\affiliation{Jodrell Bank Center for Astrophysics, School of Physics and Astronomy, University of Manchester, Oxford Road, Manchester, M13 9PL, UK}
\author{E.~S.~Rykoff}
\affiliation{Kavli Institute for Particle Astrophysics \& Cosmology, P. O. Box 2450, Stanford University, Stanford, CA 94305, USA}
\affiliation{SLAC National Accelerator Laboratory, Menlo Park, CA 94025, USA}
\author{R.~H.~Wechsler}
\affiliation{Department of Physics, Stanford University, 382 Via Pueblo Mall, Stanford, CA 94305, USA}
\affiliation{Kavli Institute for Particle Astrophysics \& Cosmology, P. O. Box 2450, Stanford University, Stanford, CA 94305, USA}
\affiliation{SLAC National Accelerator Laboratory, Menlo Park, CA 94025, USA}
\author{A.~Alarcon}
\affiliation{Institute of Space Sciences, IEEC-CSIC, Campus UAB, Carrer de Can Magrans, s/n,  08193 Barcelona, Spain}
\author{G.~M.~Bernstein}
\affiliation{Department of Physics and Astronomy, University of Pennsylvania, Philadelphia, PA 19104, USA}
\author{J.~Blazek}
\affiliation{Center for Cosmology and Astro-Particle Physics, The Ohio State University, Columbus, OH 43210, USA}
\affiliation{Institute of Physics, Laboratory of Astrophysics, \'Ecole Polytechnique F\'ed\'erale de Lausanne (EPFL), Observatoire de Sauverny, 1290 Versoix, Switzerland}
\author{C.~Chang}
\affiliation{Kavli Institute for Cosmological Physics, University of Chicago, Chicago, IL 60637, USA}
\author{J.~Clampitt}
\affiliation{Department of Physics and Astronomy, University of Pennsylvania, Philadelphia, PA 19104, USA}
\author{M.~Crocce}
\affiliation{Institute of Space Sciences, IEEC-CSIC, Campus UAB, Carrer de Can Magrans, s/n,  08193 Barcelona, Spain}
\author{J.~De~Vicente}
\affiliation{Centro de Investigaciones Energ\'eticas, Medioambientales y Tecnol\'ogicas (CIEMAT), Madrid, Spain}
\author{M.~Gatti}
\affiliation{Institut de F\'{\i}sica d'Altes Energies (IFAE), The Barcelona Institute of Science and Technology, Campus UAB, 08193 Bellaterra (Barcelona) Spain}
\author{M.~S.~S.~Gill}
\affiliation{SLAC National Accelerator Laboratory, Menlo Park, CA 94025, USA}
\author{W.~G.~Hartley}
\affiliation{Department of Physics \& Astronomy, University College London, Gower Street, London, WC1E 6BT, UK}
\affiliation{Department of Physics, ETH Zurich, Wolfgang-Pauli-Strasse 16, CH-8093 Zurich, Switzerland}
\author{S.~Hilbert}
\affiliation{Faculty of Physics, Ludwig-Maximilians-Universit\"at, Scheinerstr. 1, 81679 Munich, Germany}
\affiliation{Excellence Cluster Universe, Boltzmannstr.\ 2, 85748 Garching, Germany}
\author{B.~Hoyle}
\affiliation{Max Planck Institute for Extraterrestrial Physics, Giessenbachstrasse, 85748 Garching, Germany}
\affiliation{Universit\"ats-Sternwarte, Fakult\"at f\"ur Physik, Ludwig-Maximilians Universit\"at M\"unchen, Scheinerstr. 1, 81679 M\"unchen, Germany}
\author{B.~Jain}
\affiliation{Department of Physics and Astronomy, University of Pennsylvania, Philadelphia, PA 19104, USA}
\author{M.~Jarvis}
\affiliation{Department of Physics and Astronomy, University of Pennsylvania, Philadelphia, PA 19104, USA}
\author{O.~Lahav}
\affiliation{Department of Physics \& Astronomy, University College London, Gower Street, London, WC1E 6BT, UK}
\author{N.~MacCrann}
\affiliation{Center for Cosmology and Astro-Particle Physics, The Ohio State University, Columbus, OH 43210, USA}
\affiliation{Department of Physics, The Ohio State University, Columbus, OH 43210, USA}
\author{T.~McClintock}
\affiliation{Department of Physics, University of Arizona, Tucson, AZ 85721, USA}
\author{J.~Prat}
\affiliation{Institut de F\'{\i}sica d'Altes Energies (IFAE), The Barcelona Institute of Science and Technology, Campus UAB, 08193 Bellaterra (Barcelona) Spain}
\author{R.~P.~Rollins}
\affiliation{Jodrell Bank Center for Astrophysics, School of Physics and Astronomy, University of Manchester, Oxford Road, Manchester, M13 9PL, UK}
\author{A.~J.~Ross}
\affiliation{Center for Cosmology and Astro-Particle Physics, The Ohio State University, Columbus, OH 43210, USA}
\author{E.~Rozo}
\affiliation{Department of Physics, University of Arizona, Tucson, AZ 85721, USA}
\author{S.~Samuroff}
\affiliation{Jodrell Bank Center for Astrophysics, School of Physics and Astronomy, University of Manchester, Oxford Road, Manchester, M13 9PL, UK}
\author{C.~S{\'a}nchez}
\affiliation{Institut de F\'{\i}sica d'Altes Energies (IFAE), The Barcelona Institute of Science and Technology, Campus UAB, 08193 Bellaterra (Barcelona) Spain}
\author{E.~Sheldon}
\affiliation{Brookhaven National Laboratory, Bldg 510, Upton, NY 11973, USA}
\author{M.~A.~Troxel}
\affiliation{Center for Cosmology and Astro-Particle Physics, The Ohio State University, Columbus, OH 43210, USA}
\affiliation{Department of Physics, The Ohio State University, Columbus, OH 43210, USA}
\author{J.~Zuntz}
\affiliation{Institute for Astronomy, University of Edinburgh, Edinburgh EH9 3HJ, UK}
\author{T.~M.~C.~Abbott}
\affiliation{Cerro Tololo Inter-American Observatory, National Optical Astronomy Observatory, Casilla 603, La Serena, Chile}
\author{F.~B.~Abdalla}
\affiliation{Department of Physics \& Astronomy, University College London, Gower Street, London, WC1E 6BT, UK}
\affiliation{Department of Physics and Electronics, Rhodes University, PO Box 94, Grahamstown, 6140, South Africa}
\author{S.~Allam}
\affiliation{Fermi National Accelerator Laboratory, P. O. Box 500, Batavia, IL 60510, USA}
\author{J.~Annis}
\affiliation{Fermi National Accelerator Laboratory, P. O. Box 500, Batavia, IL 60510, USA}
\author{K.~Bechtol}
\affiliation{LSST, 933 North Cherry Avenue, Tucson, AZ 85721, USA}
\author{A.~Benoit-L{\'e}vy}
\affiliation{CNRS, UMR 7095, Institut d'Astrophysique de Paris, F-75014, Paris, France}
\affiliation{Department of Physics \& Astronomy, University College London, Gower Street, London, WC1E 6BT, UK}
\affiliation{Sorbonne Universit\'es, UPMC Univ Paris 06, UMR 7095, Institut d'Astrophysique de Paris, F-75014, Paris, France}
\author{E.~Bertin}
\affiliation{CNRS, UMR 7095, Institut d'Astrophysique de Paris, F-75014, Paris, France}
\affiliation{Sorbonne Universit\'es, UPMC Univ Paris 06, UMR 7095, Institut d'Astrophysique de Paris, F-75014, Paris, France}
\author{S.~L.~Bridle}
\affiliation{Jodrell Bank Center for Astrophysics, School of Physics and Astronomy, University of Manchester, Oxford Road, Manchester, M13 9PL, UK}
\author{D.~Brooks}
\affiliation{Department of Physics \& Astronomy, University College London, Gower Street, London, WC1E 6BT, UK}
\author{E.~Buckley-Geer}
\affiliation{Fermi National Accelerator Laboratory, P. O. Box 500, Batavia, IL 60510, USA}
\author{A.~Carnero~Rosell}
\affiliation{Laborat\'orio Interinstitucional de e-Astronomia - LIneA, Rua Gal. Jos\'e Cristino 77, Rio de Janeiro, RJ - 20921-400, Brazil}
\affiliation{Observat\'orio Nacional, Rua Gal. Jos\'e Cristino 77, Rio de Janeiro, RJ - 20921-400, Brazil}
\author{M.~Carrasco~Kind}
\affiliation{Department of Astronomy, University of Illinois, 1002 W. Green Street, Urbana, IL 61801, USA}
\affiliation{National Center for Supercomputing Applications, 1205 West Clark St., Urbana, IL 61801, USA}
\author{J.~Carretero}
\affiliation{Institut de F\'{\i}sica d'Altes Energies (IFAE), The Barcelona Institute of Science and Technology, Campus UAB, 08193 Bellaterra (Barcelona) Spain}
\author{C.~E.~Cunha}
\affiliation{Kavli Institute for Particle Astrophysics \& Cosmology, P. O. Box 2450, Stanford University, Stanford, CA 94305, USA}
\author{C.~B.~D'Andrea}
\affiliation{Department of Physics and Astronomy, University of Pennsylvania, Philadelphia, PA 19104, USA}
\author{L.~N.~da Costa}
\affiliation{Laborat\'orio Interinstitucional de e-Astronomia - LIneA, Rua Gal. Jos\'e Cristino 77, Rio de Janeiro, RJ - 20921-400, Brazil}
\affiliation{Observat\'orio Nacional, Rua Gal. Jos\'e Cristino 77, Rio de Janeiro, RJ - 20921-400, Brazil}
\author{S.~Desai}
\affiliation{Department of Physics, IIT Hyderabad, Kandi, Telangana 502285, India}
\author{H.~T.~Diehl}
\affiliation{Fermi National Accelerator Laboratory, P. O. Box 500, Batavia, IL 60510, USA}
\author{J.~P.~Dietrich}
\affiliation{Faculty of Physics, Ludwig-Maximilians-Universit\"at, Scheinerstr. 1, 81679 Munich, Germany}
\affiliation{Excellence Cluster Universe, Boltzmannstr.\ 2, 85748 Garching, Germany}
\author{P.~Doel}
\affiliation{Department of Physics \& Astronomy, University College London, Gower Street, London, WC1E 6BT, UK}
\author{A.~Drlica-Wagner}
\affiliation{Fermi National Accelerator Laboratory, P. O. Box 500, Batavia, IL 60510, USA}
\author{E.~Fernandez}
\affiliation{Institut de F\'{\i}sica d'Altes Energies (IFAE), The Barcelona Institute of Science and Technology, Campus UAB, 08193 Bellaterra (Barcelona) Spain}
\author{B.~Flaugher}
\affiliation{Fermi National Accelerator Laboratory, P. O. Box 500, Batavia, IL 60510, USA}
\author{P.~Fosalba}
\affiliation{Institute of Space Sciences, IEEC-CSIC, Campus UAB, Carrer de Can Magrans, s/n,  08193 Barcelona, Spain}
\author{J.~Frieman}
\affiliation{Fermi National Accelerator Laboratory, P. O. Box 500, Batavia, IL 60510, USA}
\affiliation{Kavli Institute for Cosmological Physics, University of Chicago, Chicago, IL 60637, USA}
\author{J.~Garc\'ia-Bellido}
\affiliation{Instituto de Fisica Teorica UAM/CSIC, Universidad Autonoma de Madrid, 28049 Madrid, Spain}
\author{E.~Gaztanaga}
\affiliation{Institute of Space Sciences, IEEC-CSIC, Campus UAB, Carrer de Can Magrans, s/n,  08193 Barcelona, Spain}
\author{T.~Giannantonio}
\affiliation{Institute of Astronomy, University of Cambridge, Madingley Road, Cambridge CB3 0HA, UK}
\affiliation{Kavli Institute for Cosmology, University of Cambridge, Madingley Road, Cambridge CB3 0HA, UK}
\affiliation{Universit\"ats-Sternwarte, Fakult\"at f\"ur Physik, Ludwig-Maximilians Universit\"at M\"unchen, Scheinerstr. 1, 81679 M\"unchen, Germany}
\author{R.~A.~Gruendl}
\affiliation{Department of Astronomy, University of Illinois, 1002 W. Green Street, Urbana, IL 61801, USA}
\affiliation{National Center for Supercomputing Applications, 1205 West Clark St., Urbana, IL 61801, USA}
\author{J.~Gschwend}
\affiliation{Laborat\'orio Interinstitucional de e-Astronomia - LIneA, Rua Gal. Jos\'e Cristino 77, Rio de Janeiro, RJ - 20921-400, Brazil}
\affiliation{Observat\'orio Nacional, Rua Gal. Jos\'e Cristino 77, Rio de Janeiro, RJ - 20921-400, Brazil}
\author{G.~Gutierrez}
\affiliation{Fermi National Accelerator Laboratory, P. O. Box 500, Batavia, IL 60510, USA}
\author{K.~Honscheid}
\affiliation{Center for Cosmology and Astro-Particle Physics, The Ohio State University, Columbus, OH 43210, USA}
\affiliation{Department of Physics, The Ohio State University, Columbus, OH 43210, USA}
\author{D.~J.~James}
\affiliation{Astronomy Department, University of Washington, Box 351580, Seattle, WA 98195, USA}
\author{T.~Jeltema}
\affiliation{Santa Cruz Institute for Particle Physics, Santa Cruz, CA 95064, USA}
\author{K.~Kuehn}
\affiliation{Australian Astronomical Observatory, North Ryde, NSW 2113, Australia}
\author{N.~Kuropatkin}
\affiliation{Fermi National Accelerator Laboratory, P. O. Box 500, Batavia, IL 60510, USA}
\author{M.~Lima}
\affiliation{Departamento de F\'isica Matem\'atica, Instituto de F\'isica, Universidade de S\~ao Paulo, CP 66318, S\~ao Paulo, SP, 05314-970, Brazil}
\affiliation{Laborat\'orio Interinstitucional de e-Astronomia - LIneA, Rua Gal. Jos\'e Cristino 77, Rio de Janeiro, RJ - 20921-400, Brazil}
\author{M.~March}
\affiliation{Department of Physics and Astronomy, University of Pennsylvania, Philadelphia, PA 19104, USA}
\author{J.~L.~Marshall}
\affiliation{George P. and Cynthia Woods Mitchell Institute for Fundamental Physics and Astronomy, and Department of Physics and Astronomy, Texas A\&M University, College Station, TX 77843,  USA}
\author{P.~Martini}
\affiliation{Center for Cosmology and Astro-Particle Physics, The Ohio State University, Columbus, OH 43210, USA}
\affiliation{Department of Astronomy, The Ohio State University, Columbus, OH 43210, USA}
\author{P.~Melchior}
\affiliation{Department of Astrophysical Sciences, Princeton University, Peyton Hall, Princeton, NJ 08544, USA}
\author{F.~Menanteau}
\affiliation{Department of Astronomy, University of Illinois, 1002 W. Green Street, Urbana, IL 61801, USA}
\affiliation{National Center for Supercomputing Applications, 1205 West Clark St., Urbana, IL 61801, USA}
\author{R.~Miquel}
\affiliation{Instituci\'o Catalana de Recerca i Estudis Avan\c{c}ats, E-08010 Barcelona, Spain}
\affiliation{Institut de F\'{\i}sica d'Altes Energies (IFAE), The Barcelona Institute of Science and Technology, Campus UAB, 08193 Bellaterra (Barcelona) Spain}
\author{J.~J.~Mohr}
\affiliation{Excellence Cluster Universe, Boltzmannstr.\ 2, 85748 Garching, Germany}
\affiliation{Faculty of Physics, Ludwig-Maximilians-Universit\"at, Scheinerstr. 1, 81679 Munich, Germany}
\affiliation{Max Planck Institute for Extraterrestrial Physics, Giessenbachstrasse, 85748 Garching, Germany}
\author{A.~A.~Plazas}
\affiliation{Jet Propulsion Laboratory, California Institute of Technology, 4800 Oak Grove Dr., Pasadena, CA 91109, USA}
\author{A.~Roodman}
\affiliation{Kavli Institute for Particle Astrophysics \& Cosmology, P. O. Box 2450, Stanford University, Stanford, CA 94305, USA}
\affiliation{SLAC National Accelerator Laboratory, Menlo Park, CA 94025, USA}
\author{E.~Sanchez}
\affiliation{Centro de Investigaciones Energ\'eticas, Medioambientales y Tecnol\'ogicas (CIEMAT), Madrid, Spain}
\author{V.~Scarpine}
\affiliation{Fermi National Accelerator Laboratory, P. O. Box 500, Batavia, IL 60510, USA}
\author{M.~Schubnell}
\affiliation{Department of Physics, University of Michigan, Ann Arbor, MI 48109, USA}
\author{I.~Sevilla-Noarbe}
\affiliation{Centro de Investigaciones Energ\'eticas, Medioambientales y Tecnol\'ogicas (CIEMAT), Madrid, Spain}
\author{M.~Smith}
\affiliation{School of Physics and Astronomy, University of Southampton,  Southampton, SO17 1BJ, UK}
\author{R.~C.~Smith}
\affiliation{Cerro Tololo Inter-American Observatory, National Optical Astronomy Observatory, Casilla 603, La Serena, Chile}
\author{M.~Soares-Santos}
\affiliation{Fermi National Accelerator Laboratory, P. O. Box 500, Batavia, IL 60510, USA}
\author{F.~Sobreira}
\affiliation{Instituto de F\'isica Gleb Wataghin, Universidade Estadual de Campinas, 13083-859, Campinas, SP, Brazil}
\affiliation{Laborat\'orio Interinstitucional de e-Astronomia - LIneA, Rua Gal. Jos\'e Cristino 77, Rio de Janeiro, RJ - 20921-400, Brazil}
\author{M.~E.~C.~Swanson}
\affiliation{National Center for Supercomputing Applications, 1205 West Clark St., Urbana, IL 61801, USA}
\author{G.~Tarle}
\affiliation{Department of Physics, University of Michigan, Ann Arbor, MI 48109, USA}
\author{D.~Thomas}
\affiliation{Institute of Cosmology \& Gravitation, University of Portsmouth, Portsmouth, PO1 3FX, UK}
\author{V.~Vikram}
\affiliation{Argonne National Laboratory, 9700 South Cass Avenue, Lemont, IL 60439, USA}
\author{A.~R.~Walker}
\affiliation{Cerro Tololo Inter-American Observatory, National Optical Astronomy Observatory, Casilla 603, La Serena, Chile}
\author{J.~Weller}
\affiliation{Excellence Cluster Universe, Boltzmannstr.\ 2, 85748 Garching, Germany}
\affiliation{Max Planck Institute for Extraterrestrial Physics, Giessenbachstrasse, 85748 Garching, Germany}
\affiliation{Universit\"ats-Sternwarte, Fakult\"at f\"ur Physik, Ludwig-Maximilians Universit\"at M\"unchen, Scheinerstr. 1, 81679 M\"unchen, Germany}
\author{Y.~Zhang}
\affiliation{Fermi National Accelerator Laboratory, P. O. Box 500, Batavia, IL 60510, USA}

\collaboration{DES Collaboration}


\date{\today}


\label{firstpage}
\begin{abstract}
We derive cosmological constraints from the probability distribution function (PDF) of evolved large-scale matter density fluctuations. We do this by splitting lines of sight by density based on their count of tracer galaxies, and by measuring both gravitational shear around and counts-in-cells in overdense and underdense lines of sight, in Dark Energy Survey (DES) First Year and Sloan Digital Sky Survey (SDSS) data. Our analysis uses a perturbation theory model \citep{Oliver} and is validated using $N$-body simulation realizations and log-normal mocks. It allows us to constrain cosmology, bias and stochasticity of galaxies w.r.t. matter density and, in addition, the skewness of the matter density field.

From a Bayesian model comparison, we find that the data weakly prefer a connection of galaxies and matter that is stochastic beyond Poisson fluctuations on $\leq20$~arcmin angular smoothing scale. The two stochasticity models we fit yield DES constraints on the matter density $\Omega_m=0.26^{+0.04}_{-0.03}$ and $\Omega_m=0.28^{+0.05}_{-0.04}$ that are consistent with each other. These values also agree with the DES analysis of galaxy and shear two-point functions (3x2pt, \citeauthor{keypaper}) that only uses second moments of the PDF. Constraints on $\sigma_8$ are model dependent ($\sigma_8=0.97^{+0.07}_{-0.06}$ and $0.80^{+0.06}_{-0.07}$ for the two stochasticity models), but consistent with each other and with the 3x2pt results if stochasticity is at the low end of the posterior range.

As an additional test of gravity, counts and lensing in cells allow to compare the skewness $S_3$ of the matter density PDF to its $\Lambda$CDM prediction. We find no evidence of excess skewness in any model or data set, with better than 25 percent relative precision in the skewness estimate from DES alone.
\end{abstract}

\pacs{Valid PACS appear here}

\preprint{DES-2017-0254}
\preprint{FERMILAB-PUB-17-444-AE}

\maketitle



\section{Introduction}

Measurements of the two-point correlation function of the evolved matter density field have provided competitive constraints on fundamental cosmological parameters. In combination with cosmic microwave background (CMB) and other geometric data, they are stringent tests of $\Lambda$CDM predictions on the evolution of structure over cosmic time \citep{2013MNRAS.432.1544M,2017MNRAS.465.1454H,2017arXiv170605004V,shearcorr,keypaper}. Ostensibly, larger studies of this kind are the primary goal of the upcoming ambitious ground-based and space-based surveys by \emph{Euclid}, LSST and WFIRST.

On a given smoothing scale, one can describe a field locally by its PDF.  An example of this is the PDF of fluctuations of the mean matter density inside spherical or cylindrical volumes. The variance, or second moment, of the PDF is measured by two-point statistics. For Gaussian distributions, this captures all the information in all the moments of the PDF. But when the field is non-Gaussian, the third moment (skewness) can take any value -- therefore it contains information that is \emph{not contained} in two-point statistics.

Unlike the primordial CMB, which is extremely close to a Gaussian random field, the density distribution in the evolved Universe has been driven away from Gaussianity by gravitational collapse. Third and higher order moments arise on any scale. Two-point measurements are therefore inherently very incomplete pictures of the matter density field. Even the full hierarchy of $N$-point correlations ceases to fully describe its statistics \citep{1991MNRAS.248....1C,Carron2012,2012ApJ...750...28C}. This is unfortunate in two ways: (1) A lot of information on cosmology, and (2) a lot of opportunities to test additional, independent $\Lambda$CMB predictions of the growth of structure beyond its variance, are lost by looking at two-point functions alone.

There are other reasons that make two-point correlations a somewhat blunt tool. First, the information to be gained from galaxy auto- or cross-correlations must be related to the clustering of matter by a \emph{bias model}, i.e.~a description of how galaxies trace matter density. Yet the information on the bias model that is available from two-point functions alone is limited. The primary reason for the success of joint probes is that they can partially break these degeneracies. For instance, the joint analysis of galaxy clustering and galaxy-galaxy lensing can constrain two combinations of $\sigma_8$, galaxy bias, and galaxy stochasticity. As one pushes to smaller scales where a lot of the cosmological constraining power resides, a linear bias model without stochasticity is not sufficient. The resulting degeneracies thus largely annihilate the information that is gained. Second, the information on the variance of the matter density field can be used only if that variance can be modeled -- on nonlinear scales, complex physics that involve baryons and neutrinos begin to influence any moment of the matter density field. Using two-point functions, these complex physics can be constrained \citep{2016MNRAS.463.3326F,2017MNRAS.465.2567M}, although (for the same reason as for the bias model) only with limited discriminating power. If we could recover small scale information with models that can be trusted, the ability of presently and imminently available data sets to confirm or reduce tensions between the CMB and evolved power spectrum would immediately be boosted. 

For these reasons, studies of the cosmic density PDF, which address these problems from a different and complementary direction, have gained interest over the last years. The full shape of the joint matter and galaxy density PDF depends on moments of the matter density field and parameters of the bias model that are degenerate in correlation function measurements. Numerical simulations \citep{Takahashi2011,2013MNRAS.435.2968P,Klypin2017}, tree-level perturbation calculations \citep{Bernardeau2000,Valageas2002}, and extensions of theory beyond that \citep{2016MNRAS.460.1529U}, have been shown to provide accurate predictions for the matter density PDF. Parameter forecasts show that PDF measurements on data are promising \citep{2016MNRAS.460.1549C,2016PhRvD..94j3501L,2017MNRAS.472..439P} due to the complementary information, different degeneracies, and different dependence on observational systematics -- factor-of-two improvements in constraining power can be achieved in joint measurements of PDF and two-point functions. While the galaxy count PDF alone can be used to break degeneracies of cosmological and bias parameters \citep{2008ApJ...682..737R}, gravitational lensing greatly complements this by measuring the actual matter density PDF. Practical application of shear PDF statistics to data has been made with DES \citep{2017MNRAS.466.1444C,2017arXiv170801535C}, yet so far with limited use for quantitative constraints.

In this paper, we use the smoothed, joint, projected galaxy count and matter density PDF to constrain cosmological parameters and a galaxy bias model. Our basic concept is to (1) split the sky by the count of tracer galaxies in a top-hat aperture and extended redshift range into quantiles of density, and to (2) measure the gravitational shear around each of the quantiles to reconstruct the matter density PDF. These measurements are a generalization of trough lensing, introduced in \citet{2016MNRAS.455.3367G} (see also \cite{2016MNRAS.459.2762H,2017JCAP...02..031B,2017arXiv171001730C}). They are also closely related to the galaxy-matter aperture statistics of \citet{2013MNRAS.430.2476S}. We make them on Dark Energy Survey Year 1 (DES Y1) and SDSS DR8 data. 

The measurements are analyzed with a tree-level perturbation theory prediction for the joint statistical properties of lensing convergence and density contrast and galaxy bias models of varying complexity (cf. our companion paper \citeauthor{Oliver}). We use the analysis not just to provide an independent measurement of cosmological parameters, but also to confront the $\Lambda$CDM prediction for the \emph{skewness} of the matter density field with data, in a model independent test of structure formation. That is, we measure the asymmetry of the low and high density tails of the distribution of matter density in the Universe. Two-point statistics, which only measure the \emph{width} of the matter density distribution, discard this information.

This paper is structured as follows. We describe the data we use and our measurement methodology in \autoref{sec:measurement}. Our modeling of these measurements, based on \citet{Oliver}, is summarized in \autoref{sec:model}. The covariance matrix we estimate is described in \autoref{sec:covariance}. We combine measurements, model and covariance into an inference framework in \autoref{sec:likelihood}. Results are presented in \autoref{sec:constraints}, and we conclude in \autoref{sec:conclusions}. Several tests and technical aspects of this work are detailed in the appendix.

\section{Measurement}
\label{sec:measurement}
In the following section, we first describe our method of splitting lines of sight by density based on counts of a tracer galaxy sample (\autoref{sec:troughfinding}). The \textsc{redMaGiC} tracer catalogs we use to do this in DES and SDSS and their redshift distribution calibration are presented in \autoref{sec:des_redmagic} and \autoref{sec:sdss_redmagic}. Details on our DES and SDSS source shape and photometric redshift catalogs are given in \autoref{sec:des_shapes} and \autoref{sec:sdss_shapes}. The measurement of shear and counts-in-cells signals is described in \autoref{sec:signals}.

\subsection{Splitting the sky by density}
\label{sec:troughfinding}

The basic idea of this study is to split the sky into lines of sight of different density. 

To this end, we use a sample of foreground galaxies as tracers of the matter field 
(the \textsc{redMaGiC} galaxies at $0.2<z_T<0.45$ described in \autoref{sec:des_redmagic} and \ref{sec:sdss_redmagic}).
We count these galaxies within circular top-hat apertures with a range of radii $\theta_T=10',20',30',60'$,
centered on a regular \textsc{healpix} \citep{2005ApJ...622..759G} grid of $N_{\rm side}=1024$ (3.4~arcmin grid spacing).

We then assign each line of sight to one of five density quintiles by sorting all lines of sight by galaxy count. The 
20 per-cent of lines of sight with the lowest galaxy count are what we will call the lowest density quintile 1 (or troughs, cf. \cite{2016MNRAS.455.3367G}). The
20 per-cent of lines of sight with the highest galaxy count (quintile 5) we will denote as overdense lines of sight.

Compared to \citet{2016MNRAS.455.3367G}, we apply a more elaborate scheme of accounting for varying fractions of masked area
within the respective survey region. A mask accompanying the \textsc{redMaGiC} \citep{wthetapaper} catalog that we will use as our tracers (\autoref{sec:des_redmagic}) indicate what fraction of the area inside each pixel in a $N_{\rm side}=4096$ \textsc{healpix} map is covered by DES Y1 Gold photometry \citep{y1gold} to sufficient depth for detecting \textsc{redMaGiC} galaxies out to at least $z=0.45$.  For each line of sight, we estimate the fraction of masked area $f_{\rm mask}$ within the corresponding top-hat aperture from the \textsc{redMaGiC} masks. Centers with more than a fraction $f_{\rm mask}^{\rm max}$ of area within the aperture lost to masking are discarded. 

The depth of SDSS is very uniform, 
with the \textsc{redMaGiC} sample being complete to $z=0.45$ everywhere. In this case, we use $f^{\rm max}_{\rm mask,SDSS}=0.1$.
Despite its greater overall depth, DES Y1~\citep{2014SPIE.9149E..0VD} is generally more inhomogeneous than the final SDSS imaging data. 
Where the \textsc{redMaGiC} sample is not complete to $z=0.45$, we remove all
tracer galaxies and define the area to be fully masked. Due to the larger fraction of masked area, we use 
$f^{\rm max}_{\rm mask,DES}=0.2$, above which we discard lines of sight.

To account for residual differences in $f_{\rm mask}$ we apply the following probabilistic scheme of quintile assignment.
For each line of sight $i$ with masking fraction $f_{\mathrm{mask},i}$ and raw tracer galaxy count $N_{\mathrm{raw}, i}$, we define $N_i$ as 
a draw from a binomial distribution with $N_{\mathrm{raw}, i}$ repetitions and success probability $p_i=1-\left(f_{\rm mask}^{\rm max}-f_{\mathrm{mask},i}\right)$,
\begin{equation}
P(N_i|N_{\mathrm{raw}, i},f_{\mathrm{mask},i})=\left(\begin{array}{c} N_{\mathrm{raw}, i}\\ N_i\end{array}\right) p_i^{N_i} (1-p_i)^{N_{\mathrm{raw}, i}-N_i} \; .
\end{equation}
This emulates the masking of a fixed fraction $f_{\rm mask}^{\rm max}$ of area within each aperture. It preserves the expectation value of galaxy count in an aperture, regardless of its masking fraction. Under the assumption that galaxies or masked pixels do not cluster, and that galaxy count is not stochastic beyond Poissonian noise, this masking procedure would preserve the full distribution of galaxy counts at fixed matter density (see Appendix~\ref{app:bernoulli}). The latter conditions are not true in practice, which is why the degree and spatial distribution of masking still affects the width of $P(N_i)$ at fixed expectation value. Tests of likelihood runs and the masked $P(N)$ in the \emph{Buzzard} simulations (see Appendix~\ref{app:mock} and \cite{Oliver}) indicate that this is not a major concern for our analysis.

We assign a line of sight $i$ to a density quintile based on many random realizations of $N_i$. Different realizations of $N_i$ can cause different quintile assignments. To account for this, we define a weight $w_i^q$, proportional to the number of times $N_i$ is in quintile $q$. This weight is assigned to line of sight $i$ when measuring the signal, e.g.~the mean tangential shear, of quintile $q$.

\begin{figure*}
  \includegraphics[width=\linewidth]{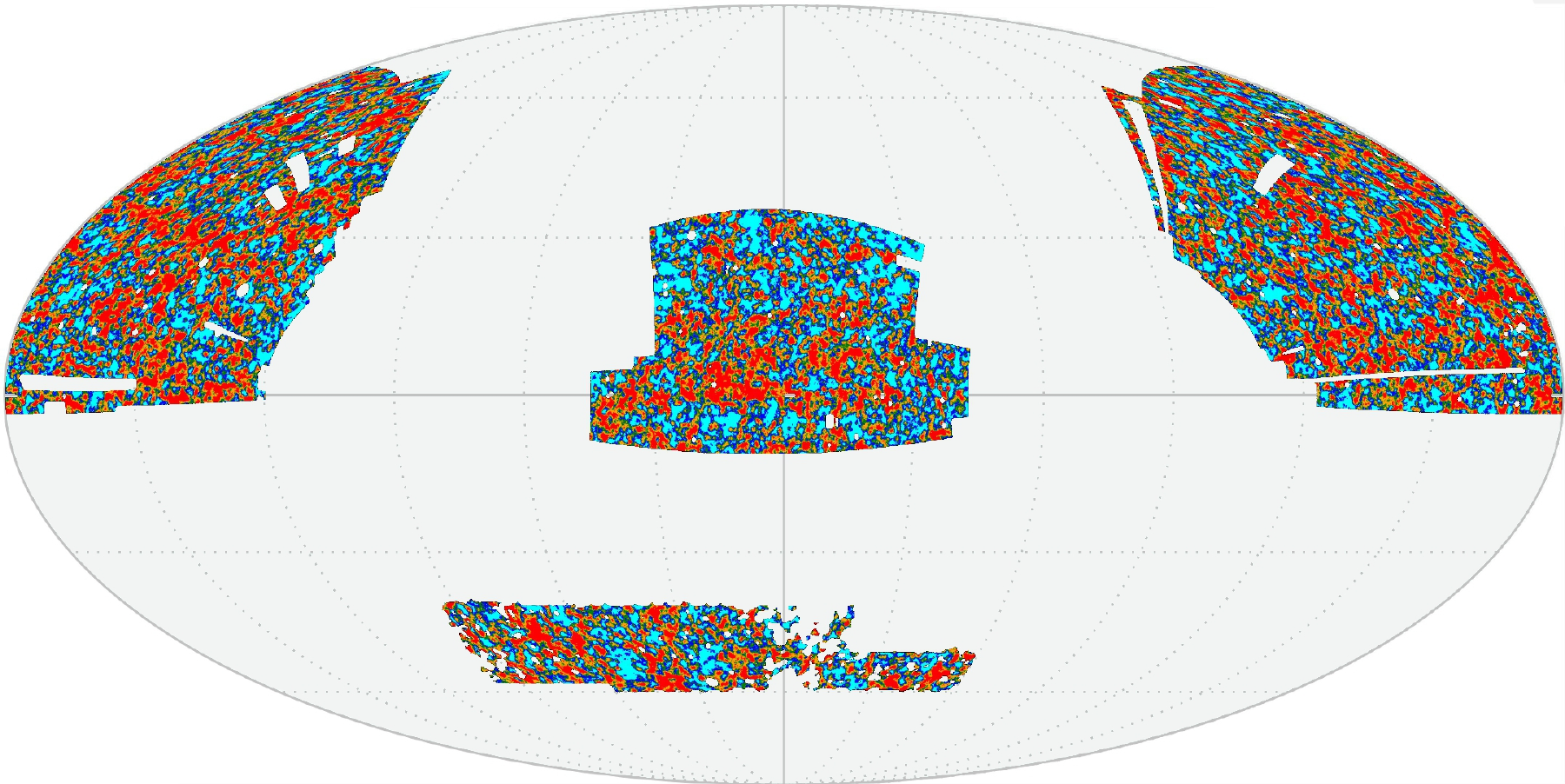}
  \caption{Overdense (red) and underdense (cyan) lines of sight in the DES Y1 (south/bottom) and SDSS (north/top) survey regions. Five quintiles of density of the $z_T=0.2-0.45$ \textsc{redMaGiC} tracer galaxy sample smoothed within a $\theta_T=30'$ radius are shown in the same color scheme as in \autoref{fig:tomosignal}. Pixels are drawn in the color of the quintile with the highest probability/weight $w^q$ (see \autoref{sec:troughfinding}). Graticule shows lines of $\Delta\mathrm{RA},\Delta\mathrm{dec}=\pm30^{\circ}$, centered on $(\mathrm{RA},\mathrm{dec})=(0,0)$.}
\label{fig:skyplots}
\end{figure*}

\autoref{fig:skyplots} shows the result of this quintile assignment procedure for the joint region covered by DES Y1 and SDSS.

\subsection{Dark Energy Survey Y1 data}

The Dark Energy Survey data we use in this work is from the SPT region of the first year of science observations (Y1) performed between 31 August 2013 and 9 February 2014. Details of the data and photometric pipeline are described in \citet{y1gold}. 

We use catalogs of luminous red galaxies (\textsc{redMaGiC} galaxies) as tracers of the foreground matter density field and galaxy shape and photometric redshift catalogs for measuring its gravitational shear signal, all of which are described briefly below and in detail in \citet{wthetapaper,shearcat,photoz}.

In all likelihood analyses run on data in this work, we propagate the three most relevant calibration uncertainties  of these catalogs:
\begin{itemize}
\item the multiplicative bias of the shear signal, characterized as $m=\gamma^{\rm obs}/\gamma^{\rm true}-1$,
\item the bias in mean redshift of each source bin $i$, characterized by a $\Delta z_s^i$ which we use to evaluate $n_s^i(z)=n_s^{i,PZ}(z-\Delta z_s^i)$, where $n_s^{i,PZ}$ is the photometric estimate of the source redshift distribution, and
\item the bias in mean redshift of the tracer galaxy sample, characterized by a $\Delta z_l$ which we use to evaluate $n_l(z)=n_l^{\rm redMaGiC}(z-\Delta z_l)$.
\end{itemize}
The derivation of priors on these calibration uncertainties is described or referenced in \autoref{sec:nuisance}.

\subsubsection{Tracer catalog}
\label{sec:des_redmagic}

The \textsc{redMaGiC} \citep{redmagic} algorithm identifies a sample of red galaxies with constant comoving density and fixed luminosity threshold. This is done by fitting the DES photometry of each galaxy in the survey to find its maximum likelihood luminosity and redshift under the assumption of the redMaPPer \citep{2014ApJ...785..104R} red sequence template. Galaxies are removed from the \textsc{redMaGiC} catalog if their fitted luminosity falls below a threshold ($0.5L^{\star}$ for the \emph{high density} run used in this work). The catalog is further pruned to retain a fixed number density of galaxies per comoving volume element, keeping those that are best fit (in terms of photometric $\chi^2$) by the red sequence template. The resulting galaxy density is $10^{-3} h^{3} \mathrm{Mpc}^{-3}$ in the case of the \emph{high density} catalog.

This procedure was run on two different photometric measurements of DES Y1 galaxies, one with the \textsc{SExtractor} \texttt{MAG\_AUTO} method and one performing a joint fit to the multi-epoch data of multiple overlapping objects (MOF). Potential correlation of the surface density of \textsc{redMaGiC} galaxies with observational systematics in DES Y1 have been extensively tested in \citet{wthetapaper} for both versions of the catalog. They found that in the redshift range used for the tracer galaxies, the \texttt{MAG\_AUTO} version of the \textsc{redMaGiC} catalog shows smaller correlations with observational systematics. 

We hence adopt \texttt{MAG\_AUTO} \textsc{redMaGiC} with high density as our fiducial tracer catalog. In a trade-off of signal and noise, we choose $z_T=0.2-0.45$ as the tracer redshift range. We derive weights for the correction of \textsc{redMaGiC} density for the effect of systematics as in \citet{wthetapaper}. We find significant correlations of \textsc{redMaGiC} density with $r$ band exposure time and seeing, and with $i$ band sky brightness. In the algorithm described in \autoref{sec:troughfinding}, we have applied these by dividing the fraction of good area in each pixel by the systematics weight that decorrelates \textsc{redMaGiC} density with these survey properties.

We do, however, test whether the choice of photometry pipeline (\texttt{MAG\_AUTO} or MOF) and the choice of whether we apply the systematics weight in our density splitting procedure makes a difference to our analysis. These tests are detailed in Appendix \ref{app:sysmaps} and show that the effect on the amplitude of our measured signals is negligible.

The redshift distribution of the tracer galaxy population, estimated by convolving the photometric redshift of each \textsc{redMaGiC} galaxy with its error estimate $\sigma_z\approx0.017\times(1+z)$ \citep{wthetapaper}, is shown as the grey contour in \autoref{fig:tofz}. Note that due to scatter in photo-$z$ this extends beyond the redshift range $z_T=0.2-0.45$ inside which these galaxies were selected.

As for other uses of \textsc{redMaGiC} for cosmology \citep{wthetapaper,gglpaper,keypaper}, we limit the catalog to the contiguous DES-SPT area of 1321 deg$^2$.

\subsubsection{Lensing source catalogs}
\label{sec:des_shapes}

Detailed descriptions and tests of the DES Y1 lensing source catalogs are presented in \citet{shearcat}, \citet{shearcorr} and \citet{gglpaper}, and the redshift distributions of source galaxies are estimated and calibrated in \citet{photoz,xcorr,xcorrtechnique}. We only give a brief summary of the two independent shape catalogs from DES Y1 here. 

The fiducial catalog with the larger number of source galaxies is based on the \textsc{metacalibration} method \citep{2017arXiv170202600H,2017ApJ...841...24S}. In this scheme, a Gaussian, convolved with the individual exposure point-spread function, is fit jointly to all single-epoch $r$, $i$, and $z$-band images of each galaxy. Galaxies are selected by the size and signal-to-noise ratio of the best fit, and the ellipticity of the Gaussian is used as an estimate of shear. Multiplicative biases in mean shear are caused by both the galaxy selection (selection bias) and the use of a maximum likelihood estimator with a simplified model (noise and model bias). In \textsc{metacalibration}, these are calibrated and removed using a repetition of the Gaussian fit on versions of the galaxy images that have been artificially sheared by a known amount. 

As a second catalog, we use \textsc{im3shape}, which produces a maximum likelihood estimate of shear based on a bulge or disc fit to all DES Y1 $r$ band images of each galaxy. Multiplicative biases in these estimators are calibrated using realistic images simulations of DES Y1 \citep{shearcat, des_sim_2017}.

Our estimator of tangential shear around overdense and underdense lines of sight, including the bias corrections, is described in \autoref{sec:measure_shear}. We use the galaxy selection criteria recommended in \citet{shearcat}. We split galaxies into redshift bins using the mean $z$ of the individual galaxy $p(z)$ as estimated by BPZ \citep{2000ApJ...536..571B,photoz}. We note that for the \textsc{metacalibration} catalog, we run BPZ on \textsc{metacalibration} measurements of galaxy fluxes (both on the original and artificially sheared images) to be able to correct for photo-$z$ related shear selection biases (see also section 3.3 of \citealt{photoz} and section IV.A.1 of \citealt{gglpaper}). The three source redshift bins we use are identical to the three highest redshift bins of \citet{photoz}, i.e.~with sources at mean $z=0.43-0.60,0.60-0.93,0.93-1.30$. Their redshift distributions, as estimated by BPZ using MOF photometry, are shown in \autoref{fig:tofz}. 

Uncertainties on residual multiplicative shear bias and on the mean values of the binned redshift distributions \citep{shearcat,photoz,xcorr,xcorrtechnique} are marginalized over in our analysis (see \autoref{sec:nuisance}).

\subsection{SDSS DR8 data}

\subsubsection{\textsc{redMaGiC} tracer catalog}
\label{sec:sdss_redmagic}

The tracer population in SDSS is the \textsc{redMaGiC} \citep{redmagic} high density sample, selected by SDSS photometry and cut to the same redshift range $z_T=0.2-0.45$. Despite this similarity, we will not assume in this work that SDSS and DES \textsc{redMaGiC} galaxies are the exact same populations.

SDSS has the benefit of an overlapping sample of galaxies with spectroscopic redshifts. We use this to calibrate the mean of the redshift distribution with clustering redshifts, independent of the photometric estimate, in \autoref{sec:redmagiczbias}. We find no significant bias, yet marginalize over the uncertainty in the analyses presented herein (see \autoref{sec:nuisance}).

\subsubsection{Lensing source catalogs}
\label{sec:sdss_shapes}
We use the shape and photometric redshift $p(z)$ catalog of \citet{2009ApJ...703.2217S} with minor modifications, identical to those in \citet{2015MNRAS.454.3357C}. We refer to these papers for details, but describe our source selection and priors on systematic uncertainties of shears and photometric redshifts below.

\begin{figure}
  \includegraphics[width=\linewidth]{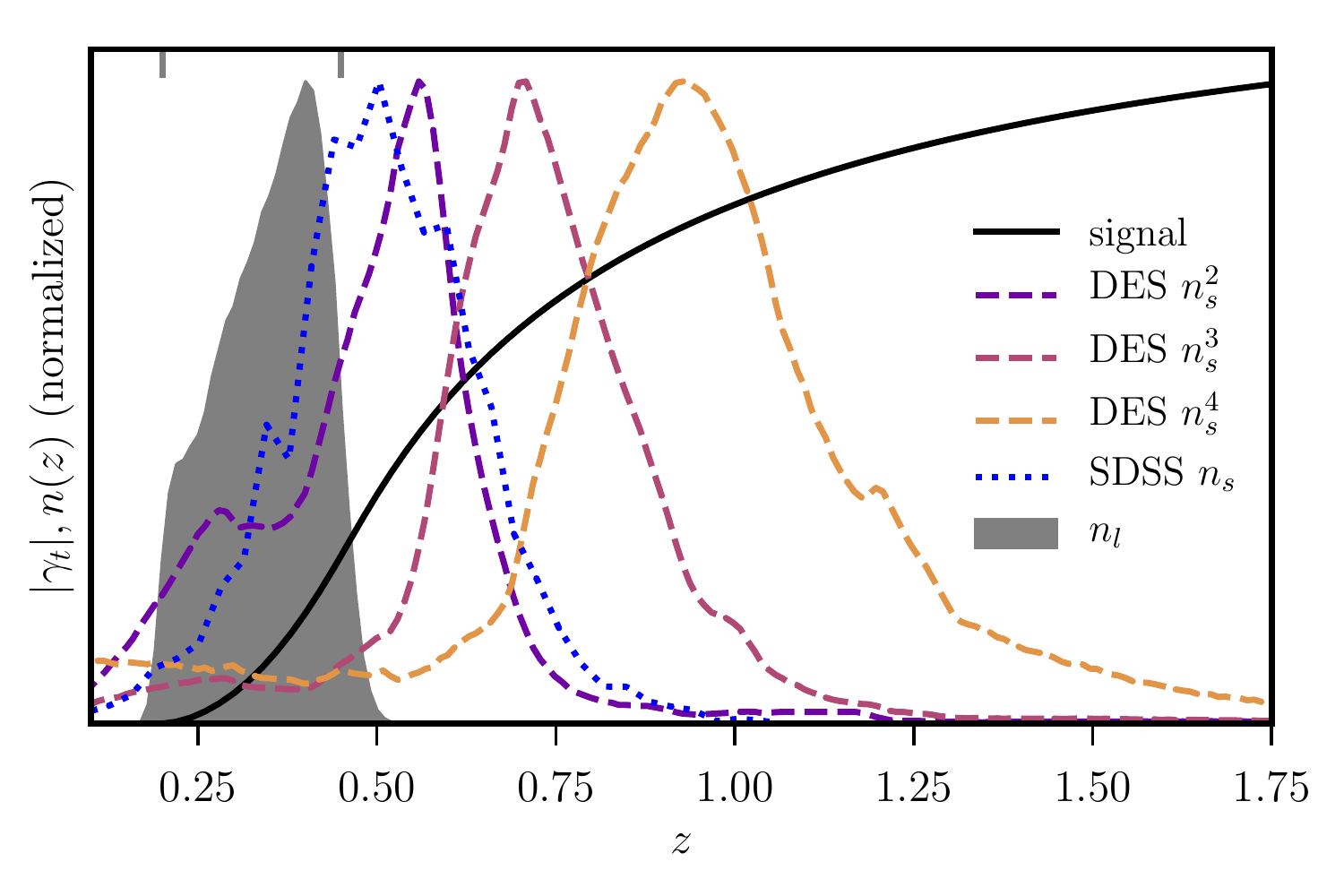}
  \caption{The redshift distributions of the DES source redshift bins (violet, red, yellow dashed lines) and SDSS sources (blue, dotted line), and of the \textsc{redMaGiC} tracer galaxies (grey shaded area) including scatter in their photometric redshift estimates. The dependence of predicted amplitude of the trough lensing signal on source redshift is shown by the black curve. Grey marks on upper axis indicate nominal redshift range of tracers galaxies $z_T=0.2-0.45$. Lines are normalized to match maxima and the trough signal is evaluated at $\theta=2\theta_T$, although the dependence of source redshift scaling on angular distance is minor.}
\label{fig:tofz}
\end{figure}

Due to the lower observational depth, the SDSS shape catalog peaks at much lower redshift than the one from DES Y1.
The source redshift dependence of the trough lensing signal (cf. \autoref{fig:tofz}) and complications arising from significant overlap of sources with the tracer redshift range
lead us to only use sources with a mean redshift estimate of $0.45\leq z<1.0$. We split these into four bins of $z=0.45-0.5, 0.5-0.55, 0.55-0.6$
and $0.6-1.0$. Within each of these bins, each individual source is assigned a minimum-variance relative weight (cf. \cite{2015MNRAS.454.3357C}) of
\begin{equation}
 w_i=[\sigma_{i, \rm shape, meas}^2+0.32^2]^{-1} \; .
 \label{eqn:sdssweights}
\end{equation}

Due to the moderate signal-to-noise ratio and redshift range of sources, we combine the four source redshift bins into one for the purpose of our final data vector. In this, we apply an optimal relative weighting of the bins as follows.

The predicted amplitude of shear around our troughs at $z_T=0.2-0.45$ (see black line in \autoref{fig:tofz}) scales with source redshift approximately as the amplitude of gravitational shear $\Sigma_{\rm crit}^{-1}$ (see \autoref{eqn:sigmacrit}) due to a lens at $z_d=0.36$.
We use the value of $\left\langle\Sigma_{\rm crit}^{-1}\right\rangle$ estimated for $z_d=0.36$ and the stacked $p(z)$ of each of these four bins to apply a relative weight of $\mathcal{W}_{\rm bin}=1,1.30,1.56$ and $1.81$ to each
of them. Because the number density of sources is steeply falling with source redshift in this range, the effective total relative weights of the four bins (equal to this $\mathcal{W}_{\rm bin}$ times the sum of all source $w_i$) are $1,0.697,0.453$ and $0.202$. 
We use these effective weights to combine the measured shear signal and the $n_s(z)$ from each of the four bins into a single source sample. 

As a calibration of the photometric estimate, the mean redshift of the sources is constrained by their angular cross-correlation with galaxies with known spectroscopic redshift (\autoref{sec:sourcezbias}).

\subsection{Measured signals}
\label{sec:signals}

Our data vector in this work contains two components, the modeling of which was extensively tested in \citet{Oliver}. In \autoref{sec:measure_shear}, we describe the measurement of gravitational shear signals around overdense and underdense lines of sight. Section \ref{sec:measure_cic} details our measurement of mean counts-in-cells in each density quintile in the presence of masking.

All measurements are made in jackknife resamplings of the survey. The covariance model constructed in \autoref{sec:covariance} can therefore be compared to an jackknife covariance. These were made based on 100 and 200 patches in the DES and SDSS footprint, respectively, defined by $k$-means clustering\footnote{https://github.com/esheldon/kmeans\_radec} of the tracer galaxies, an algorithm that splits the tracer galaxies into spatially compact subsets by their distance to the nearest among a set of centers, optimizing the center positions to minimize these distances. 

\subsubsection{Shear}
\label{sec:measure_shear}

The ellipticity of a galaxy is a pseudovector with two components, $e_1$ and $e_2$ that, for any lens position, can equivalently be described by a component tangential to a circle around a lens ($e_{\rm t}$) and by a component rotated by $\pi/4$ relative to that ($e_{\times}$). 

\begin{figure*}
\subfigure[split by density, $z_s=0.63-0.90$, $\theta_T=20'$]{
  \includegraphics[width=0.48\linewidth]{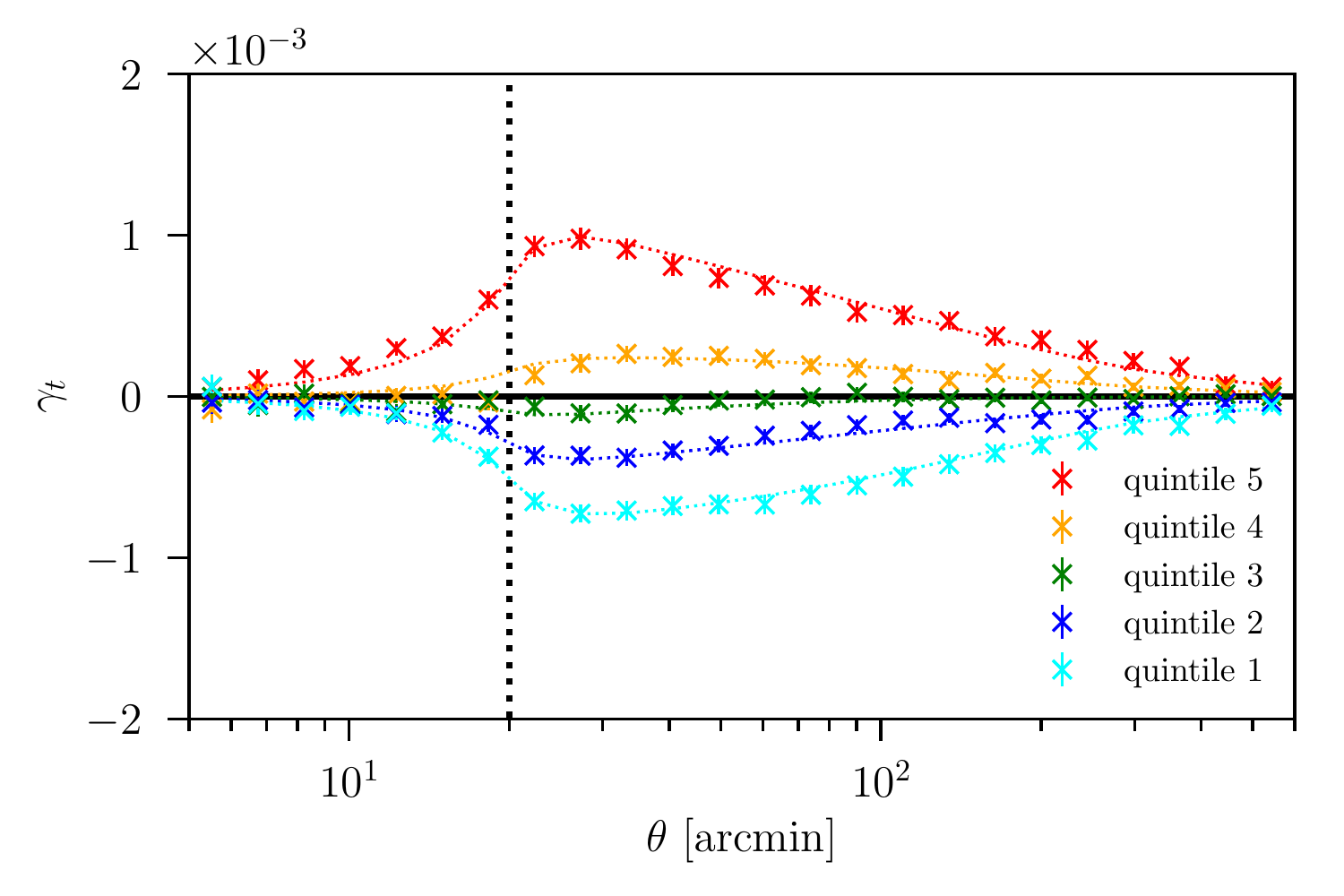}  
  \includegraphics[width=0.48\linewidth]{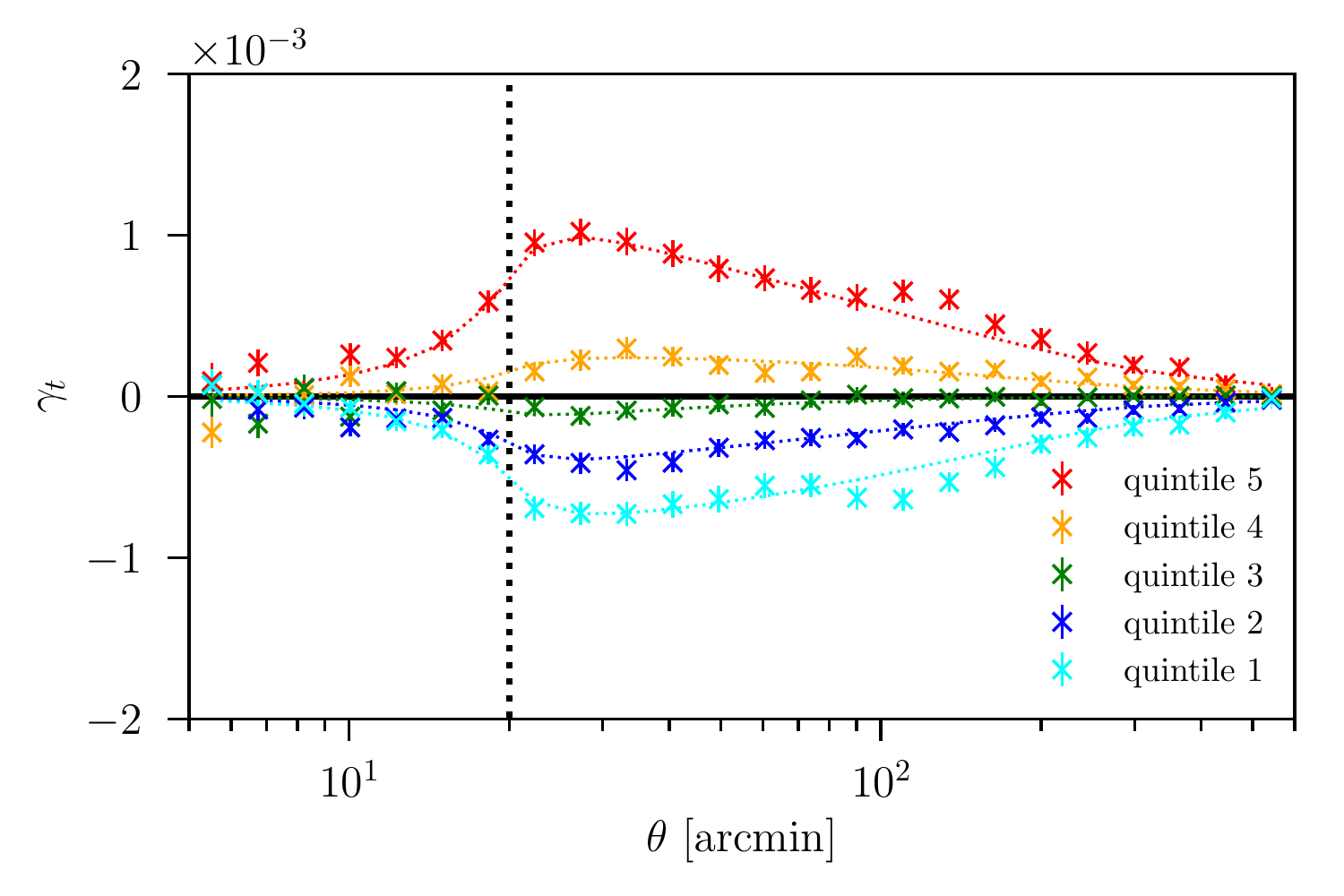} 
}
\subfigure[split by source redshift, $\theta_T=20'$, top/bottom quintile only]{
  \includegraphics[width=0.48\linewidth]{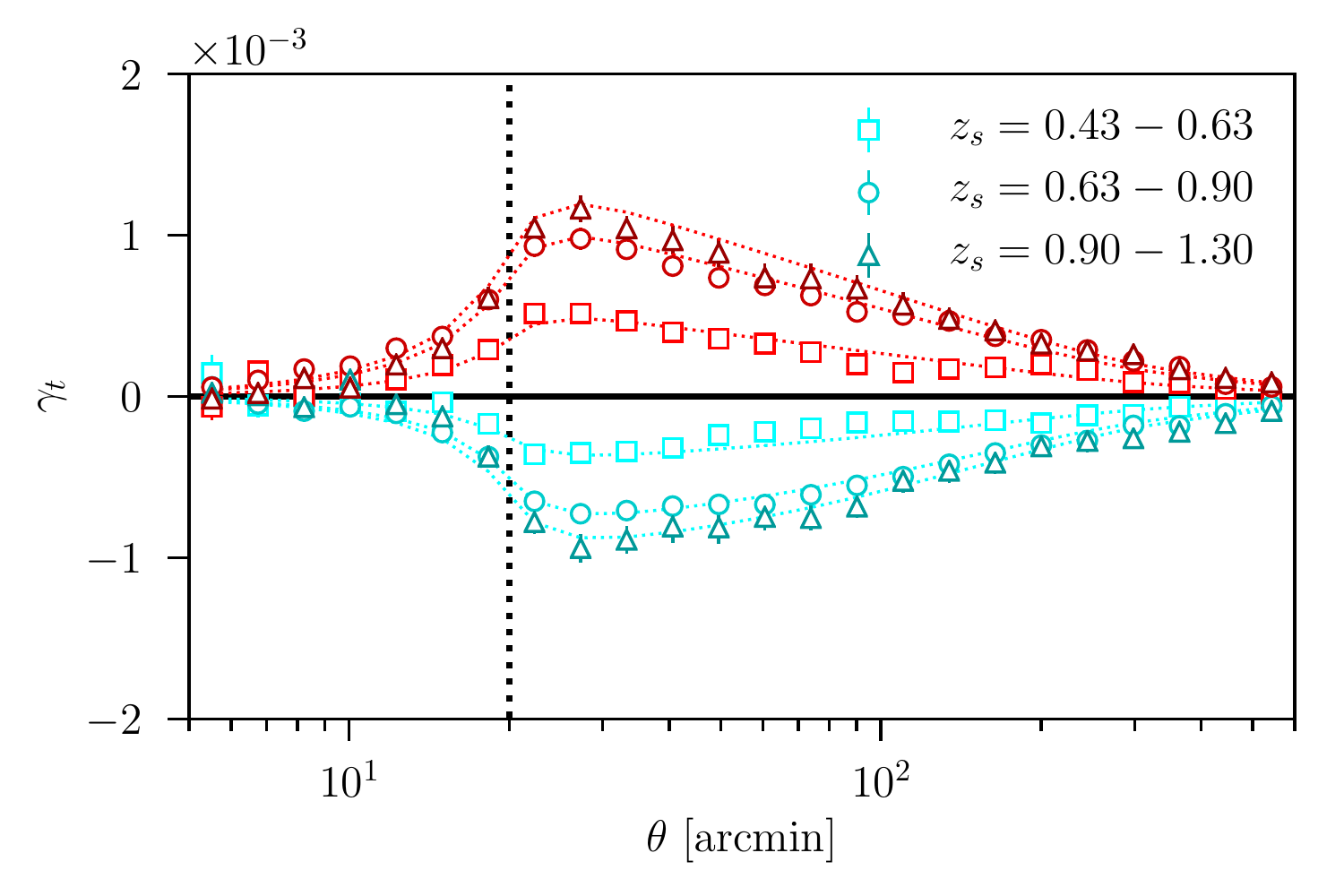} 
  \includegraphics[width=0.48\linewidth]{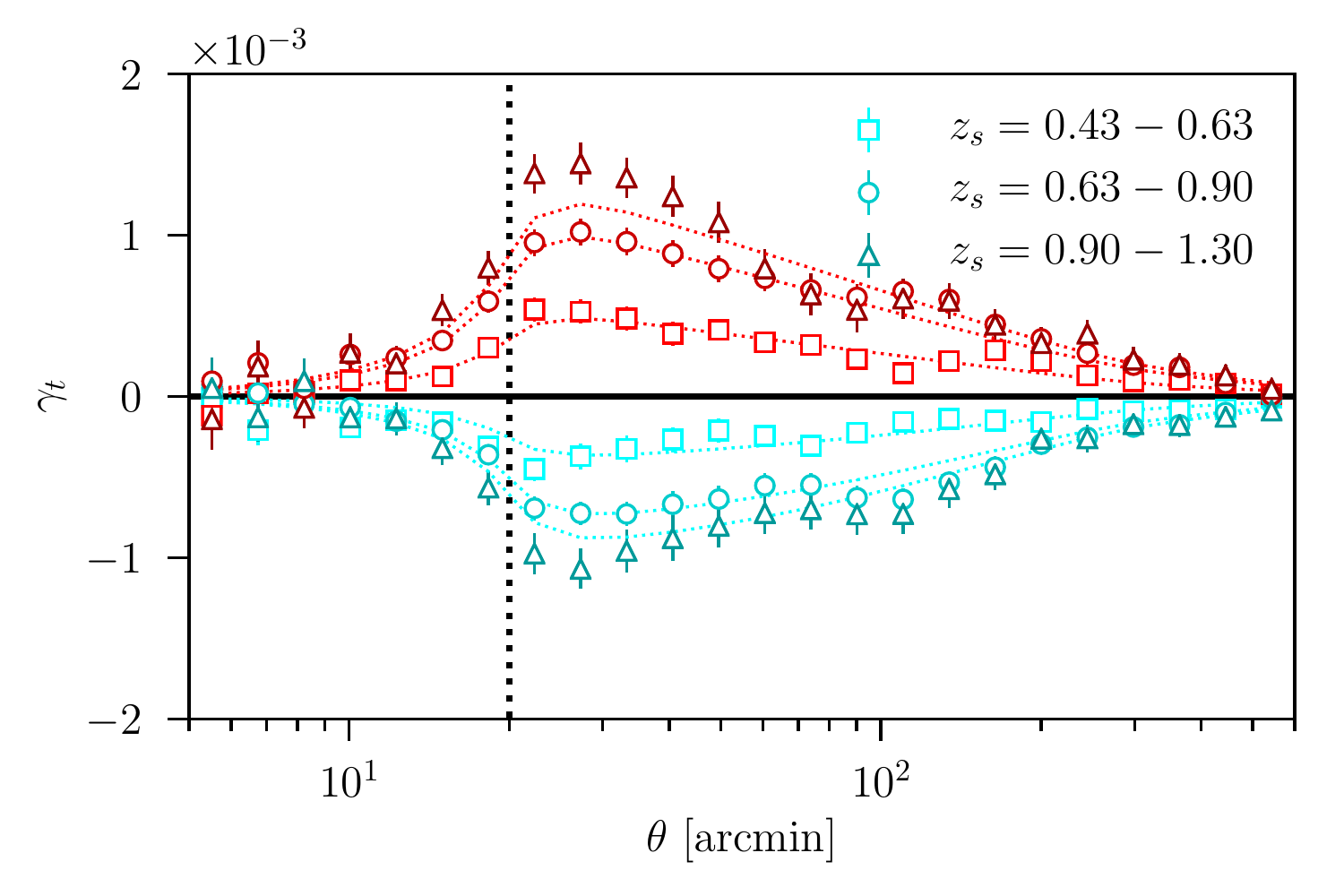}
}
\subfigure[split by smoothing scale, $z_s=0.63-0.90$, top/bottom quintile only]{  
  \includegraphics[width=0.48\linewidth]{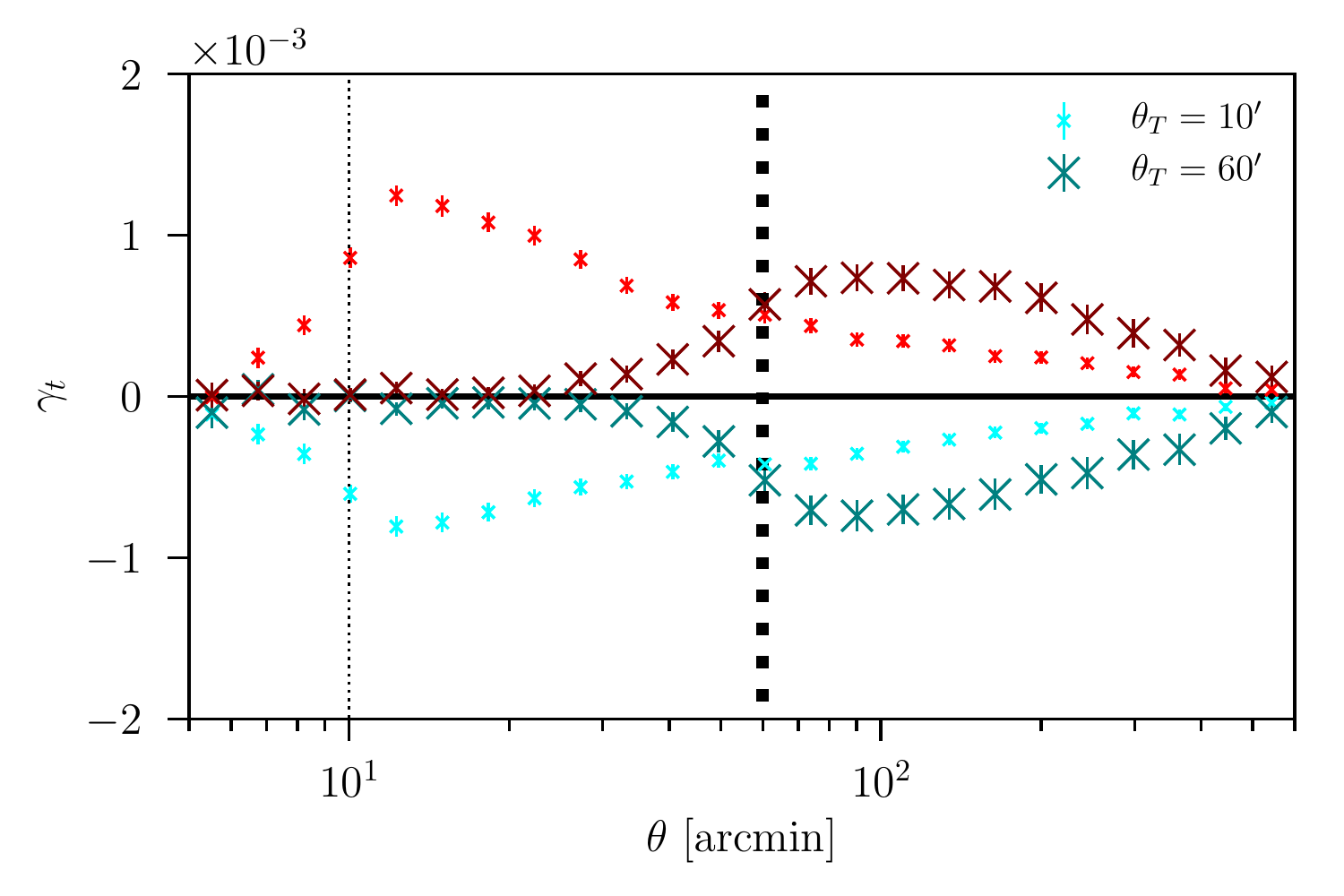}
  \includegraphics[width=0.48\linewidth]{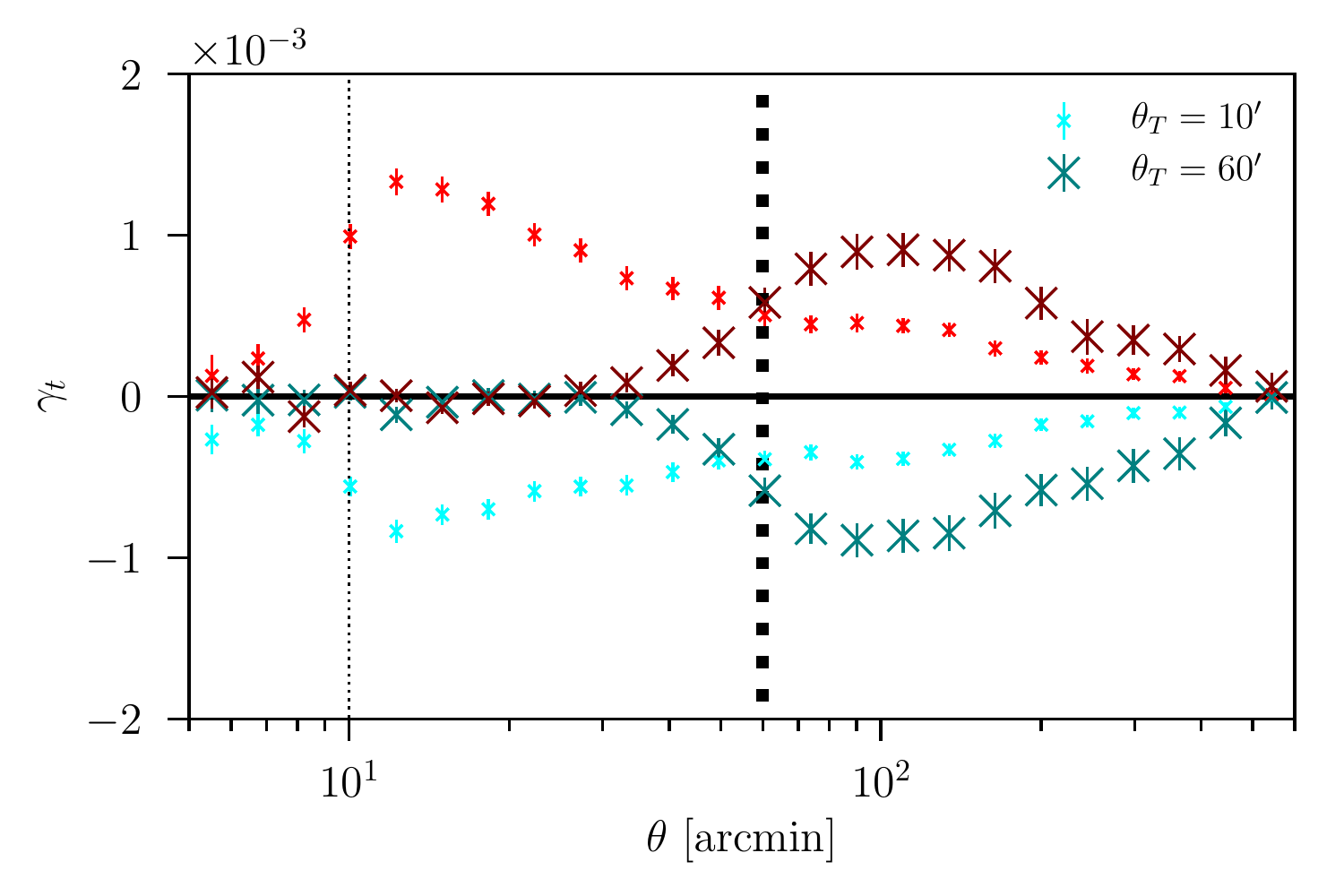}
}
  \caption{Shear signal around overdense and uncerdense lines of sight in DES Y1, split by line-of-sight density quintile (top row), source redshift $z_s$ (central row) from BPZ run on \metacal\ (left) and MOF (right) photometry, and aperture radius (bottom panel). Our fiducial data vector is the shear around the most underdense and most overdense quintile of $\theta_T=20'$ lines of sight as seen by sources in $z_s=0.63-0.90$. Left-hand panels: measurements with \metacal\ shears. Right-hand panels: measurements with \imshape. Error bars are from 100 jackknife resamplings of the survey, consistent with our model covariance (\autoref{sec:covariance}). Dotted lines indicate model prediction at maximum likelihood parameters (\autoref{sec:constraints}) and are a good fit to the data ($\chi^2=171$ (\metacal) and $201$ (\imshape) for $\approx200$ d.o.f., as determined only after unblinding).}
\label{fig:tomosignal}
\vspace{30pt}
\end{figure*}

Gravitational shear due to any single lens only affects the mean component of $e_{\rm t}$ for an ensemble of sources sampling a full annulus around the lens. As a function of angular separation $\theta$ from the lens, this effect is described by the tangential shear profile $\gamma_{\rm t}(\theta)$. For a single lens, the tangential shear profile is directly related to the azimuthally averaged, projected surface mass density $\Sigma(\theta)$ of the lens, i.e.~the projected mass per physical area, as
\begin{equation}
\gamma_{\rm t}(\theta) = \left[\langle\Sigma\rangle(<\theta) - \Sigma(\theta)\right]\times\Sigma_{\rm crit}^{-1} \equiv \langle\kappa\rangle(<\theta)-\kappa(\theta)\; ,
\label{eqn:gammasigma}
\end{equation}
where, in a flat universe, 
\begin{equation}
\Sigma_{\rm crit}^{-1}=
\frac{4\pi G}{c^2}\frac{\chi_d\left(\chi_s-\chi_d\right)}{\chi_s\left(1+z_d\right)}
\label{eqn:sigmacrit}
\end{equation}
is the inverse of the critical surface mass density and $\chi_{d,s}$ is the comoving distances to the deflector at redshift $z_d$ and the lensed source, respectively. For a set of lenses along the line of sight, the signal on any source is close to the sum of the effects of all lenses. One can still define a convergence $\kappa$ related to mean gravitational shear as in \autoref{eqn:gammasigma}, although it is no longer relatable to a uniformly weighted surface mass density (see \autoref{eqn:kappa}).

The relation between tangential shear and measured tangential ellipticity is less straightforward and depends on the implementation of the selection and measurement of source ellipticities. For the two schemes used on DES Y1, the responsivity $R=\mathrm{d}\langle e_{t}\rangle/\mathrm{d}\gamma_{\rm t}$ of observed ellipticity to applied tangential shear is calibrated very differently: for \metacal, it is estimated from versions of the actual galaxy images sheared with image manipulation algorithms; for \imshape, it is estimated from realistic simulations of DES imaging data (and usually defined as $m=R-1$). Both types of calibration contain an explicit or implicit correction for selection biases, i.e.~the shear dependence of the choice of whether to include a galaxy in the source sample.

\begin{figure}
  \includegraphics[width=\linewidth]{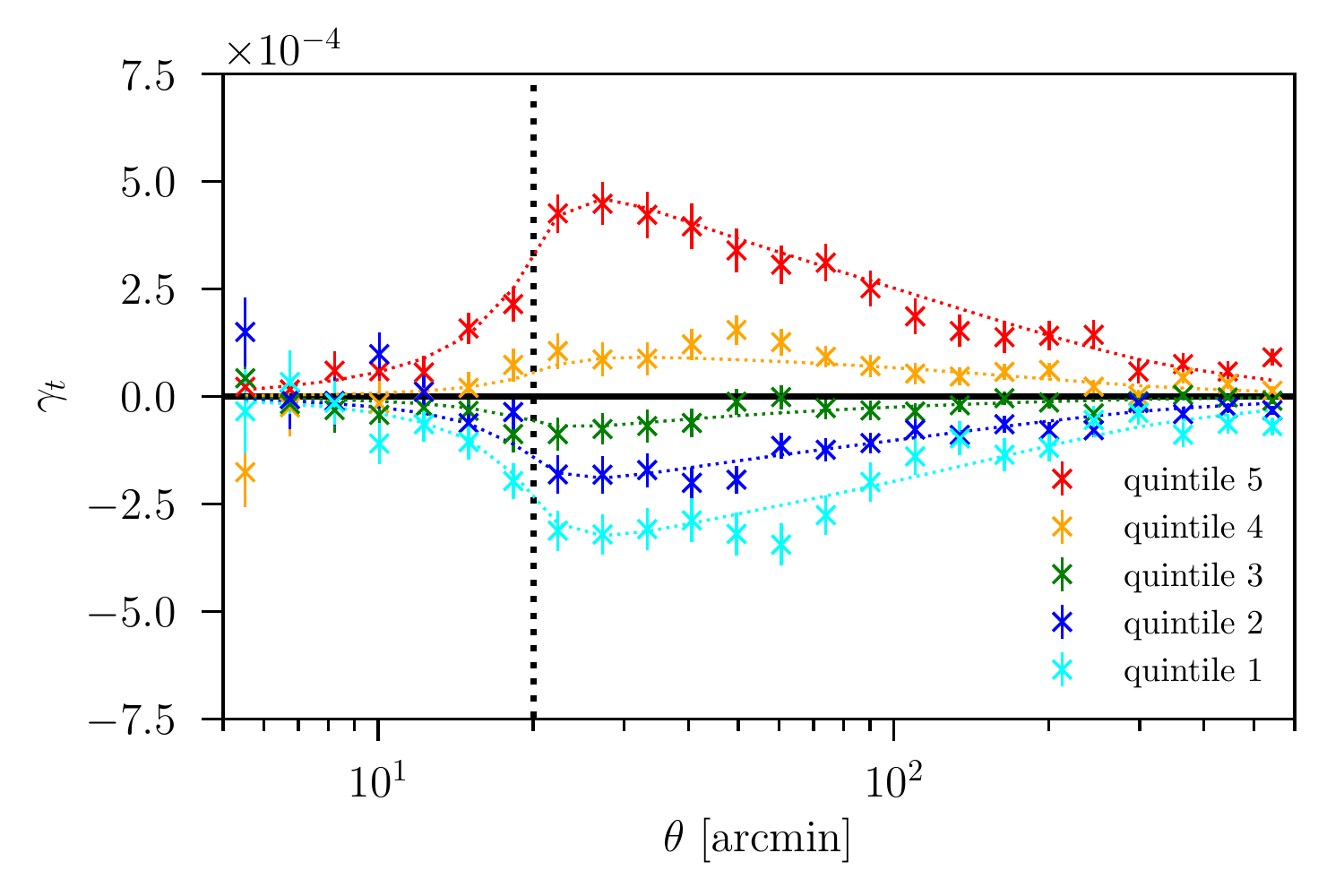}
  \caption{Shear signal around overdense and uncerdense lines of sight in SDSS, shown for all quintiles, with $\theta_T=20'$ and a single source bin of $z_s=0.45-1$. Error bars are from 200 jackknife resamplings of the survey. Dotted lines indicate model prediction at maximum likelihood parameters (\autoref{sec:constraints}) and are a good fit to the data ($\chi^2=81$ for $\approx70$ d.o.f., as determined only after unblinding). For comparison with \autoref{fig:tomosignal}, note the changed $\gamma_t$ axis scale.}
\label{fig:sdsssignal}
\end{figure}

For \metacal, we define the estimator $\hat{\gamma}_{\rm t}^q$ of mean tangential shear around lines of sight $i$ with probability $w_i^q$ to be in a given density quintile $q$ as
\begin{equation}
\hat{\gamma}_{\rm t}^q=\hat{\gamma}_{\rm t}^{q, \rm signal}-\hat{\gamma}_{\rm t}^{\rm random}=\frac{\sum_{i,j} w_i^q e_{ij, \rm t}}{R\sum_{i,j} w_i^q} - \frac{\sum_{i,j} e_{ij, \rm t}}{R\sum_{i,j} 1}\; ,
\end{equation}  
where $e_{ij, \rm t}$ is the ellipticity of source $j$ in the tangential direction around line of sight $i$, the sums run over all lines of sight $i$ in the mask of the density-split sky and all sources $j$ in an angular bin around each line of sight. The second term subtracts shear around random lines of sight -- for our statistic, these are all healpix pixels around which the masked fraction of area $f_{\rm mask}<f_{\rm mask}^{\rm max}$ (cf. \autoref{sec:troughfinding}). $R$ is the sum of shear and selection responsivity,
\begin{eqnarray}
R=R_{\gamma}+R_{\rm S}&=&
\frac{1}{2}\left\langle\frac{e_1^+-e_1^-}{2\Delta\gamma_1}+\frac{e_2^+-e_2^-}{2\Delta\gamma_2}\right\rangle \nonumber \\ &+&\frac{1}{2}\left(\frac{\langle e_1\rangle^+-\langle e_1\rangle^-}{2\Delta\gamma_1}+\frac{\langle e_2\rangle^+-\langle e_2\rangle^-}{2\Delta\gamma_2}\right)\; ,\nonumber \\
\end{eqnarray}
where superscripts $\pm$ on $e$ indicate an ellipticity measured on an image artificially sheared by $\Delta\gamma$ in the same component and superscripts $\pm$ on $\langle\ldots\rangle$ indicate an average taken on an ensemble of source \emph{selected} by quantities measured on an image artificially sheared by $\Delta\gamma$.

We note that this is identical to the methodology for DES Y1 galaxy-galaxy lensing employed in \citet{gglpaper}, except that we estimate the responsivity separately for the source galaxies in each radial bin around the cluster, rather than as a global scalar. The scale dependence, however, is negligible -- the \metacal\ R is equal to within 0.5 per-cent for any two angular bins. As in other DES Y1 lensing analyses \citep{gglpaper,shearcorr}, we weight all sources in a bin uniformly -- using the inverse variance of the shape measurement underlying the \textsc{metacalibration} scheme as a weight would require a re-derivation of the redshift calibration \citep{photoz} and additional bookkeeping for selection bias correction, yet increases signal-to-noise ratio only mildly.

For \imshape, we use the source weights $W_j$ defined in \citet{shearcat} to first measure the weighted mean $R$ of the source sample, then define the estimator for tangential shear as
\begin{equation}
\hat{\gamma}_{\rm t}^q=\frac{\sum_{i,j} w_i^q W_j e_{ij,\rm t}}{R\sum_{i,j} w_i^q W_j} - \frac{\sum_{i,j} W_j e_{ij,\rm t}}{R\sum_{i,j} W_j}\; .
\end{equation}
The \imshape\ $e$ are defined with the calibration correction for additive bias already applied.

For SDSS, multiplicative bias is already corrected in the source catalog. We therefore measure tangential shear with the above equation by setting $R=1$, and use weights $W_j=w_j\times \mathcal{W}_{\rm bin}$ (see \autoref{eqn:sdssweights} and subsequent description).

The measured shear signals are shown for DES in \autoref{fig:tomosignal}, slicing the data by density percentile, source redshift, and smoothing scale of the tracer galaxy field. SDSS signals are in \autoref{fig:sdsssignal}.

\subsubsection{Counts-in-cells}
\label{sec:measure_cic}

\begin{figure}
  \includegraphics[width=\linewidth]{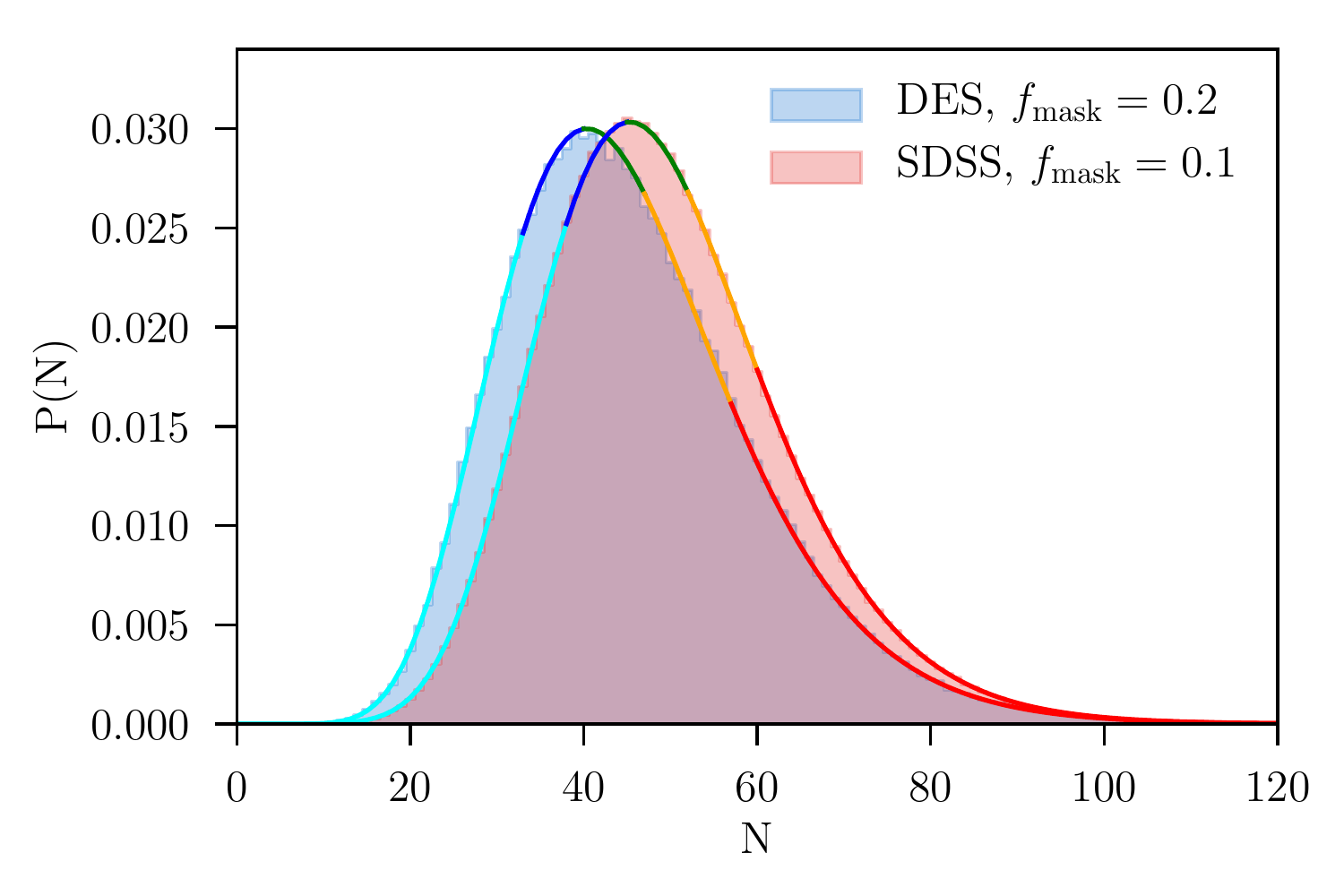}
\caption{Counts-in-cells distribution for \textsc{redMaGiC} $z_T=0.2-0.45$ galaxies in circular $\theta_T=20'$ top-hat apertures for DES with 20 percent masking fraction (blue) and SDSS with 10 percent masking fraction (red). Lines indicate prediction of $P(N)$ for the maximum likelihood model fit to the lensing and counts-in-cells data in the $(b,\alpha_0,\alpha_1)$ bias model. The color of the line denotes quintiles, in the color scheme of \autoref{fig:sdsssignal}, i.e.~the integral under each colored segment is 0.2. The mean count in each quintile is part of the data vector and consistent with our best-fit model, as checked via its $\chi^2$ after unblinding.}
\label{fig:pofn}
\end{figure}

The discriminating power of density split lensing signals for cosmological parameters and parameters describing the connection of galaxies and matter is greatly improved by adding some degree of information of galaxy clustering or bias. Here, we use a very basic statistic, the mean tracer galaxy overdensity in our density quintiles, that was extensively tested in \citet{Oliver} -- other signals could significantly improve the constraining power in the future.

Operationally, we define the mean tracer galaxy overdensity in all quintiles $q$ as follows. We convert the raw tracer galaxy count $N^{\rm raw}_i$ within the aperture radius around each line of sight $i$ to a stochastically masked count $N_i$ with fixed masking fraction by a Bernoulli draw (\autoref{sec:troughfinding} and \autoref{app:bernoulli}). We then order lines of sight by $N_i$ and take the mean of $N_i$ in each quintile $q$ of that list as $\langle N_i\rangle^q$.  The mean tracer galaxy overdensity in quintile $q$ is
\begin{equation}
C^q=\frac{\langle N_i\rangle^q}{\langle N_i\rangle} \; ,
\end{equation}
where the average in the denominator runs over all lines of sight.

We note that this does account for the fact, in a stochastic fashion, that a given line of sight can end up in different density quintiles depending on the realization on masking that decides the galaxy count.

\autoref{fig:pofn} shows the full $P(N)$ distribution in both DES and SDSS, alongside a model evaluated at the maximum likelihood parameter values fit to the shear signal and mean tracer galaxy overdensity in quintiles. The model not only fits these mean overdensities, but also the full $P(N)$ extremely well: absolute differences in probabilities of finding $N$ galaxies in a random line of sight, $|P^{\rm model}-P^{\rm data}|$, are below $10^{-3}$ and $3\times10^{-4}$ for any $N$ in DES and SDSS, respectively. The bias model used for the plot is one with two-parametric stochasticity (called $b,\alpha_0,\alpha_1$ in \autoref{sec:modeltracers}), although even a simpler model can reproduce the $P(N)$ well.

\section{Model}
\label{sec:model}

In order to describe our signal as a function of
\begin{itemize}
\item cosmological parameters,
\item parameters that connect galaxy counts to the matter (over)density, and
\item nuisance parameters,
\end{itemize}
we use the model developed and tested in \citet{Oliver}. We only briefly summarize it here, with an emphasis on required extensions for the use on observational data, and refer the reader to that paper for details.

Let $\hat{\mathbf{n}}$ be a unit vector on the sky. The signal we have to predict in this work is the shear profile around lines of sight that fall into a certain quintile of foreground tracer density. Also, our data vector includes the average tracer density contrast in each of those density quintiles. To model these two parts of our data vector, we have to consider the following fields on the sky:
\begin{trivlist}
\item[$\bullet$]\underline{$\delta_{m,2D}(\hat{\mathbf{n}})$:} the line-of-sight density contrast underlying our tracer galaxies. Given the redshift distribution $n_l(z)$ of our tracer sample, this is given by
\begin{equation}
\delta_{m,\mathrm{2D}}(\hat{\mathbf{n}}) = \int \mathrm{d}\chi\ q_l(\chi)\ \delta_{m,\mathrm{3D}}(w\hat{\mathbf{n}}, \chi)\ ,
\end{equation}
where $\chi$ is comoving distance and the projection kernel $q_l(\chi)$ is given in terms of $n_l(z)$ as
\begin{equation}
q_l(\chi) = n_l(z[\chi]) \frac{\mathrm d z[\chi]}{\mathrm d \chi}\ .
\end{equation}

\item[$\bullet$]\underline{$\delta_{m,T}(\hat{\mathbf{n}})$:}
the result of smoothing the field $\delta_{m,\mathrm{2D}}(\hat{\mathbf{n}})$ with a circular top-hat aperture $T$.

\item[$\bullet$]\underline{$N_{T}(\hat{\mathbf{n}})$:}
the number of tracer galaxies in the aperture $T$ around the line-of-sight $\hat{\mathbf{n}}$

\item[$\bullet$]\underline{$\kappa_{<\theta}(\hat{\mathbf{n}})$:} the convergence inside an angular radius $\theta$ around the line-of-sight $\hat{\mathbf{n}}$.
\end{trivlist}
Because the Universe is isotropic, we will omit the dependence on $\hat{\mathbf{n}}$, i.e.~only consider a single line of sight.

As detailed in \citet{Oliver}, the density split lensing signal can be calculated from the convergence profile around lines of sight with a fixed value of $N_T$. This profile can be computed as
\begin{equation}
\langle \kappa_{<\theta} | N_T\rangle = \int \mathrm{d}\delta_{m,T}\; \langle \kappa_{<\theta} | \delta_{m,T} \rangle \; p(\delta_{m,T}|N_T) 
\end{equation}
where Bayes' theorem can be used to express the PDF of $\delta_{m,T}$ at fixed $N_T$ as
\begin{equation}
p(\delta_{m,T}|N_T) = \frac{P(N_T|\delta_{m,T})\; p(\delta_{m,T})}{P(N_T)}\ .
\end{equation}
Here $p(\delta_{m,T})$ is the overall PDF of $\delta_{m,T}$, $P(N_T|\delta_{m,T})$ is the probability of finding $N_T$ in a line-of-sight with fixed $\delta_{m,T}$ and
\begin{equation}
P(N_T)=\int \mathrm{d}\delta_{m,T} \; P(N_T|\delta_{m,T}) \; p(\delta_{m,T}) \; .
\end{equation}
From the convergence profile $\langle \kappa_{<\theta} | N_T\rangle$ the corresponding shear profile can be computed as (cf. \citeauthor{Oliver})
\begin{equation}
\label{eq:gamma_in_terms_of_kappa}
\langle \gamma_t(\theta)| N_T\rangle = \frac{\cos\theta - 1}{\sin \theta} \frac{\mathrm d}{\mathrm d \theta} \langle \kappa_{<\theta} | N_T\rangle
\end{equation}
and the shear profile around a certain quintile of tracer density is given by the average of $\langle \gamma_t(\theta)| N_T\rangle$ over the values $N_T$ occurring in that quintile (cf. \citeauthor{Oliver} for further details). Fundamentally, we therefore have to model
\begin{trivlist}
\item[$\bullet$]\underline{$p(\delta_{m,T})$,} the PDF of matter density smoothed inside our aperture,
\item[$\bullet$]\underline{$\langle \kappa(<\theta) | \delta_{m,T} \rangle$,} the expectation value of convergence inside an angular radius $\theta$ around a line-of-sight with given density contrast $\delta_{m,T}$ inside our aperture, and
\item[$\bullet$]\underline{$P(N_T|\delta_{m,T})$,} the probability of finding $N_T$ galaxies in an aperture, given its density contrast is $\delta_{m,T}$.
\end{trivlist}

We describe our approaches on each of these ingredients in the following subsections, and close with a description of how we account for biases in source redshift and shear estimates, and overlap of the source redshift distribution with the tracer redshift distribution.

In all these steps, in order to predict the nonlinear 3D matter power spectrum, we use the \citet{2012ApJ...761..152T} halofit approximation with the \citet{1998ApJ...496..605E} transfer function with baryonic features, which is sufficiently accurate given our large scale binning.

\subsection{PDF of matter density contrast}

The PDF $p(\delta_{m,T})$ can be computed from its cumulant generating function (CGF). This function can be derived at tree-level in perturbation theory with the help of the the cylindrical collapse model (\cite{Oliver}, see also pioneering work on the computation of the CGF in \cite{Bernardeau1994,Bernardeau2000,Valageas2002}).

The computations are numerically involved and, at least in our implementation, too slow for application in a likelihood analysis. We however show in \citeauthor{Oliver} that, on the scales used in this work, the perturbation theory computation of $p(\delta_{m,T})$ is well approximated by a log-normal distribution that matches the second and third moments $\langle\delta_{m,T}^2\rangle$ and $\langle\delta_{m,T}^3\rangle$ of the perturbation theory approach. We use this log-normal model \citep[section 4.1.1 in ][]{Oliver} for the smoothed, projected matter density field in this work.

\subsection{Mean convergence around apertures with fixed density contrast}
\label{sec:modelconvergence}
We now turn to the convergence field $\kappa$, defined as 
\begin{equation}
\kappa(\hat{\mathbf{n}}) = \int \mathrm{d}\chi\ W_{s}(\chi)\ \delta_{m,\mathrm{3D}}(\chi\hat{\mathbf{n}}, \chi)\ ,
\label{eqn:kappa}
\end{equation}
where  the lensing efficiency $W_{s}$ is given by
\begin{equation}
W_{s}(\chi) = \frac{3 \Omega_m H_0^2}{2c^2} \int_\chi^\infty \mathrm d \chi'\ \frac{\chi (\chi'-\chi)}{\chi'\ a(\chi)}\ q_s(\chi')\ ,
\end{equation}
and 
\begin{equation}
q_s(\chi)=n_s(z[\chi])\frac{\mathrm{d}z[\chi]}{\mathrm{d}\chi}
\end{equation}
is the line-of-sight density of the sources. As before, denote by $\kappa_{<\theta}$ the result of smoothing the convergence field over circles of angular radius $\theta$.

As described in \citet{Oliver} \citep[see also][and references therein]{Bernardeau2000}, the expectation value of $\kappa_{<\theta}$ around lines of sight with fixed values of $\delta_{m, T}$ is mostly determined by the moments
\begin{equation}
\langle \delta_{m,T}^2\rangle, \langle \delta_{m,T}^3\rangle
\end{equation}
as well as the mixed moments
\begin{equation}
\langle \delta_{m,T}\ \kappa_{<\theta} \rangle, \langle \delta_{m,T}^2\ \kappa_{<\theta}\rangle\ .
\end{equation}
In a similar way as for the projected density PDF, a full tree-level computation of $\langle \kappa_{<\theta} | \delta_{m,T}\rangle$ can be replaced by a log-normal approximation that involves the above moments (cf. \citeauthor{Oliver} for details of this). We want to stress, that this does \emph{not} mean that we employ a log-normal approximation for the joint PDF of $\delta_{m,T}$ and $\kappa_{<\theta}$. E.g. \citet{2016MNRAS.459.3693X} have shown that such an approximation can be inaccurate if the lensing kernel $W_{s}(\chi)$ and the line-of-sight distribution of tracers $q_l(\chi)$ have strongly different widths in comoving distance. Rather, we model the convergence field as a sum of two fields, one of which is a log-normal random field and one of which is Gaussian and uncorrelated to $\delta_{m,T}$. Also, unlike for a joint log-normal distribution, we allow the log-normal parameter of $\kappa_{<\theta}$, i.e.~the minimum allowed value of $\kappa_{<\theta}$, to depend on the scale $\theta$. In \citet{Oliver} we have shown that this indeed gives a good approximation to the joint statistical properties of convergence and density contrast.

\subsection{Probability of galaxy counts in apertures with fixed density contrast}
\label{sec:modeltracers}

Finally, we need to model the probability of finding $N_T$ galaxies inside an aperture given the matter density contrast $\delta_{m,T}$. As defined in \citet{Oliver}, we consider three models of increasing complexity. All of them assume bias to be linear, i.e.~the mean count of galaxies to be proportional to the overdensity of matter in the large aperture volumes we consider. They differ, however, in their parametrization of stochasticity \citep{1999ApJ...520...24D}. We note that the latter may arise arise from nonlinear biasing on scales smaller than our apertures or from truly non-Poissonian noise in galaxy density at fixed matter density that is present in subhalo distributions \citep{2010MNRAS.406..896B,2015ApJ...810...21M,2017MNRAS.472..657J}.

In all equations below, $\bar{N}$ denotes the mean count of tracer galaxies inside apertures after masking a fraction $f_{\rm mask}^{\rm max}$ of area, and the generalized Poisson distribution that is also defined for noninteger arguments is 
\begin{equation}
\mathrm{Poisson}(N,\bar{N})=\exp[N\ln\bar{N}-\bar{N}-\ln \Gamma(N+1)] \; ,
\end{equation}
with the Gamma function $\Gamma$. 

Our three models are:
\begin{itemize}
\item \textbf{bias only: $b$ model} -- as in \citet{2016MNRAS.455.3367G}, one could assume $P(N_T)$ to be a Poisson distribution of a nonstochastic tracer population with bias $b$, \begin{equation}P(N_T|\delta_{m,T})=\mathrm{Poisson}(N_T,\bar{N}(1+b\delta_{m,T}))\; .\end{equation}
\item \textbf{bias and stochasticity: $b,r$ model} -- in this case, the galaxy count is assumed to be distributed as \begin{equation}P(N_T|\delta_{g,T})=\mathrm{Poisson}(N_T,\bar{N}(1+\delta_{g,T}))\; ,\end{equation} where $\delta_{g,T}$ is an auxiliary galaxy density field with \begin{equation}\left\langle\delta_{g,T}^n\right\rangle=b^n\left\langle\delta_{m,T}^n\right\rangle\; .\end{equation} The auxiliary field is correlated with the smoothed matter density field with a correlation coefficient $r$. Setting $r=1$ reduces this to the $b$ model with no stochasticity.
\item \textbf{bias and density dependent non-Poissonianity: $b,\alpha_0,\alpha_1$ model} -- because it introduces independent scatter, stochasticity with $r<1$ boosts the shot noise in galaxy count at fixed matter density; yet a dependence of this super-Poissonianity on matter density that may be present in the data need not be fully described by the $b,r$ model; to account for this, we use a more general model defined in \citet{Oliver}. Here, 
\begin{eqnarray}
P(N_T|\delta_{m,T})=\alpha^{-1}(\delta_{m,T})\times \nonumber \\ \mathrm{Poisson}[N_T/\alpha(\delta_{m,T}),\bar{N}(1+b\delta_{m,T})/\alpha(\delta_{m,T})]\;  .
\end{eqnarray}
We note that this model can be related to the halo count and occupation distributions \citep{2011MNRAS.415..153F}. Our ansatz can be thought of as a model of Poisson-distributed haloes with $\alpha$ \textsc{redMaGiC} galaxies in each one of them, similar to e.g.~the relation of Poissonian photon and non-Poissonian electron shot noise in CCD detectors, described by a gain factor $\alpha$.  It could similarly accommodate non-Poissonianity in halo counts \citep{2014MNRAS.441..646N}. We allow for $\alpha$ to be different in higher and lower density regions, e.g.~because more massive haloes might be more common in the former, by means of a linear dependence of $\alpha$ on $\delta_{m,T}$ as \begin{equation}\alpha(\delta_{m,T})=\alpha_0+\delta_{m,T}\alpha_1\;.\end{equation}
\end{itemize}

We note that a bias model without stochasticity is a common assumption made for the galaxy distribution on large scales \citep[e.g.][]{keypaper}. \citet{Niall} show that in the \emph{Buzzard} simulations, large scale stochasticity is present. From the combination of probes with different sensitivity to $b$ and $r$, such as the three galaxy and convergence auto- and cross-correlation functions, the two parameters could be disentangled. Density split statistics, in addition, are sensitive to differences in higher moments of the galaxy and matter density field, and can test and, potentially, constrain, more complex models such as $b,\alpha_0,\alpha_1$. 

\subsection{Nuisance effects on data}

In all runs on data, biases $\Delta z$ in the means of redshift distributions in DES and SDSS are accounted for at the level of the model: we marginalize over lens redshift and (multiple, in the case of a tomographic analysis) source redshift bias parameters $\Delta z$ by shifting the tracer galaxy and source galaxy redshift distributions accordingly before computing predictions for the signals. Likewise, we scale the predicted shear signal by $(1+m)$ to account for multiplicative shear biases $m$.

A more complex issue arises from the clustering of sources with the overdense and anticorrelation of sources with the underdense lines of sight. This is a common problem in cluster lensing or galaxy-galaxy lensing, accounted for by so-called \emph{boost factors} \citep{Sheldon04.1,Mandelbaum2005}. 

In the case of density split lensing, we apply the assumption of linear bias to predict the radius dependence of boost factors and their effect, given the non-thin lenses, on our model predictions. For a given tracer redshift distribution and a the matter field at redshift $z_s$, the angular clustering $w_q(\theta,z_s)$ of quintile $q$ with matter can be calculated with the same formalism as the convergence in \autoref{sec:modelconvergence}. Assuming a linear bias of source galaxies $b_s$, their redshift distribution at separation $\theta$ from quintile $q$ changes due to clustering to
\begin{eqnarray}
n_s(z) \rightarrow n_{s,q}(z, \theta) = \left[1+b_s w_q(\theta,z)\right] \, n_s(z) \times \nonumber \\ \left[\int\mathrm{d}z_s\; \left[1+b_s w_q(\theta,z_s)\right] n_s(z_s)\right]^{-1} \; .
\end{eqnarray}
The lowest redshift bin in DES Y1 or the \emph{Buzzard} simulations and the sources in SDSS have sufficiently strong overlap with the lens redshift distribution that we include this effect in the modeling and marginalize over $b_s$ in the analysis. This means that we use a different source redshift distribution for predicting each point of the density split, radially binned shear signal data vector. While $b_s$ is in reality a function of $z_s$, one can very accurately describe the deboosting of the lensing signals by an effective $b_s$ because the radial profile shape of the shear signal is almost independent of source redshift.

We note that in this derivation we neglect a second, but likely subdominant effect: the source redshift dependence of the probability of failing to include a source in the DES shape catalogs due to blending, that might cause a similar density dependence of source $n_s(z)$.

A potential spurious signal is due to intrinsic alignment of physical source galaxy shapes with the underdense or overdense lines of sight due to gravitational interactions (see \cite{Troxel20151,2015SSRv..193....1J} for a review).
For cross-correlations between the positions of object and gravitational shear, such as counts and lensing in cells, intrinsic alignments affect only the signal from source galaxies that are physically associated with the lensing objects, i.e.~if redshift distributions of source galaxies and lensing objects overlap. This is the case primarily in the lowest redshift bin for DES Y1. Hence test (4) in Sect.~\ref{sec:tests}, which demonstrates the robustness of the results to removing the lowest source redshift bin from the data vector, indicates that the current analysis is at most weakly affected by intrinsic alignments. This is in agreement with our expectation that the tidal alignment of galaxies with the comparatively small mean over- and underdensities of our density quintiles is small at $\theta>20'$ separation.

On small scales, baryonic effects can modify the matter power spectrum from its dark matter only prediction, primarily by affecting overdense regions. For our statistic, this could be absorbed by the bias model on scales smaller than the top-hat aperture $\theta_T$. The shear signal is used on scales larger than $\theta_T=20'$ only, and parameter constraints are robust to a more conservative scale cut of $\theta>40'$ (\autoref{sec:tests}). We hence do not expect a significant impact of baryonic effects on the parameter constraints at the accuracy level of the current analysis, but note that these effects require further study for future, more constraining analyses.

\section{Covariance}
\label{sec:covariance}

In order to interpret our measurements, we need an accurate description of their covariance. We construct this covariance from a large number of mock realizations of our data vectors. In that, we make use of the fact that the noise in our measurements can be separated into two components: a contribution from shape noise and a contribution from large scale structure and shot noise in the galaxy catalog. This approach is similar to the one of \citet{2017arXiv170701907M}.

In the following, we describe how we measure these contributions, and how we combine them into a covariance matrix. 

We assume in all following analyses that the signals measured in SDSS and DES Y1 are uncorrelated, justified by the fact that the survey footprints (using only the contiguous SPT region of DES) are well separated.

\subsection{Shape noise}
\label{sec:shapenoise}
The primary contribution to the shape we measure for any individual galaxy in our survey is the sum of its intrinsic shape and measurement noise, not the weak gravitational shear that distorts the galaxy image.

Because of this dominance of the noise over the signal, and because the intrinsic shapes of neighboring galaxies are almost uncorrelated, we can measure shape noise by rotating each galaxy in our shape catalog by an independent random angle. The shear signal around our actual underdense and overdense lines of sight as measured from these rotated source catalogs represents a random realization of the shape noise (cf., e.g., \cite{2016MNRAS.463.3653K}, for a similar technique for shear peak statistics, \cite{2017MNRAS.465..746S} for void lensing, and \cite{2017arXiv170701907M} for cluster lensing).

\begin{figure}
  \includegraphics[width=\linewidth]{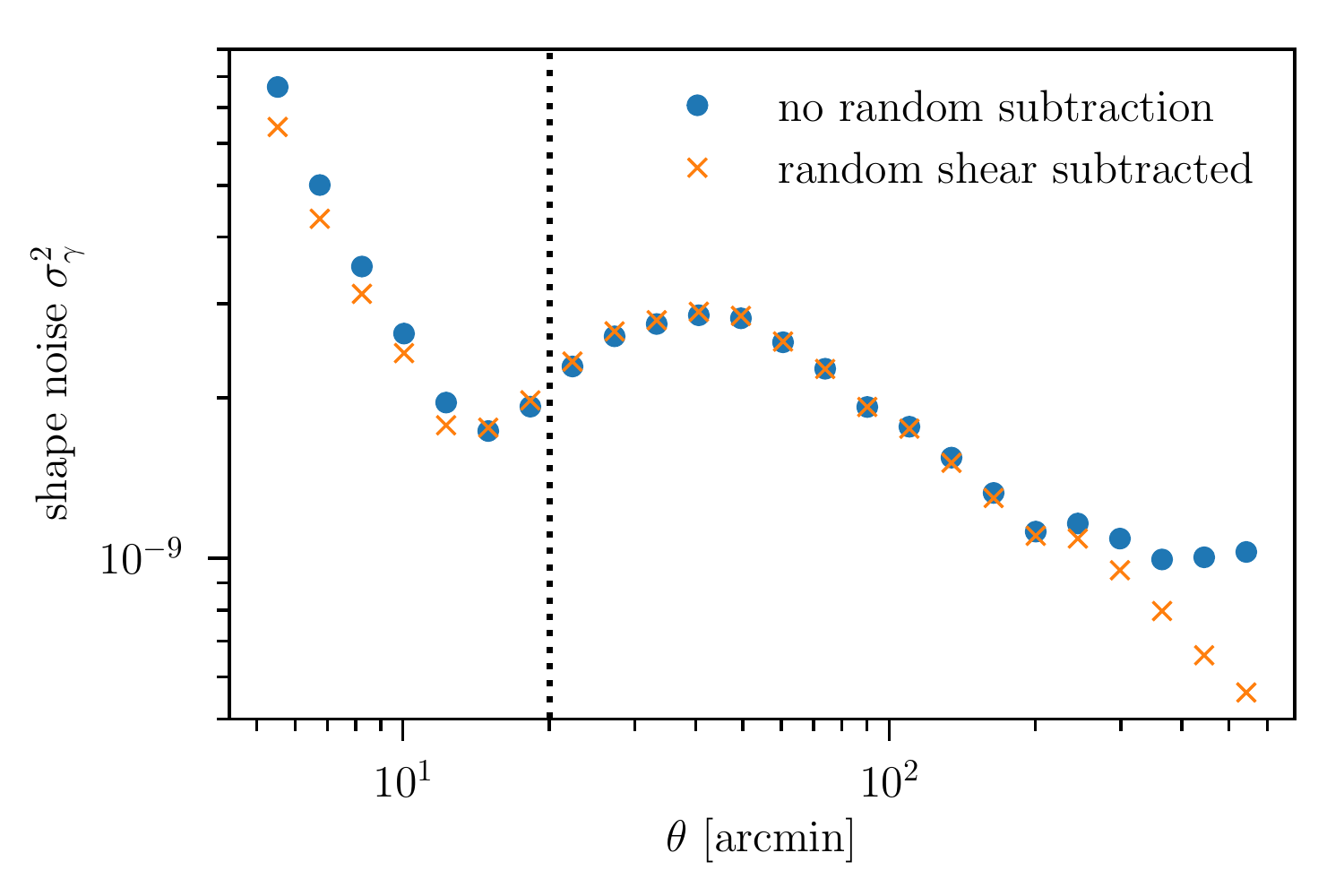}
  \caption{Variance of the shear signal around troughs due to shape noise in DES Y1 \metacal\ $z_s=0.63-0.9$ galaxies. Subtracting the shape noise around random points (cross symbols), as we also do in our data, lowers the variance considerably on scales above 300 arcmin.}
\label{fig:shapenoise_random}
\end{figure}

In measuring the signals on the rotated catalogs, we take care to use the same methodology as for the measurements on data. That means we use each randomly rotated source catalog for cross-correlation with all of the maps contributing to our data vector. As on the data, we subtract the mean shears measured around random points, for which we simply use the centers of \emph{all} \textsc{healpix} pixels that are used as lines of sight in any density quintile. The subtraction of shear around random points considerably reduces shape noise on large scales (see \autoref{fig:shapenoise_random}, and refer to \cite{2017MNRAS.471.3827S} for a detailed study of the effect).

\subsection{Cosmic variance and shot noise}
\label{sec:cosmicvariance}

Two additional effects cause our signal to deviate from its expectation value:
\begin{itemize}
\item The cosmic density field present in our survey volume is a random realization. This is true both for the volume in which our tracer galaxies are located (and in which the signal of troughs and overdense lines of sight originates) and for the redshift range along the line of sight in between us and the source galaxies that is not contained in that volume. This causes there to be noise in the true convergence around the lines of sight we identify, and in the counts-in-cells distribution.
\item In a given realization of the matter density field, tracer galaxies could be placed differently (for instance, according to Poisson noise around their expectation value in any given volume). Which of these possible galaxy catalogs is realized causes there to be a different true shear signal around what we identify as troughs and overdense lines of sight, and a different counts-in-cells distribution.
\end{itemize}

On the scales we care about in this work, we can measure the sum of both contributions to the covariance, to good approximation, from log-normal simulations of the related matter and convergence fields and Poissonian realizations of the tracer galaxy catalog. We do this by generating a large number of realizations of these fields and catalogs with \textsc{flask} \citep{2016MNRAS.459.3693X}.

We note that this part of the covariance is dependent on cosmology and the parameters describing the connection of galaxies and matter. For the covariance in this work, we will assume the settings of the \textit{Buzzard} simulations, namely a fiducial flat $\Lambda$CDM cosmology with $\Omega_{m,0}=0.286$, $\sigma_8=0.82$, $\Omega_b=0.047$, $n_s=0.96$ and $H_0=h\times100$km s$^{-1}$ with $h=0.7$. 

For the matter and associated galaxy field in the tracer redshift range, we use the power spectrum with a linear bias of $b=1.54$, a redshift distribution, and a mean density of the tracer galaxy population as in the \emph{Buzzard-v1.1} suite of simulations. We assume Poissonianity of the galaxy count at fixed density, i.e.~the $b$ model (\autoref{sec:modeltracers}). Note that the relation of galaxies and matter in the \emph{Buzzard} simulations \citep{Niall} and, potentially, the actual Universe is more complex than that. We ensure, using mock likelihood runs on \emph{Buzzard}, that this does not mean our covariance from the log-normal mocks is significantly underestimated (see Appendix \ref{app:mock}). To set the log-normal parameter of the projected matter field (i.e., the minimum allowed value of $\delta_m\geq-\delta_0$), we use the methodology of \citet{Oliver}. In the \emph{Buzzard} cosmology and at a top-hat smoothing scale of 20~arcmin, this yields $\delta_0=0.669$. Details are given in Appendix \ref{sec:lognormal_params}.

For the source redshift distributions of the simulated convergence fields, we use those estimated for the source samples in our data. 

We separate the convergence field into two parts: a component correlated with the matter field that our tracer galaxies populate, and an uncorrelated component (mostly comprised of the parts of the lensing kernel in front and behind our tracer galaxies).  The correlated component is modeled as a log-normal field with cross-power spectrum and $\kappa_0$ set to match the perturbation theory predictions for $\left\langle\delta\kappa\right\rangle$ and $\left\langle\delta_T^2 \kappa\right\rangle$ at a fiducial smoothing scale. This constrains the auto power spectrum of this component to be only a fraction of the total convergence power spectrum. The uncorrelated component is then simulated as a Gaussian random field that is uncorrelated to all other fields and whose power spectrum is chosen to give the correct total convergence power spectrum (see Appendix \ref{sec:lognormal_params} for the details of the procedure).

We apply the same mask to the tracer galaxies as in our data (or in our simulations, for the mock analysis described in Appendix \ref{app:mock}), and the same prescription for splitting the survey into lines of sight of different density. 

We then measure tangential shear signals, as in our data, yet on the noiseless shear maps output by \textsc{flask} at $N_{\rm side}=4096$ resolution. In order to sample the density fields as in our data, we use the sum of weights of sources in our actual shear catalogs situated in a pixel as the weight of the shear signal in that pixel. We do this both for the correlated and the uncorrelated part of the convergence field (see above) and coadd the two signals. In addition, as in our data, we measure the mean tracer galaxy overdensity in our density quintiles.

On scales much smaller than the aperture radius $\theta_{T}=20'$, a checkerboard pattern in the off-diagonal shape noise covariance is apparent (see \autoref{fig:sdss_cov}). We find that this is due to an interference of the \textsc{healpix} grid we use to sample the density field and the angular binning scheme for our shear signal -- for adjacent \textsc{healpix} pixels, sources move from one angular bin to the next and their intrinsic shape orientation changes relative to the pixel centers. Since these effects are present in the data as well (as seen from the jackknife covariance) and only significant on angular scales below our scale cut, we do not attempt to address them further.

\subsection{Constructing the covariance matrix}
\label{sec:constructingcovariance}
We create 1000 realizations of both the shape noise and the large scale structure and shot noise contributions to the covariance. Despite this relatively large number, there is noise in our estimated covariance matrix. When inverting the covariance matrix to calculate $\chi^2$ values and run a likelihood analysis, this noise has two consequences.

First, the inverse of a noisy estimate of the covariance matrix is a biased estimate of the inverse covariance matrix. We follow the correction described in \citet{2007A&A...464..399H} to correct for this effect, i.e.~we multiply the $\chi^2$ calculated with the inverse of our estimated covariance matrix by a factor
\begin{equation}
f^{\rm AH} = \frac{N^{\rm cov}-N^{\rm data} - 2}{N^{\rm cov}-1} \; .
\label{eqn:fah}
\end{equation}
The number of entries in our data vector is at most $N^{\rm data}=208$ in the fiducial DES analysis and we use $N^{\rm cov}=960$ realizations of the log-normal field to estimate the covariance, which means $f^{\rm AH}$ is 0.78 or larger for all our likelihood runs. We confirm, using independent log-normal mock realizations of our data vector, that the inverse covariance matrix rescaled such does lead to a consistent $\chi^2$ distribution of residuals (Appendix \ref{sec:lognormallikelihood}).

Second, the noise in the inverse variance leads to additional scatter in the best fit we find \citep{2013PhRvD..88f3537D,2016MNRAS.456L.132S,2017arXiv170307786F}. Under the assumption that the model is linear in all parameters within the range probed, this can be compensated by multiplying $\chi^2$ with a factor
\begin{equation}
f^{\rm DS} = \left[1+\frac{(N^{\rm data}-N^{\rm par})(N^{\rm cov}-N^{\rm data}-2)}{(N^{\rm cov}-N^{\rm data}-1)(N^{\rm cov}-N^{\rm data}-4)}\right]^{-1} \; ,
\end{equation}
where $N^{\rm par}$ is the number of free parameters in the model.

These corrections are only appropriate for a monolithic covariance estimated from a fixed number of independent realizations. From the previous subsections, we can get independent, unbiased estimates of the two contributions, $\Cov^{\rm shape}$ and $\Cov^{\rm LSS}$. The sum of $\Cov^{\rm shape}+\Cov^{\rm LSS}$ would be an unbiased and less noisy estimate of the total covariance. But to apply the above corrections, we need to resort to coadding shape noise and cosmic variance realizations before estimating the full covariance matrix.

In addition to the 960 realizations used to estimate the covariance, we use 40 independent realizations to confirm that our prediction  indeed matches the mean signal measured from the log-normal simulations at the expected $\chi^2\approx N_{\rm data}$. This is a test of both the numerical scheme employed by \textsc{FLASK} and the implementation of the analytical calculations of \citet{Oliver}. We find that the two are in excellent agreement, except for a small offset of the predicted and measured counts-in-cells statistic. The mean tracer galaxy overdensities (\autoref{sec:measure_cic}) we measure in log-normal mocks are offset from the predictions at most at the $10^{-3}$ level. We hypothesize that this is due to resolution effects of the simulations, but cannot exclude that similar effects could also present in the data.\footnote{We confirm, however, that the mean tracer galaxy overdensities in our data are well fit by the model at its maximum likelihood parameters.} To compensate for this, we boost the variance of each of the four counts-in-cells entries in our data vector by $0.002^2$. Using this covariance and $f^{\rm AH}$ (but not $f^{\rm DS}$) as defined above, the mean of all realizations with no shape noise matches the predicted signal at the true input parameters at total $\chi^2=0.19$ with 208 degrees of freedom, proving the numerical accuracy of the prediction code at a sufficient level. Additional tests of our likelihood pipeline run on the 40 log-normal realizations are shown in Appendix~\ref{sec:lognormallikelihood}.

\subsection{Comparison with jackknife covariances}

\begin{figure}
\begin{centering}
 \includegraphics[width=0.88\linewidth]{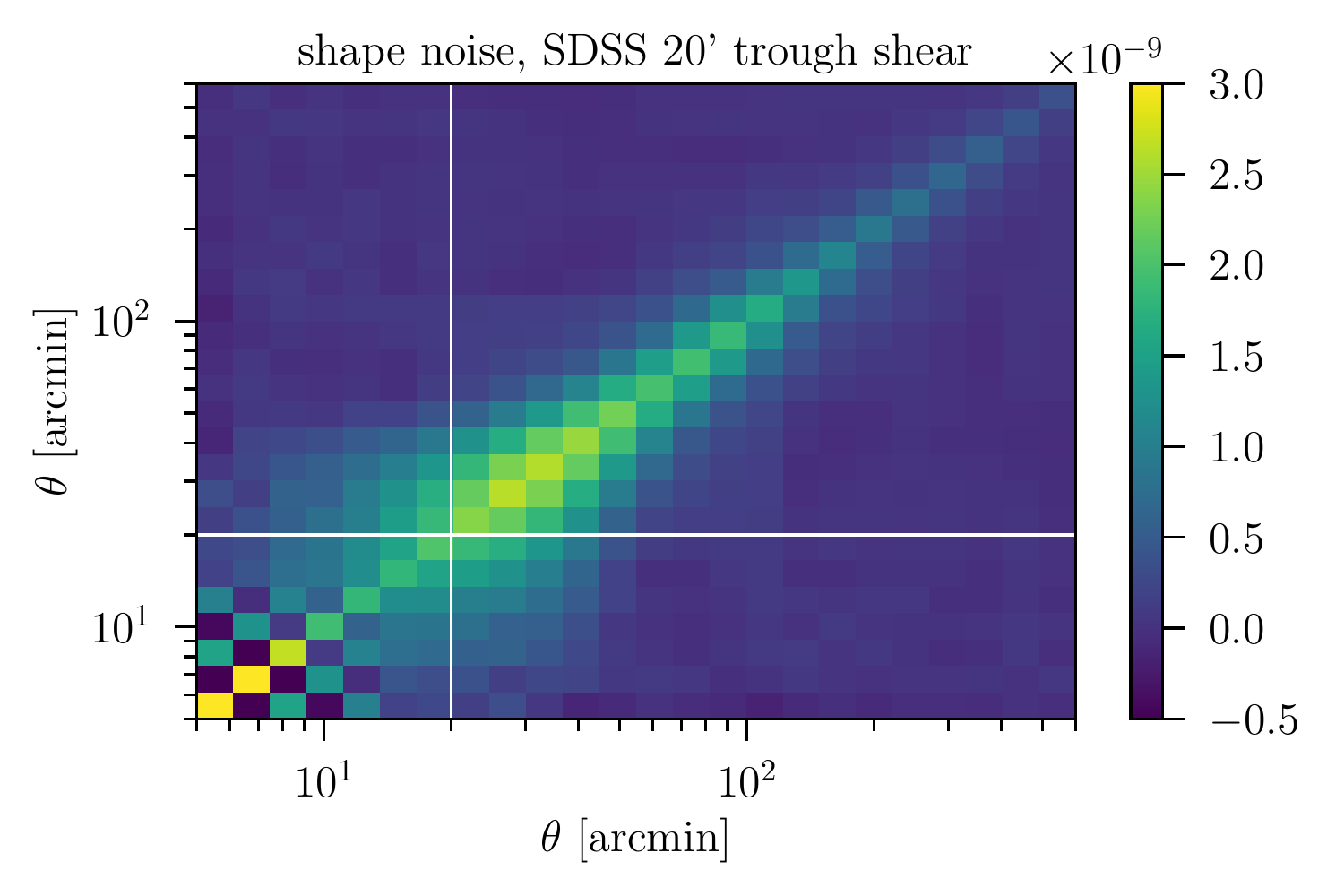}
 
 +
 
 \includegraphics[width=0.88\linewidth]{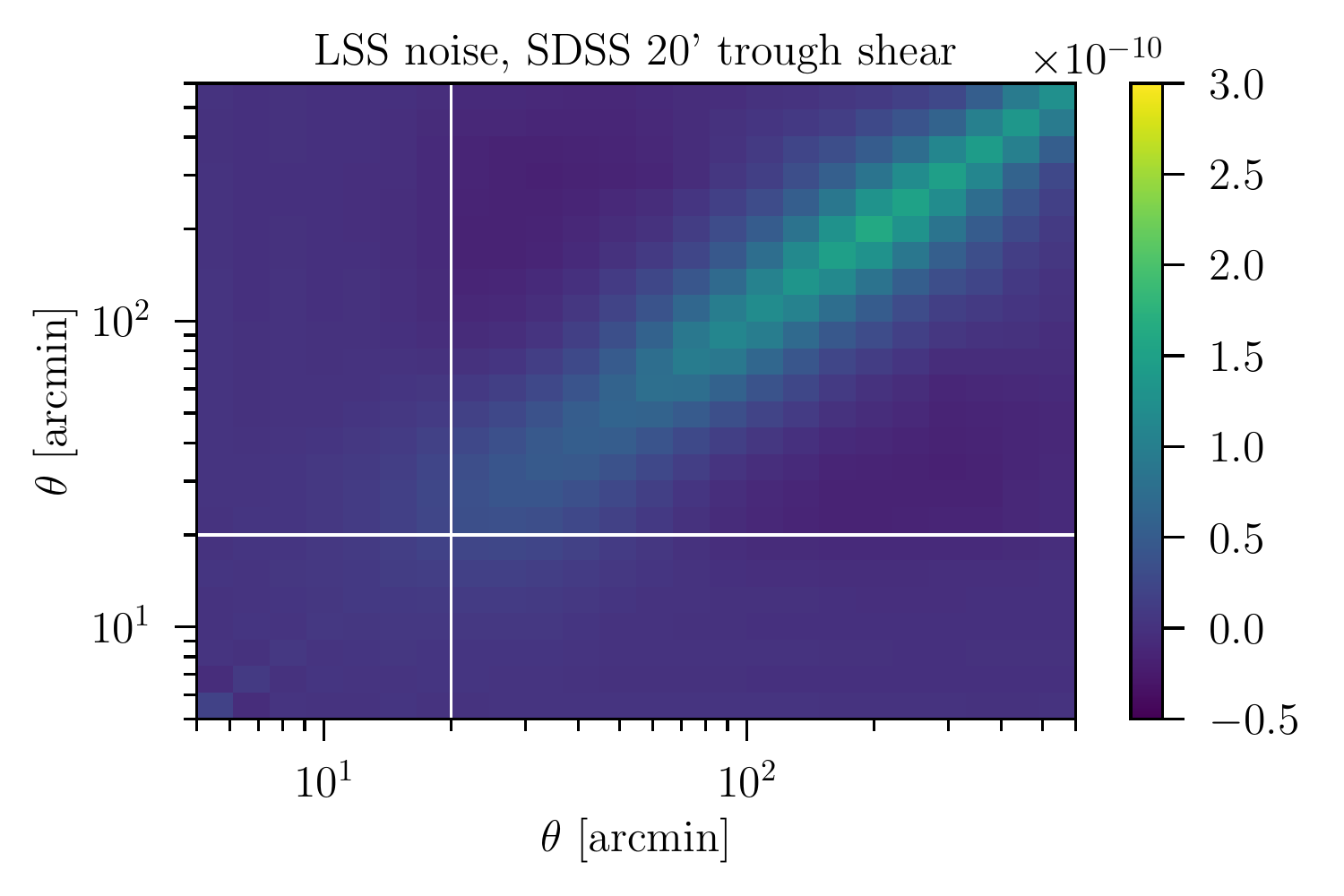}
 
 =
 
 \includegraphics[width=0.88\linewidth]{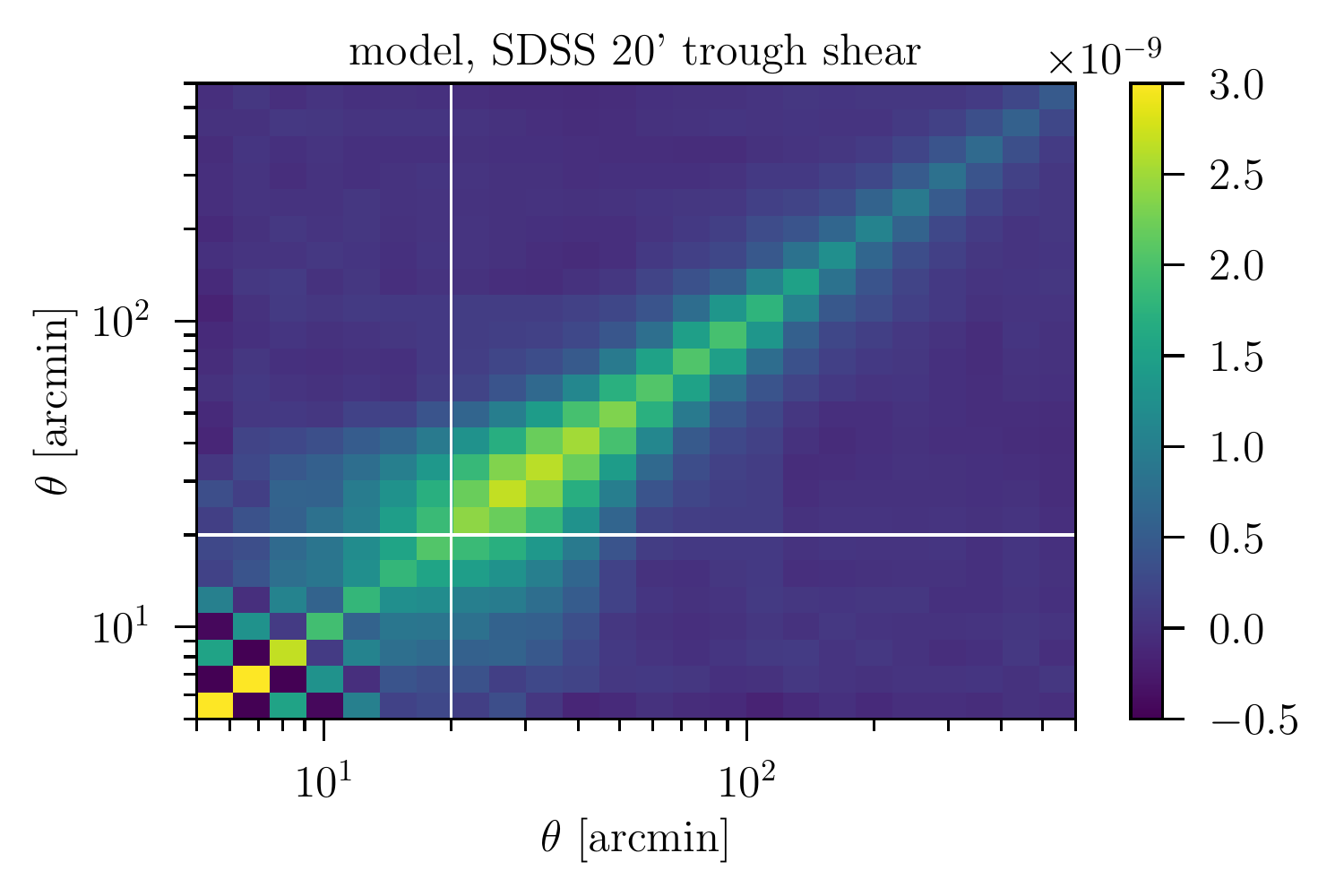}
 
 $\approx$
 
 \includegraphics[width=0.88\linewidth]{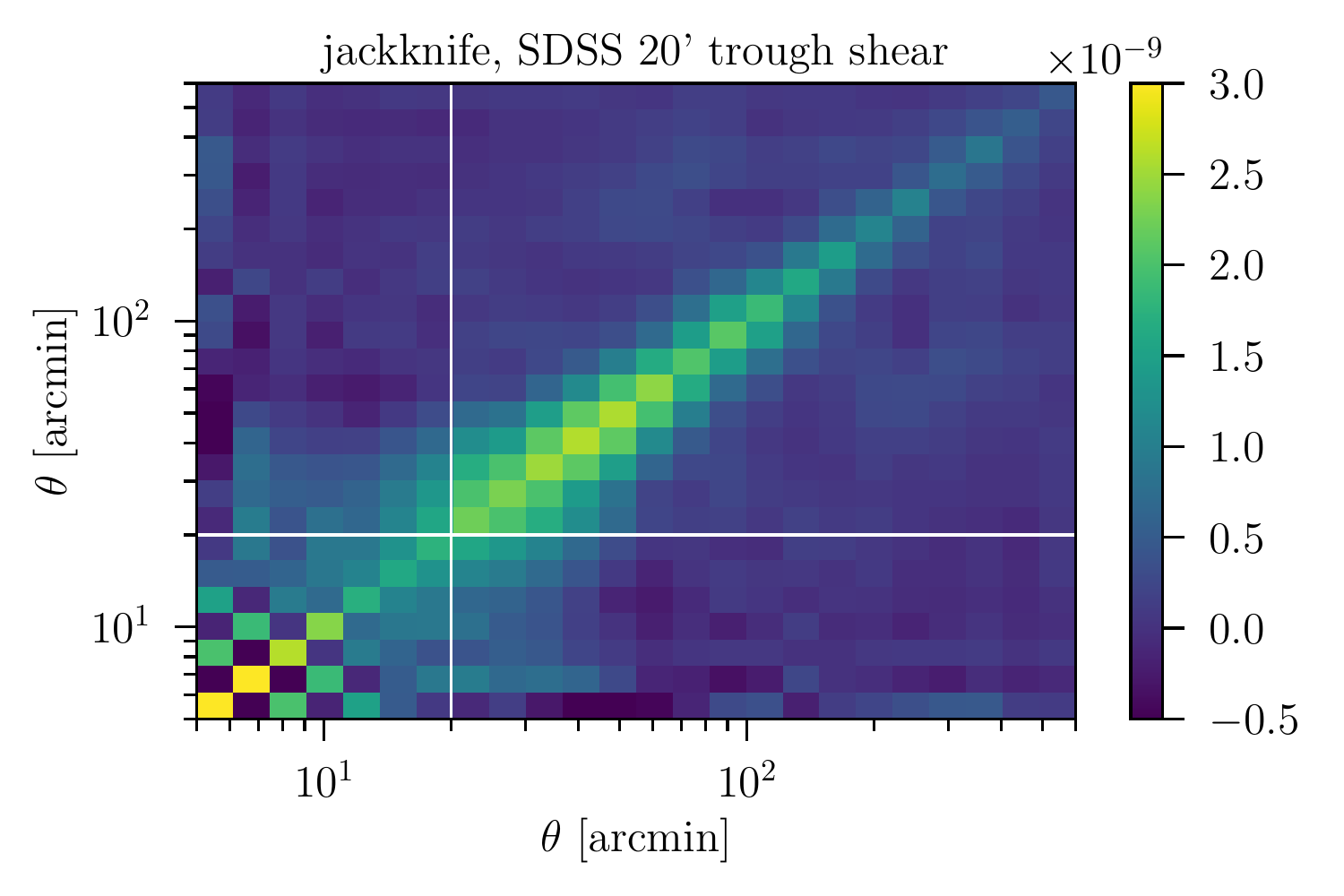}
  \caption{Covariance of shear signal around SDSS troughs from shape noise (top), large scale structure and shot noise (second from top, note the different scale). The model (third from top) is the sum of these two and closely matches the jackknife covariance (bottom). White lines indicate aperture radius $\theta_T=20'$ -- only data above that is used in the likelihood.}
\label{fig:sdss_cov}
\end{centering}
\end{figure}
\begin{figure}
\begin{centering}
 \includegraphics[width=0.84\linewidth]{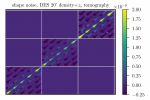}
 
 +
 
 \includegraphics[width=0.84\linewidth]{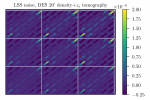}
 
 =
 
 \includegraphics[width=0.84\linewidth]{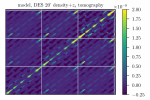}
 
 diagonal compared to jackknife:
 
 \includegraphics[width=0.84\linewidth]{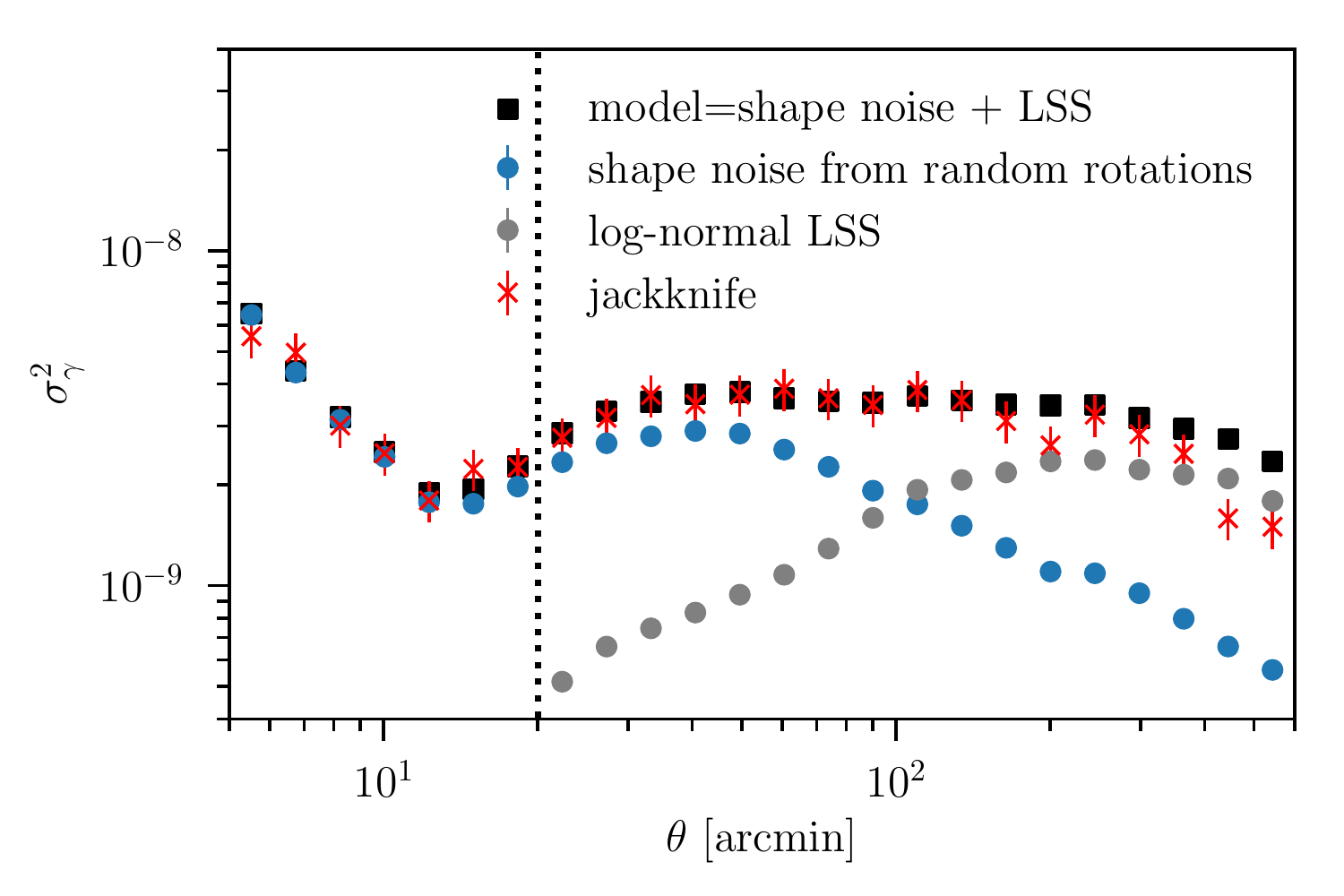}\hspace{30pt}
  
  \caption{Covariance of tomographic shear signal around DES density quintiles (20', lowest to highest density quintile, then lowest to highest redshift source bin, division indicated by white lines) and counts in cells (last 5 bins) from shape noise (top), large scale structure and shot noise (second from top, note the different scale), and the full model from the sum of these two (third from top). Bottom panel: diagonal of shear around troughs in intermediate source redshift bin. Counts in cells residuals have been rescaled by factor $1/50$ to match a common color scale. }
\label{fig:des_cov}
\end{centering}
\end{figure}

From jackknife resamplings of our data, we can internally estimate the covariance matrix. While more care would have to be taken for applying this estimate of the covariance matrix in a likelihood analysis \citep{2016MNRAS.456.2662F}, it does provide confirmation of our scheme to compare the jackknife estimate to the covariance estimated above. 

\autoref{fig:sdss_cov} shows the shape noise and cosmic variance + shot noise components of the covariance matrix and compares their sum to the jackknife covariance, for the shear signal around underdense lines of sight in SDSS. The same for the full density and source-redshift tomographic covariance matrix of DES, including counts-in-cells, is displayed in \autoref{fig:des_cov}.

\section{Likelihood}
\label{sec:likelihood}

We compare our data $\bm{D}$ to model predictions $\bm{M}$ in a Bayesian fashion, i.e.~we sample the posterior distribution of model parameters $\bm{p}$ with a Monte Carlo Markov Chain (MCMC) run on the likelihood
\begin{equation}
-2\ln\mathcal{L}=f^{\rm AH}f^{\rm DS}\left[\bm{D}-\bm{M}(\bm{p})\right]^{\rm T} C^{-1} \left[\bm{D}-\bm{M}(\bm{p})\right] + \mathrm{Prior}(\bm{p})\; .
\end{equation}

For our fiducial likelihood analysis, we remove the following parts of the full data vector:
\begin{itemize}
\item lensing and counts-in-cells signal for any aperture radii other than $\theta_T=20'$ -- on smaller smoothing scales, small but significant deviations of our model and measurements in $N$-body simulations appear \citep{Oliver}. Smoothing on larger scales than 20' yields signals with errors that are highly correlated to the 20' measurements, thus adding little independent information.
\item lensing signal on scales smaller than  $\theta_T=20'$ -- small but significant deviations of our model and measurements in $N$-body simulations are present on scales smaller than the aperture radius $\theta_T=20'$. The lensing signal on these scales has low signal-to-noise ratio. In addition, shape noise in adjacent small-scale bins is anticorrelated, visible as the checkerboard pattern in the lower left of \autoref{fig:sdss_cov}. This is due to interference of the radial binning scheme with the \textsc{healpix} grid of lines of sight: when we measure the contribution of a source galaxy to the shear signal around two adjacent lines of sight, its intrinsic orientation relative to a line of sight and its distance from the line of sight change coherently. While the effect is consistently seen in jackknife and model covariance, it makes these small scale lensing signals numerically redundant. This leaves 17 angular bins in each shear profile.
\item signal for quintile 3 -- the signals we use are \emph{not} linearly independent between all quintiles; we therefore discard the signal in the median quintile, which is close to zero by construction anyway.
\end{itemize}

Therefore, in all of the following, unless otherwise noted, $\bm{D}$ contains the shear signals measured at $\theta=20-600'$ and the relative overdensity of tracer galaxy count for the lower two and upper two quintiles of galaxy count, measured in $\theta_{\rm T}=20'$ apertures. For the source tomographic DES Y1 analysis, these are 208 entries (72 for SDSS).

The precision matrix $C^{-1}$ is estimated as detailed in \autoref{sec:covariance}.

In the following subsections, we describe our choice of parametrization, the nuisance parameters and associated  priors, and the consistency tests we perform before unblinding the estimated cosmological parameters.

\subsection{Cosmological parameters}
\label{sec:parameters}

Since this is our first cosmological analysis of counts and lensing in cells, we choose to only vary a minimal set of cosmological parameters, adopting fixed priors for ones that the density split lensing and counts signal is not very sensitive to. For the fiducial run of our likelihood, we validate this approach by marginalizing over these parameters with informative external priors.

All likelihoods assume a flat $\Lambda$CDM cosmology. The main parameters we wish to constrain are the matter density in units of the critical density $\Omega_{\rm m}$, and the amplitude of structure in the present day Universe, parametrized as the RMS of overdensity fluctuations on $8 h^{-1}$Mpc scale, $\sigma_8$. 

In an alternate run of our likelihoods, we will also leave free the parameter $S_3$ that describes the skewness of the matter density field when smoothed over the given aperture and redshift range,
\begin{equation}
S_3 = \langle \delta^3 \rangle / \langle \delta^2\rangle^2 \; .
\end{equation}
$S_3$ was first defined by \citet{1980lssu.book.....P}, who derived a perturbation theory prediction $S_3=34/7$ for the unsmoothed matter density field, and later generalized to top-hat smoothed fields and higher orders \citep{1984ApJ...279..499F,1993ApJ...412L...9J,Bernardeau1994}. Perturbation theory predicts also the smoothed $S_3$ to be almost independent of $\Omega_{\rm m}$ and $\sigma_8$ and to only vary slowly with redshift or scale. A skewness $S_3$ that is inconsistent with these predictions could be caused either by non-Gaussian initial density fluctuations (although CMB limits set tight constraints on these \cite{2016A&A...594A..17P}) or by physics beyond gravity that affect collapse either in the overdense or underdense regime.

We assume wide, flat priors for $\Omega_{\rm m}$, $\sigma_8$ and, in the likelihood runs that vary it, $S_3$ that do not limit the range sampled by the likelihoods. We fix the Baryon density $\Omega_{\rm b}=0.047$, the spectral index of primordial density fluctuations $n_s=0.96$, and a dimensionless Hubble parameter $h=0.7$, equal to the values used in the \emph{Buzzard} simulations and consistent with best constraints. For the transfer function of primordial to initial matter power spectrum, we assume a radiation density $\Omega_r h^2 = 4.15\times10^{-5}$. The evolution of expansion and growth of structure in the late universe assumes only matter and a cosmological constant.

An overview of these choices is given in \autoref{tab:params}.

\subsection{Nuisance parameters}
\label{sec:nuisance}

In our likelihoods, we apply three different models to describes the distribution of \textsc{redMaGiC} galaxy count $N_T$ inside an aperture at given mean matter overdensity $\delta_{m,T}$, $P(N_T|\delta_{m,T})$. Details of this are described in \autoref{sec:modeltracers}, and sampling ranges for the parameters $b$, $(b,r)$, or $(b,\alpha_0,\alpha_1)$, designed to span any physically sensible configurations, are listed in \autoref{tab:params}.

Similar to previous cosmological constraints derived from DES Y1 data, we assume and always marginalize over nuisance parameters describing photometric redshift and shear biases in our measurements. 
The nuisance parameter for the redshift bias of \textsc{redMaGiC} sources in the $z_T=0.2-0.45$ redshift range that is constrained from cross-correlations with a sample of galaxies with spectroscopic redshifts as in \citet{redmagicpz}. Specifics of this are described in Appendix \ref{sec:redmagiczbias}.

The photometric redshift biases and multiplicative shear biases of source galaxies are described by two parameters in each redshift bin. The three bins we use, i.e.~all but the lowest redshift bin of \citet{keypaper}, are labeled as $i=2,3,4$ in \autoref{tab:params}. Priors on the redshift biases are taken from the combination of the redshift distributions of a matched sample of galaxies in the COSMOS survey and angular cross-correlation with \textsc{redMaGiC} galaxies \citep{2017arXiv170708256D,xcorr,xcorrtechnique} as described in detail in \citet{photoz}. The priors on multiplicative shear bias in DES Y1 are described in detail in \citet{shearcat}. Both of these priors are widened in our analysis to account for their potential correlation between bins \citep[see appendices of][]{photoz,shearcat}, conservatively assuming comparable signal-to-noise ratio in each bin.

Multiplicative bias in an independent SDSS shear catalog that is consistent with the one we use \citep{2017MNRAS.466.3103S} was investigated in detail in \citet{2013MNRAS.432.1544M}. The authors in that paper find a Gaussian uncertainty related to multiplicative shear calibration of $\sigma_m=0.037$, in addition to photometric redshift biases over which we marginalize separately. We assume a slightly more conservative Gaussian uncertainty of $\sigma_m=0.05$ for the multiplicative shear bias in the SDSS catalog used in this work.

These priors are also summarized in \autoref{tab:params}.

\begin{table}
\caption{Priors for likelihood runs}
\label{tab:params}
\begin{center}
\begin{tabular}{| c  c |}
\hline
\hline
Parameter & Prior \\  
\hline 
\multicolumn{2}{|c|}{{\bf Cosmology}} \\
$\Omega_m$  &  flat (0.1, 0.9)  \\ 
$\sigma_8$ &  flat (0.2, 1.6)  \\
$S_3$ & fixed to PT / flat \\
\hline
$\Omega_b$ &  fixed (0.047)  \\
$h$  &  fixed (0.70)   \\
$\Omega_r h^2$  & fixed ($4.15\times10^{-5}$) \\
\hline
\multicolumn{2}{|c|}{{\bf Tracer Galaxies}} \\
$b$   & flat (0.8, 2.5) \\
$r$    & flat (0, 1)\\
$\alpha_0$   & flat (0.1, 3.0) \\
$\alpha_1$   & flat (-1.0, 4.0) \\
\hline
\multicolumn{2}{|c|}{{\bf Tracer galaxy photo-$z$ shift}} \\
$\Delta z_{\rm l,\rm{DES}}$  & Gauss ($0.003, 0.008$) \\
$\Delta z_{\rm l,\rm{SDSS}}$  & Gauss ($0.002, 0.006$) \\
\hline
\multicolumn{2}{|c|}{{\bf Source photo-$z$ shift}} \\
$\Delta z^2_{\rm s, \metacal}$  & Gauss ($-0.019, 0.018$) \\
$\Delta z^3_{\rm s, \metacal}$  & Gauss ($+0.009, 0.016$) \\
$\Delta z^4_{\rm s, \metacal}$  & Gauss ($-0.018, 0.031$) \\
$\Delta z^2_{\rm s, \imshape}$  & Gauss ($-0.024, 0.018$) \\
$\Delta z^3_{\rm s, \imshape}$  & Gauss ($-0.003, 0.016$) \\
$\Delta z^4_{\rm s, \imshape}$  & Gauss ($-0.057, 0.031$) \\
$\Delta z_{\rm s, SDSS}$  & Gauss ($-0.014, 0.011$) \\
\hline
\multicolumn{2}{|c|}{{\bf Shear calibration}} \\
$m^{i}_{\metacal} (i=2,3,4)$ & Gauss ($0.012, 0.023$)\\
$m^{i}_{\rm \imshape} (i=2,3,4)$ & Gauss ($0.0, 0.029$)\\
$m_{\rm SDSS}$ & Gauss($0.0,0.05$)\\
\hline
\end{tabular}
\end{center}
\end{table}

\subsection{Sampling and evidence}
\label{sec:evidence}
To sample the posterior likelihoods efficiently, we employ both the \textsc{emcee} \citep{2013PASP..125..306F} and the \textsc{MultiNest} \citep{2009MNRAS.398.1601F} algorithm. The latter has the advantage of also estimating Bayesian evidences $E$,
\begin{equation}
E\propto p(\bm{D}|\mathrm{model})=\int \mathrm{d}\bm{\mu} \; p(\bm{D}|\mathrm{model},\bm{\mu}) \, p(\bm{\mu}|\mathrm{model}) \; ,
\end{equation}
where $\bm{\mu}$ are the parameters of the model.

Knowing the evidence of two models $1$ and $2$ allows comparing them with the Bayes factor, $E_1/E_2$. If the latter ratio exceeds 3.2 or 10, the evidence for model 1 over model 2 can be considered \emph{substantial} or \emph{strong} in the nomenclature of \citet{jeffreys61}.

We confirm in the tests performed in Appendices~\ref{sec:lognormallikelihood} and \ref{app:mock} that both sampling algorithms and the analysis of their outputs with \textsc{ChainConsumer} \citep{Hinton2016} and custom codes yield reliable parameter constraints and test results.

\subsection{Blinding and tests}
\label{sec:tests}
\begin{figure}
\begin{centering}
  \includegraphics[width=0.9\linewidth]{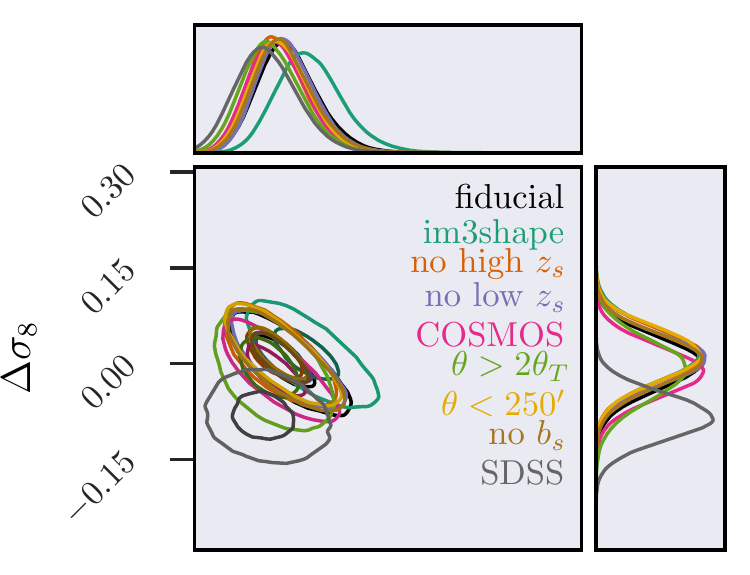}
  \includegraphics[width=0.9\linewidth]{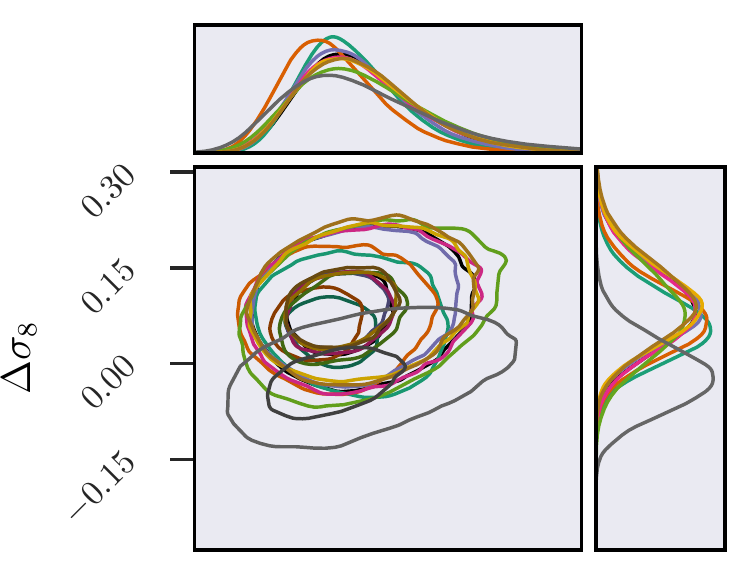}
  \includegraphics[width=0.9\linewidth]{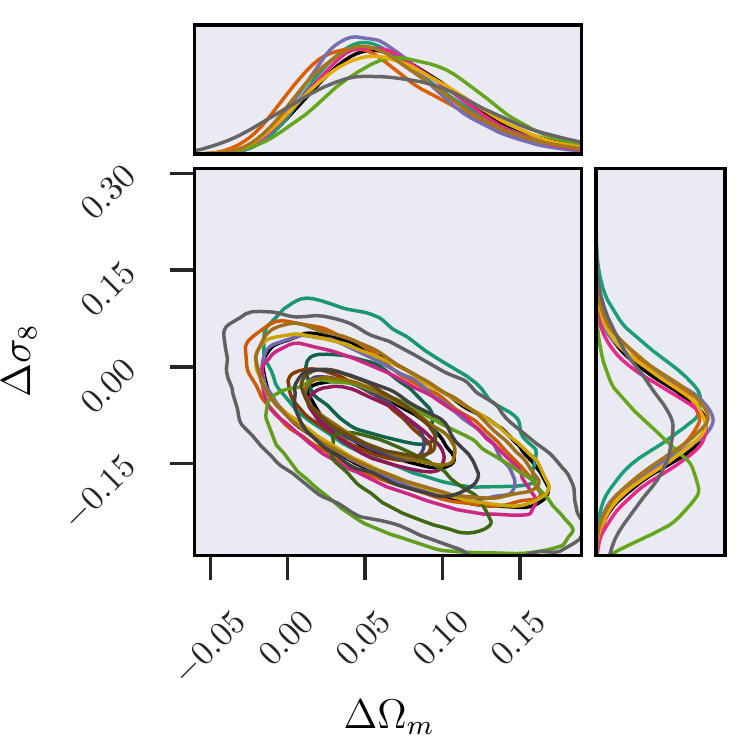}
\end{centering}
  \caption{Likelihoods run on variants of the data to test robustness to analysis choices. Use of \imshape\ shape catalogs (turquoise), removal of the highest (orange) or lowest (violet) source redshift bin, use of a direct estimate of source redshift distributions from COSMOS instead of BPZ (magenta), removal of small (green) or large scales (yellow) in the lensing data or neglect of the clustering of sources in the lowest redshift bin with the lenses (brown) do not have an effect on recovered parameters beyond their statistical uncertainty. Contours are centered on the fiducial result in the $b$ model. The same likelihood run on fully independent SDSS data vectors (grey) yields consistent results. Panels show different models for the connection of galaxies and matter (top: $b$ only, center: $b,r$, bottom: $b,\alpha_0,\alpha_1$).}
\label{fig:testlike_data}
\end{figure}

Since most of the tests in this paper were performed after the scaling factors of the initial, blinded shear catalogs had been revealed \citep{shearcat}, we primarily rely on parameter level blinding. This means that we do not compare measurements on data to predictions in a known cosmology before the following tests are passed:

\begin{enumerate}
\item Log-normal simulations show $\chi^2$ values of data vs. model at the input set of parameters consistent with a fiducial $\chi^2$ distribution with the appropriate number of d.o.f.. Likelihood runs on mock data have a coverage within expectations (i.e., the input cosmology lies within the confidence interval the expected fraction of times). Results are unremarkable and described in Appendix~\ref{sec:lognormallikelihood}.
\item Bayesian model comparisons run on log-normal simulations without stochasticity do not provide evidence for more complex models. We find that this requirement is met, both for the $b,r$ model of stochasticity and models with free skewness $S_3$, in Appendix~\ref{sec:lognormallikelihood}.
\item 21 independent \emph{Buzzard} $N$-body realizations of our data vector give consistent $\chi^2$ relative to the model evaluated at the input cosmology and independently measured nuisance parameters. Their coverage in likelihood runs, i.e.~the number of times the input cosmology is within derived confidence limits, is within expectations only for the $b,\alpha_0,\alpha_1$ model of bias and stochasticity (Appendix~\ref{app:mock}). The fact that the most complex bias model is required may be particular to these mock galaxy catalogs, which may have different relations to matter density than real \textsc{redMaGiC} galaxies. We still take this as evidence that the most general stochasticity model, unless disfavored by the data, needs to be considered in our analysis.
\item Likelihood runs on $N$-body realizations are insensitive to replacing true source redshift distributions with source redshift distributions estimated from BPZ and marginalizing over $\Delta z$ uncertainties. Results: we find that the mean shifts in cosmological parameters are at or below the ten percent level of their statistical uncertainty, and that the statistical uncertainty increases by less than five percent due to marginalization over $\Delta z$, both tested with the $b,\alpha_0,\alpha_1$ model for the galaxy-matter connection.
\end{enumerate}

Once these tests are successful, we continue to make tests on likelihood analyses run on the data itself. To ensure that these do not introduce experimenter bias, before looking at any chains we shift all cosmological parameters by a constant unknown vector, uniformly distributed between $+2$ and $-2$ standard deviations of the parameters as found from $N$-body simulations. We then proceed with the following tests, the results of which are shown in \autoref{fig:testlike_data}:

\begin{enumerate}
\setcounter{enumi}{4}
\item Cosmological constraints from the data are consistent between the fiducial \metacal\ and additional \imshape\ measurements. For models including galaxy stochasticity (lower two panels of \autoref{fig:testlike_data}) this is the case to a fraction of the statistical uncertainty. For the $b$ only model (top panel), \imshape~constraints on $\Omega_{m}$ are offset by $\approx\sigma/2$. Accounting for the fact that shape noise is largely uncorrelated between the two catalogs, is is possible that this is simply a statistical fluctuation. We note, however, that \citet{keypaper} found a similar discrepancy, likely attributed to the multiplicative bias or source redshift calibration of the \imshape\ highest source redshift bin. 
\item Cosmological constraints are robust to removing the lowest or highest source redshift bin from the data vector. This is the case for all models, indicating that the calibration of \metacal\ catalogs is consistent between bins. 
\item Cosmological constraints are robust to replacing the source redshift distributions estimated by BPZ by ones directly estimated from COSMOS. Again, this is the case for all models, with a noticeable but insignificant offset in the $b$ only model.
\item Cosmological constraints are robust to cutting scales smaller than $2\times\theta_{\rm T}$ or larger than $250$' from the shear signal. Removal of small scale shear information shifts $\sigma_8$ in the $b,\alpha_0,\alpha_1$ model by approximately $1\sigma$. Given the cosmic variance in large-scale modes and the unremarkable result of all other variants of the scale cut test, this does not pose a significant issue. 
\item Cosmological constraints are robust to not correcting for clustering of the overlapping source redshift bins with the matter distributions around overdense and underdense lines of sight. While this is not necessary a null test -- it could be possible that we need to account for the effect -- we find that marginalizing over $b_{s}$, the source bias in the lowest redshift source bin, neither significantly widens nor shifts the contours in either model.
\item Cosmological constraints are consistent between DES and SDSS. We note that these are completely independent data sets, i.e.~have no cross-covariance, and thus we \emph{a priori} expect larger offsets between the two than in the other tests. We find that constraints on $\Omega_m$ are very similar and $\sigma_8$ is offset by $\approx1\sigma$, both consistent with these expectations.
\end{enumerate}

In addition, we confirm that the central value of the nuisance parameter priors (multiplicative shear bias, tracer and source galaxy redshift biases as defined in \citet{shearcat,photoz}, \autoref{sec:redmagiczbias} and \autoref{sec:sourcezbias}) is within the $1\sigma$ confidence interval of the posteriors for both DES and SDSS. 

Only after unblinding do we test whether the model at its maximum likelihood parameters is a good fit to the data. For the tomographic data vector of DES Y1, there are 208 elements fit with 13 parameters in the $(b,\alpha_0,\alpha_1)$ model. Because the model is not linear in the parameters, the number of degrees of freedom and expectation value for the $\chi^2$ distribution is not known precisely \citep{chi2}, but likely between $208-13$ and $208$. Its standard deviation is $\sigma_{\chi^2}=\sqrt{2N_{\rm d.o.f}}\approx20$. The data vectors for the two shear pipelines have a $\chi^2_{\metacal}=171$ and $\chi^2_{\imshape}=201$, respectively, both consistent with expectations for multivariate Gaussian noise around a signal that is correctly described by our model. The $b$ only and $(b,r)$ models give equally acceptable fits. For the SDSS single source bin data vector with 72 entries and 9 parameters in the $(b,\alpha_0,\alpha_1)$ model, we find $\chi^2=81$, and equally acceptable results for the other models.

We also perform a run of the fiducial DES Y1 data vector with a full cosmological model that marginalizes, in addition, over baryon density $\Omega_b$, spectral index of primordial density fluctuations $n_s$ and Hubble parameter $h$ with the flat priors also used in \citet{keypaper}. We find that this does not shift or increase the uncertainty on the reported parameters at a discernible level in any of the models for the connection of galaxies and matter.

\section{Cosmological constraints}
\label{sec:constraints}

\begin{figure*}
  \includegraphics[width=0.83\linewidth]{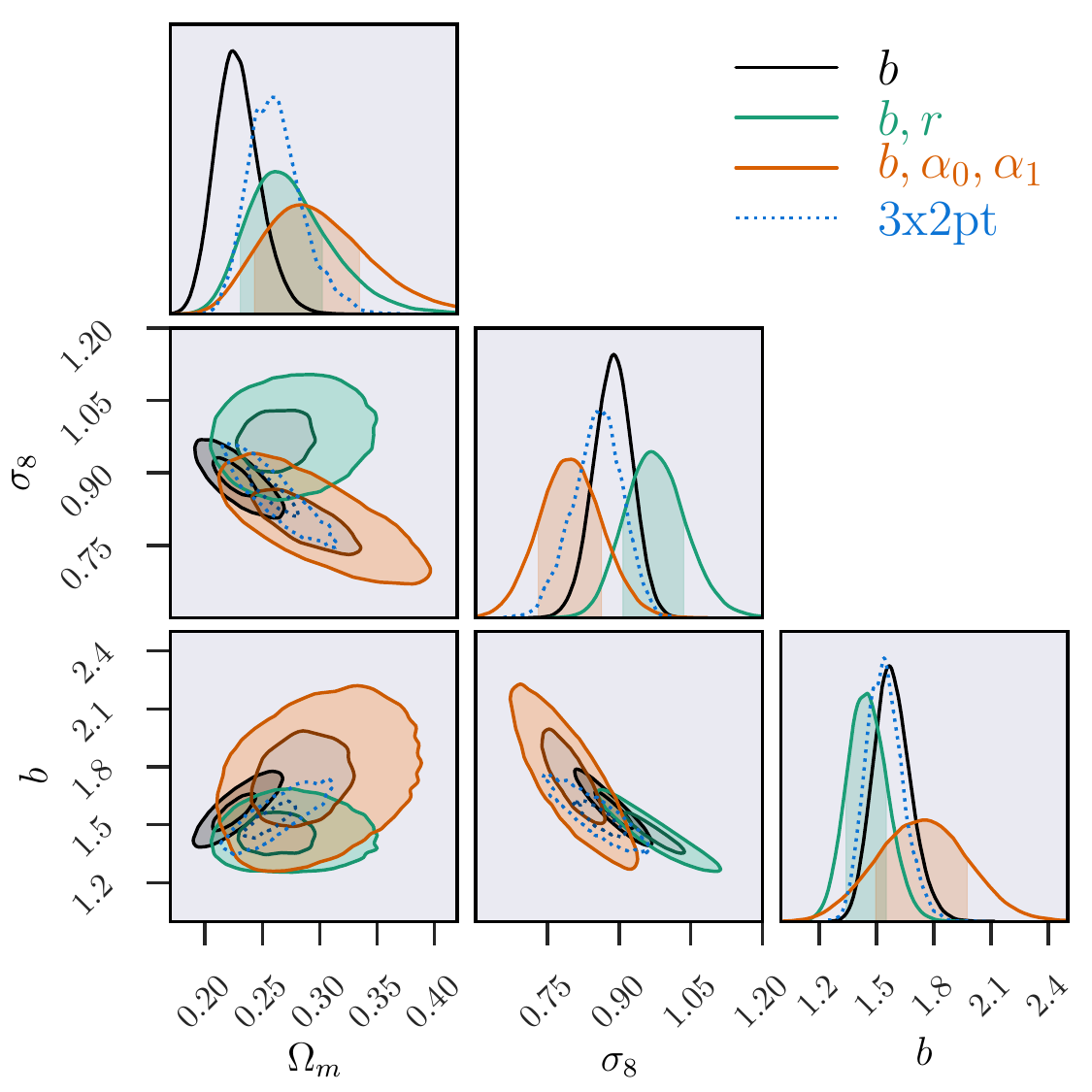}
  \caption{Constraints on matter density $\Omega_m$, amplitude of late-time structure formation $\sigma_8$, and galaxy bias of the \textsc{redMaGiC} tracer galaxies. Shown are results for three different models for the connection of matter density and galaxy density on the scales of our apertures: linearly biased tracers with Poissonian noise (black, $b$); biased, stochastic tracers (green, $b,r$); and biased tracers with density-dependent stochasticity (orange, $b,\alpha_0,\alpha_1$). The Bayes factors of these different models indicate substantial evidence that stochasticity of galaxy count is required to model the measurements. Constraints from DES galaxy and shear auto- and cross-two-point correlation functions for fixed neutrino mass are shown as blue, dotted contours \citep{keypaper}.}
\label{fig:data_models}
\end{figure*}
\begin{table*}
\caption{Constraints from counts and lensing in cells likelihood runs. Bayes factors are quoted relative to the $b,r$ model. Results for the $b$ only likelihoods are not shown, since this simpler model is disfavored by the data. Constraints on cosmological parameters and \textsc{redMaGiC} bias in $z\approx0.2-0.45$ from two-point functions are reproduced from \citet{keypaper} for comparison.}
\label{tab:results}
\begin{center}
\begin{ruledtabular}
\begin{tabular}{| l  l | r | r | r | r | r | r | r | r |}
\hline
Data   & Model & Bayes  & $S_8$ & $\Omega_m$ & $\sigma_8$ & $b$ & $r$ & $\alpha_0$ & $\alpha_1$ \\ 
& & factor & & & & & & & \\  \hline

DES   & $b,r$  & $\equiv1.0$ & $0.90^{+0.11}_{-0.08}$ & $0.26^{+0.04}_{-0.03}$ & $0.97^{+0.07}_{-0.06}$ & $1.45^{+0.10}_{-0.11}$ & $0.77^{+0.10}_{-0.13}$ & - & -  \\ \hline
SDSS & $b,r$  & $\equiv1.0$ & $0.78^{+0.13}_{-0.08}$ & $ 0.25^{+0.05}_{-0.04}$ & $0.86^{+0.06}_{-0.05}$& $1.48^{+0.09}_{-0.09}$ & $0.70^{+0.16}_{-0.14}$ & - & -  \\ \hline \hline

DES   & $b,\alpha_0,\alpha_1$  & 0.7 & $0.78^{+0.05}_{-0.04}$ &$0.28^{+0.05}_{-0.04}$ & $0.80^{+0.06}_{-0.07}$& $1.75^{+0.22}_{-0.26}$ & - & $1.5^{+0.4}_{-0.6}$ & $1.7^{+1.1}_{-0.9}$ \\ \hline
SDSS & $b,\alpha_0,\alpha_1$  & 1.6 & $0.76^{+0.08}_{-0.07}$ & $0.28^{+0.07}_{-0.05}$ & $0.80^{+0.08}_{-0.11}$& $1.18^{+0.37}_{-0.23}$ & - & $2.3^{+0.3}_{-0.5}$ & $2.9^{+1.1}_{-1.0}$ \\ \hline \hline \hline

\multicolumn{2}{|l}{$3\times2$pt, fixed $\nu$} & & $0.80^{+0.02}_{-0.02}$ & $0.26^{+0.02}_{-0.03}$ & $0.85^{+0.06}_{-0.05}$ & $1.54^{+0.09}_{-0.10}$ & \multicolumn{3}{l|}{\citet{keypaper}} \\ \hline
\end{tabular}
\end{ruledtabular}
\end{center}
\end{table*}

We perform likelihood analyses, i.e.~we determine the probability of finding our fiducial data vectors (\autoref{sec:measurement}) as a function of the parameters of our model (\autoref{sec:model}) and given their covariance (\autoref{sec:covariance}). We use models of different complexity for the connection of galaxies and matter density -- one with linear bias only ($b$), one adding stochasticity ($b,r$), and one allowing for density dependence of stochasticity ($b,\alpha_0,\alpha_1$) (for details see \autoref{sec:modeltracers}). Our philosophy, decided with parameters still blinded, will be to compare these models via their Bayesian evidence (\autoref{sec:evidence}) and report the results for models that are supported by the data.

\autoref{fig:data_models} shows constraints on the matter density $\Omega_m$, the amplitude of late-time structure formation $\sigma_8$, and galaxy bias of the \textsc{redMaGiC} tracer galaxies, when marginalizing over all remaining model parameters. Confidence limits are summarized in \autoref{tab:results}. The stochastic models, favored by the data (see next subsection), constrain the matter density consistently as $\Omega_m=0.26^{+0.04}_{-0.03}$ for the $(b,r)$ model and $\Omega_m=0.28^{+0.05}_{-0.04}$ for the $(b,\alpha_0,\alpha_1)$ model from DES data. The degeneracy directions of $\Omega_m$-$\sigma_8$ for the two stochasticity models are different, thus leading to a higher central value of $\sigma_8=0.97^{+0.07}_{-0.06}$ for $(b,r)$ and a lower $\sigma_8=0.80^{+0.06}_{-0.07}$ for $(b,\alpha_0,\alpha_1)$. 

Bias models and cosmology are thus interdependent: a prior, even a weak one, on the values of stochasticity parameters would significantly improve these cosmological constraints -- if galaxies have less stochasticity, the relevant regions of the green and orange contours in \autoref{fig:data_models} are closer to the black region. Likewise, external information on cosmological parameters allows to constrain bias parameters and, potentially, even choose between the bias models. If the true cosmology is $\Omega_m=0.3$ and $\sigma_8=0.8$, both the $b$ and the $b,r$ model are $\approx2\sigma$ off, while the $b,\alpha_0,\alpha_1$ model is consistent. 

For a sense of how these results compare to two-point function measurements, \autoref{fig:data_models} contains constraints from the three tomographic auto- and cross-correlation functions of DES \textsc{redMaGiC} galaxy positions and source galaxy shapes \citep[3x2pt,][]{keypaper}.\footnote{We use the version of the likelihood that does not vary neutrino mass, as in our counts and lensing in cells analysis. The galaxy bias parameter plotted is the mean bias of \textsc{redMaGiC} galaxies in the first two bins ($z=0.15-0.3$ and $0.3-0.45$), weighted 1:2, which is not quite the same as the bias of our single $z_T=0.2-0.45$ lens bin.} The 3x2pt contours are tighter than the constraints that counts and lensing in cells yield. This is due to the wide freedom on stochasticity parameters and models that we have allowed: if we \emph{could} fix the stochasticity (such as in the black contour with $r=1$), the smaller scale density PDF measurements would yield highly competitive cosmological constraints. It is clear from this and the different degeneracies that a joint analysis would result in improved constraints -- yet we are lacking a covariance matrix and inference pipeline to perform this at this point. Prima facie, the 3x2pt constraint is consistent with any of the bias models, and indicates a relatively small stochasticity, i.e.~a point in parameter space close to where the black, green, and orange contours intersect.

Finally, we compare the results from our DES Y1 and SDSS analysis. Within their mutual uncertainty, the two independent data sets provide consistent measurements of cosmological parameters. It is less clear whether the bias model of \textsc{redMaGiC} galaxies in SDSS and DES is identical, a question we turn to in the following subsection.

\subsection{Results on bias and stochasticity}

The Bayes factors for the stochastic models, i.e.~the ratio of their evidence over the evidence of the $b$ only model, are 3.6 ($b,r$) and 2.5 ($b,\alpha_0,\alpha_1$). This means that there is substantial evidence, as defined by the Jeffreys scale, for stochasticity in the count of \textsc{redMaGiC} galaxies at fixed projected matter density within 20' apertures and with a redshift range of $z_T=0.2-0.45$. Similar observations are made in SDSS, with Bayes factors 2.8 and 4.5 for the introduction of the stochastic models. The data thus prefers a model with stochasticity, but at an odds ratio of $\approx3:1$, the preference is not very conclusive. 

The DES constraint on $r$ is $r=0.77^{+0.10}_{-0.13}$. In likelihood runs of the $(b,r)$ model on log-normal mocks with no stochasticity (Appendix \ref{sec:lognormallikelihood}), we find smaller central values for $r$ than this in 3 out of 40 independent realizations.

We note that this finding is \emph{not} in conflict with the nondetection of stochasticity in the \citet{keypaper} 3x2pt analysis, and the associated explicit tests for consistency of the clustering and galaxy-galaxy lensing constraints on bias \citep{wthetapaper,gglpaper}. Those analyses use significantly larger scales ($>27'$ and $>45'$ in the lowest lens redshift bins, corresponding to $>8h^{-1}$ comoving Mpc for clustering, and $>12h^{-1}$ comoving Mpc for galaxy-galaxy lensing), on which stochasticity, if present, is expected to be small. Our statistic in sensitive to stochasticity on scales smaller or equal to the radius $\theta_T=20'$ of the apertures inside which we count tracer galaxies. Physically, this corresponds to $\lesssim3.5-7h^{-1}$ comoving Mpc in the $z_T=0.2-0.45$ redshift range. Uncertainty as to whether the nonstochastic bias model would be sufficient on these smaller scales was a primary reason for the conservative 3x2pt scale cuts \citep{methodpaper}.

\begin{figure}
  \includegraphics[width=\linewidth]{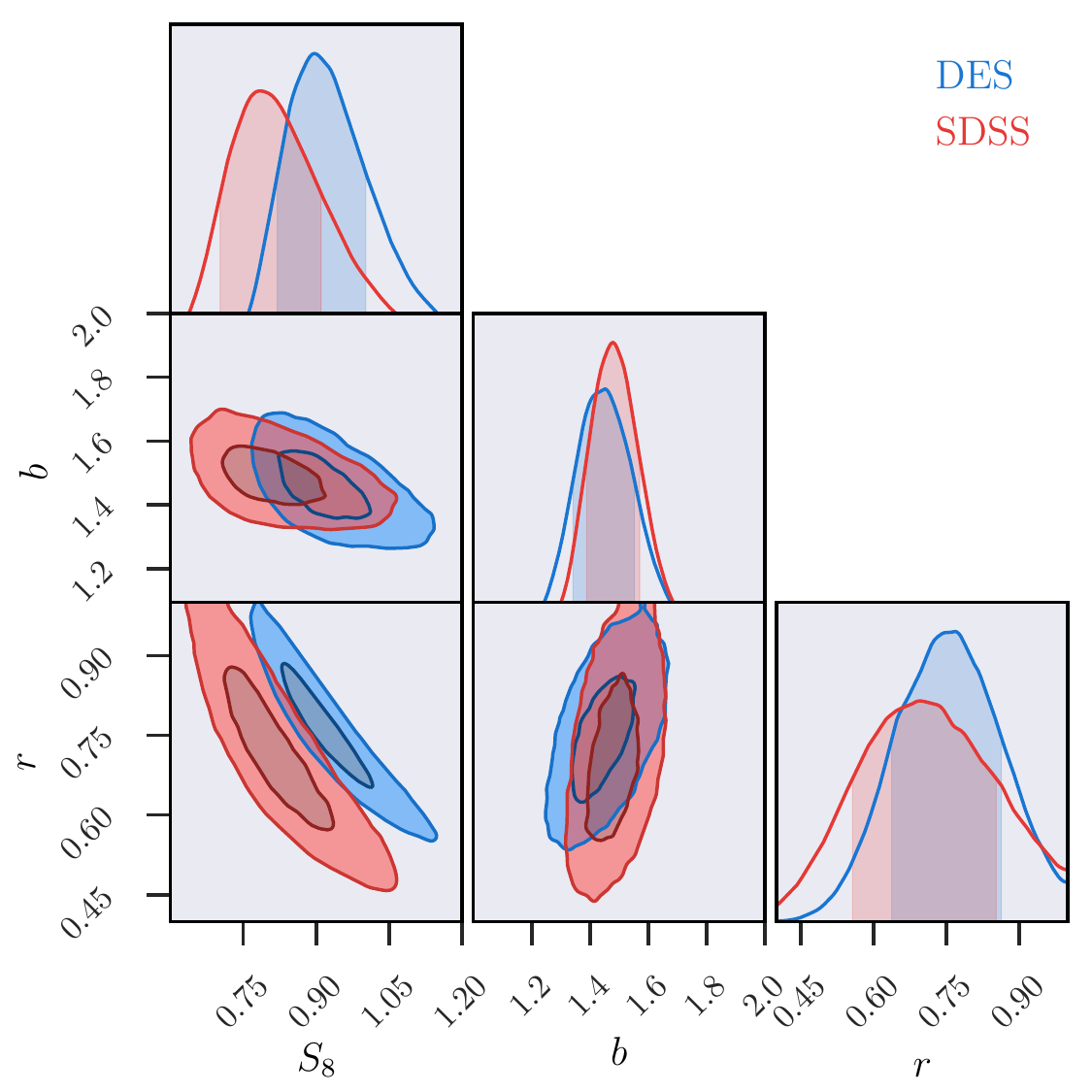}
  \caption{Constraints on $S_8=\sigma_8\sqrt{\Omega_m/0.3}$, bias and stochasticity of the \textsc{redMaGiC} tracer galaxies from density split lensing and counts-in-cells in DES Y1 (blue) and SDSS (red).}
\label{fig:rplot}
\end{figure}

\autoref{fig:rplot} shows constraints on $S_8\equiv\sigma_8\sqrt{\Omega_m/0.3}$, bias and stochasticity of the tracer galaxies, in both DES and SDSS. The deviation of $r$ from unity is at the $\approx2\sigma$ level. Individual parameter constraints are consistent, while there is a hint for a lower value of $r$ in SDSS at fixed cosmology. Note that the primary degeneracy of $S_8$ is not with bias, but with stochasticity -- even a mildly informative prior on $r$ would significantly lower uncertainty on $S_8$. For cosmological constraints close to those of the 3x2pt analysis ($S_8=0.80\pm0.02$ for the run with fixed $\nu$ mass), the counts and lensing in cells data is well fit by small stochasticity.

For the more complex $\alpha_0,\alpha_1$ model of density dependent stochasticity, which is not significantly preferred to $r$ by the data anywhere, the situation is qualitatively similar. The SDSS data does not constrain these parameters very well, especially $\alpha_1$ (cf.~\autoref{tab:results}), but there is an indication of super-Poissonian scatter in galaxy count at fixed matter density, the amplitude of which increases with density, broadly consistent with the effect of a single stochasticity parameter $r$ \citep[][their figure 6]{Oliver}.

It is difficult to compare this tentative detection of stochasticity on the $\lesssim7h^{-1}$ comoving Mpc aperture scale to the literature. Various works have found levels of stochasticity that are broadly consistent, using a range of samples and scales in numerical simulations \citep[e.g.][]{2001MNRAS.320..289S,2009MNRAS.396.1610B} and data \citep[e.g.][]{2002ApJ...577..604H,2005MNRAS.356..247W,2007ApJ...664..608W,2007A&A...461..861S,2008MNRAS.385.1635S,Leauthaud2017}. The comparison of low-$z$ galaxy clustering and galaxy-galaxy lensing in DES SV on scales above $4 h^{-1}$ comoving Mpc provided similar hints of $r<1$ (\cite{2016MNRAS.455.4301C,2016arXiv160908167P}, see also \cite{2016MNRAS.456.3213G}). Even those studies that found no evidence for $r\neq 1$ do not exclude a mild stochasticity on the relevant scales within their uncertainties \citep{2012ApJ...750...37J,2013MNRAS.433.1146C}. Most of these studies use two-point correlations, which means their results on stochasticity would have to be transformed to aperture statistics using a numerical model or simulations.

Note that we do not attempt to combine the DES and SDSS results because, without more detailed study, it is not certain that the \textsc{redMaGiC} samples trace the exact same galaxy populations. A larger stochasticity of \textsc{redMaGiC} galaxies in SDSS, if at all significant, could also be due to correlations of the \textsc{redMaGiC} density with SDSS observational systematics that, unlike in the case of DES \citep{wthetapaper}, has not been removed. 

\subsection{Test for excess skewness of matter density}

As described in \autoref{sec:parameters}, we can allow for the skewness of the projected, smoothed matter density field, $S_3$, to be a free parameter in our likelihood, rather than predicting it from perturbation theory. 

We first test whether the introduction of this additional parameter to our model is justified by the data. The Bayes factor of the extended models with $\Delta S_3$ as a free parameter, relative to any of the three models for the connection of galaxies and matter with fixed $S_3$, is smaller than unity, both on DES and on SDSS runs. This indicates no evidence that such an extension is required.

\begin{figure*}
  \includegraphics[width=0.9\linewidth]{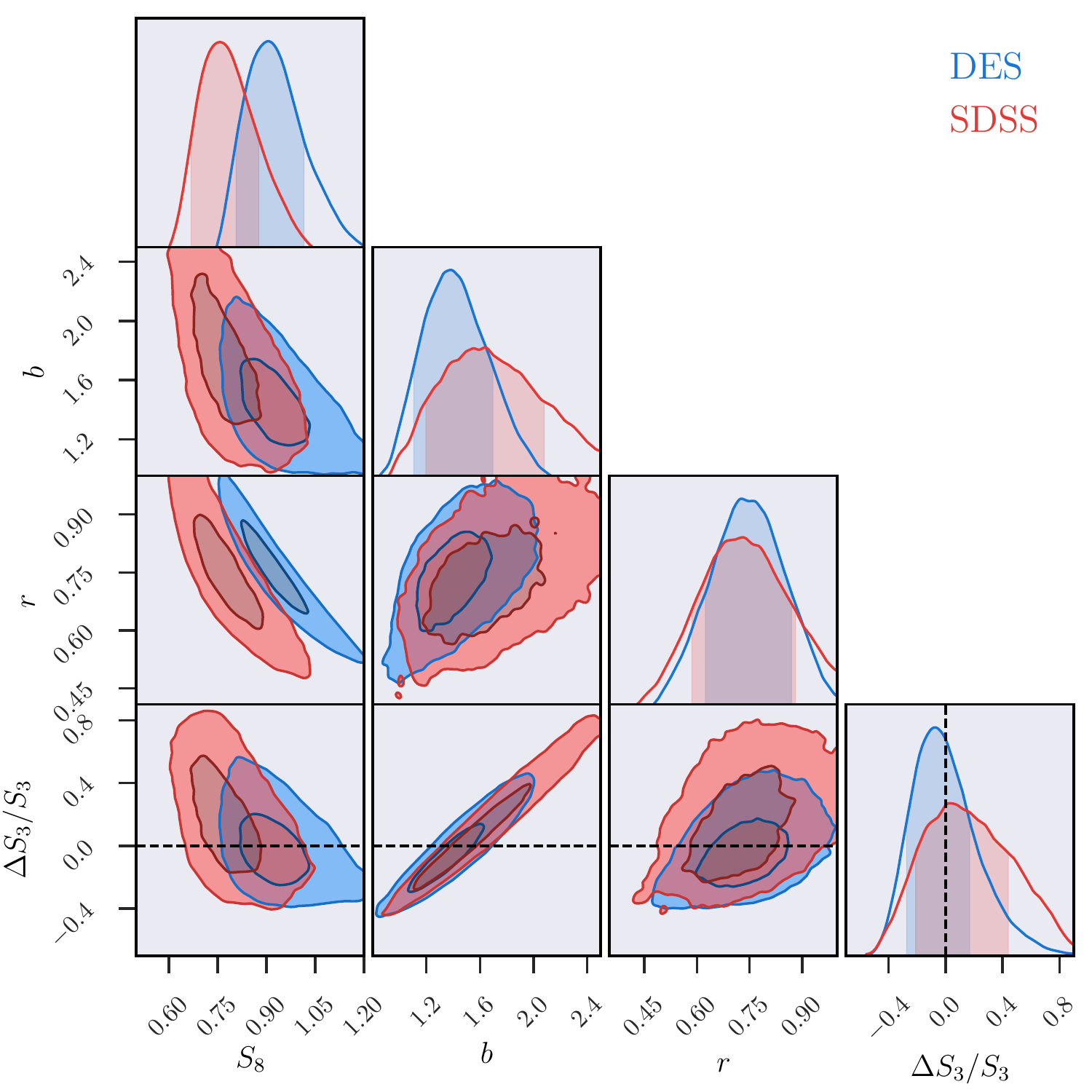}
  \caption{Constraints on excess skewness of the projected matter density field $\Delta S_3$, amplitude of structure $S_8=\sigma_8\sqrt{\Omega_m/0.3}$, bias and stochasticity of the \textsc{redMaGiC} tracer galaxies from density split lensing and counts-in-cells in DES Y1 (blue) and SDSS (red). Bayesian model comparison indicates that the introduction of $S_3$ as a free parameter, rather than fixing it from perturbation theory calculations (the $\Delta S_3=0$ indicated by dashed lines), is not necessary to describe the data.}
\label{fig:s3plot}
\end{figure*}

\begin{table*}
\caption{Constraints from counts and lensing in cells likelihood runs with skewness of the matter density field $S_3$ as a free parameter. Bayes factors are quoted relative to the $b,r$ model without free $S_3$. SDSS does not constrain the two-parametric stochasticity model $\alpha_0,\alpha_1$ within our sampling range.}
\label{tab:resultsS3}
\begin{center}
\begin{ruledtabular}
\begin{tabular}{| l  l | r | r | r | r | r | r | r | r | r |}
\hline
Data   & Model & Bayes  & $S_8$ & $\Omega_m$ & $\sigma_8$ & $b$ & $r$ & $\alpha_0$ & $\alpha_1$ & $\Delta S_3/S_3$ \\ 
& & factor & & & & & & & & \\  \hline
DES   & $b,r,\Delta S_3$  &           0.3     & $0.91^{+0.10}_{-0.10}$ & $0.26^{+0.07}_{-0.05}$ & $0.96^{+0.17}_{-0.13}$& $1.37^{+0.32}_{-0.27}$ & $0.72^{+0.14}_{-0.10}$ & - & - & $-0.08^{+0.25}_{-0.20}$ \\ \hline
SDSS & $b,r,\Delta S_3$  &            0.4 &  $0.76^{+0.12}_{-0.09}$ &$0.28^{+0.09}_{-0.07}$ & $0.72^{+0.17}_{-0.13}$& $1.64^{+0.44}_{-0.46}$ & $0.73^{+0.15}_{-0.15}$ & - & - & $0.06^{+0.40}_{-0.27}$ \\ \hline \hline

DES & $b,\alpha_0,\alpha_1,\Delta S_3$ & 0.3 &$0.80^{+0.06}_{-0.05}$ & $0.26^{+0.07}_{-0.05}$ & $0.86^{+0.10}_{-0.13}$ & $1.48^{+0.41}_{-0.32}$ & - & $1.7^{+0.4}_{-0.6}$ & $2.0^{+1.1}_{-0.9}$ & $-0.18^{+0.25}_{-0.22}$ \\ \hline
SDSS & $b,\alpha_0,\alpha_1,\Delta S_3$ & 0.6 & $0.78^{+0.07}_{-0.08}$ & $0.27^{+0.09}_{-0.05}$ & $0.80^{+0.14}_{-0.12}$ & $0.98^{+0.49}_{-0.17}$ & - & $2.5^{+0.3}_{-0.5}$ & $>2.1$ (68\% c.l.) & $-0.14^{+0.44}_{-0.39}$ \\ \hline \end{tabular}
\end{ruledtabular}
\end{center}
\end{table*}

If we still perform a likelihood analysis of the extended models despite of this, we can find constraints on $S_3$. For the $(b,r)$ model, these are shown in \autoref{fig:s3plot}. DES Y1 and SDSS provide independent constraints, both of which are consistent with $\Delta S_3=0$. The DES constraint, $\Delta S_3/S_3=-0.08^{+0.25}_{-0.20}$ is significantly tighter, primarily due to the fact that the lensing signal that breaks the degeneracy of bias and skewness is measured with higher signal-to-noise ratio. 

Generalizing the likelihood to a two-parametric $\alpha_0,\alpha_1$ model for stochasticity and leaving $S_3$ free yields similarly tight constraints on from DES data, again consistent with no excess skewness at $\Delta S_3=-0.18^{+0.25}_{-0.22}$. In SDSS, the posterior distribution of $\alpha_1$ is not constrained in this model within our sampling range.

The joint interpretation of these results is that we find no hints for an excess or deficit in skewness of the matter density relative to our $\Lambda$CDM perturbation theory prediction. This conclusion is largely independent of the bias model we choose, and tested at the 20 percent level. Future analyses with larger data sets or joint constraints from counts and lensing in cells and additional probes could provide much tighter constraints on $S_3$.

\section{Conclusions}
\label{sec:conclusions}

We perform the first cosmological analysis using counts and lensing in cells, a method that constrains the matter density PDF with the combination of counts-in-cells and gravitational lensing signals around low and high density lines of sight.  We do this by creating quintiles based on the galaxy counts in apertures and evaluating the stacked lensing for each quintile.

This analysis is tested extensively, using the perturbation theory model of \citet{Oliver}, by applying it to log-normal density fields and realistic $N$-body mock catalogs from the \textit{Buzzard} simulations. Robustness to systematics in the data and choices in the analysis is confirmed by a series of tests, performed while blind to the cosmological parameter values constrained by our data.

Applying the analysis to data vectors from DES and SDSS, we find that
\begin{itemize}
\item The data prefer stochasticity (beyond Poisson sampling) in galaxy count at fixed matter density, on the $<20'$ or $<3.5-7h^{-1}$ comoving Mpc smoothing scale of our aperture. This is indicated by a Bayesian model comparison of bias parametrizations, whose odds ratios of $\approx3-4:1$ in favor of stochastic models represent substantial evidence on the Jeffreys scale. Our data does not discriminate between the two different models of stochasticity we apply, one with the correlation coefficient of galaxy and matter density $r$ as a free parameter, and one with two parameters $\alpha_0,\alpha_1$ that allow for a density dependent deviation from Poissonian noise.
\item Either of these models yields consistent DES constraints on the cosmic matter density, $\Omega_m=0.26^{+0.04}_{-0.03}$ for the $(b,r)$ model and $\Omega_m=0.28^{+0.05}_{-0.04}$ for the $(b,\alpha_0,\alpha_1)$ model. These are consistent with, and not much less constraining than, results from the three tomographic auto- and cross-two-point correlation functions of galaxy counts and gravitational shear (3x2pt) in \citet{keypaper}, $\Omega_m=0.26^{+0.02}_{-0.03}$.
\item The degeneracy direction of $\Omega_m$ and $\sigma_8$ and best-fit values of $\sigma_8$ depend on the stochasticity model -- for $(b,r)$ we find a higher amplitude $\sigma_8=0.97^{+0.07}_{-0.06}$ and $\sigma_8=0.86^{+0.06}_{-0.05}$ from DES and SDSS than for the $(b,\alpha_0,\alpha_1)$ model. The latter has a central value of $\sigma_8=0.80$ and somewhat larger uncertainty. In turn, this means that external constraints on cosmology would yield an improved model selection and posterior on stochasticity and vice versa. For small stochasticity, both models are consistent with each other and with the 3x2pt constraint. The assumption of linear bias with no stochasticity, which is however mildly disfavored by the data, would allow constraints from the matter PDF that are competitive with 3x2pt. Thus if we \emph{could} use prior knowledge to select a particular model or a narrower range of possible bias parameter values, this would greatly improve the constraining power on cosmological parameters.
\item Because counts and lensing in cells measure the width and skewness of the matter PDF independently, they can be used to constrain $S_3=\langle\delta^2\rangle^2/\langle\delta^3\rangle$, for which $\Lambda$CDM and gravitational collapse make a very stable prediction. We find that the deviation from this prediction, $\Delta S_3/S_3$, is consistent with zero, within 20-30 percent uncertainty, in any bias and stochasticity model. 
\end{itemize}

This analysis of the density PDF opens several avenues for future research:
\begin{itemize}
\item Towards tightly constraining cosmological parameters and performing model independent tests on higher moments of the matter density PDF generated by gravity -- this could be achieved best with external priors or data on stochasticity or bias, or in the regime of larger scales or larger tracer density, where predicted signals are less sensitive to the bias model.
\item Towards discriminating between and constraining parameters of bias and stochasticity models -- optimally by analyzing smaller scales, which will require improved models of the matter density PDF and baryonic effects on it.
\item Towards joint analyses, e.g.~with two-point functions or CMB lensing data, that can break cosmological, galaxy bias parameter and nuisance parameter degeneracies.
\end{itemize}
Given a suitable model and analysis framework, counts and lensing in cells with present and imminently available data will allow tight constraints on cosmological parameters and the hierarchy of moments of the matter density field.

\section*{Acknowledgements}

D.G. thanks Yao-Yuan Mao, Cora Uhlemann, Zvonimir Vlah, and numerous members of the DES WL, LSS and Theory working groups for helpful discussions. 

Support for D.G. was provided by NASA through Einstein Postdoctoral Fellowship grant
number PF5-160138 awarded by the Chandra X-ray Center, which is
operated by the Smithsonian Astrophysical Observatory for NASA under
contract NAS8-03060. O.F. acknowledges funding by SFB-Transregio 33 `The Dark Universe'
by the Deutsche Forschungsgemeinschaft (DFG) and the DFG Cluster of Excellence `Origin
and Structure of the Universe'.

Funding for the DES Projects has been provided by the U.S. Department of Energy, the U.S. National Science Foundation, the Ministry of Science and Education of Spain, 
the Science and Technology Facilities Council of the United Kingdom, the Higher Education Funding Council for England, the National Center for Supercomputing 
Applications at the University of Illinois at Urbana-Champaign, the Kavli Institute of Cosmological Physics at the University of Chicago, 
the Center for Cosmology and Astro-Particle Physics at the Ohio State University,
the Mitchell Institute for Fundamental Physics and Astronomy at Texas A\&M University, Financiadora de Estudos e Projetos, 
Funda{\c c}{\~a}o Carlos Chagas Filho de Amparo {\`a} Pesquisa do Estado do Rio de Janeiro, Conselho Nacional de Desenvolvimento Cient{\'i}fico e Tecnol{\'o}gico and 
the Minist{\'e}rio da Ci{\^e}ncia, Tecnologia e Inova{\c c}{\~a}o, the Deutsche Forschungsgemeinschaft and the Collaborating Institutions in the Dark Energy Survey. 

The Collaborating Institutions are Argonne National Laboratory, the University of California at Santa Cruz, the University of Cambridge, Centro de Investigaciones Energ{\'e}ticas, 
Medioambientales y Tecnol{\'o}gicas-Madrid, the University of Chicago, University College London, the DES-Brazil Consortium, the University of Edinburgh, 
the Eidgen{\"o}ssische Technische Hochschule (ETH) Z{\"u}rich, 
Fermi National Accelerator Laboratory, the University of Illinois at Urbana-Champaign, the Institut de Ci{\`e}ncies de l'Espai (IEEC/CSIC), 
the Institut de F{\'i}sica d'Altes Energies, Lawrence Berkeley National Laboratory, the Ludwig-Maximilians Universit{\"a}t M{\"u}nchen and the associated Excellence Cluster Universe, 
the University of Michigan, the National Optical Astronomy Observatory, the University of Nottingham, The Ohio State University, the University of Pennsylvania, the University of Portsmouth, 
SLAC National Accelerator Laboratory, Stanford University, the University of Sussex, Texas A\&M University, and the OzDES Membership Consortium.

Based in part on observations at Cerro Tololo Inter-American Observatory, National Optical Astronomy Observatory, which is operated by the Association of 
Universities for Research in Astronomy (AURA) under a cooperative agreement with the National Science Foundation.

The DES data management system is supported by the National Science Foundation under Grant Numbers AST-1138766 and AST-1536171.
The DES participants from Spanish institutions are partially supported by MINECO under grants AYA2015-71825, ESP2015-88861, FPA2015-68048, SEV-2012-0234, SEV-2016-0597, and MDM-2015-0509, 
some of which include ERDF funds from the European Union. IFAE is partially funded by the CERCA program of the Generalitat de Catalunya.
Research leading to these results has received funding from the European Research
Council under the European Union's Seventh Framework Program (FP7/2007-2013) including ERC grant agreements 240672, 291329, and 306478.
We  acknowledge support from the Australian Research Council Centre of Excellence for All-sky Astrophysics (CAASTRO), through project number CE110001020.
SH acknowledges support by the DFG cluster of excellence \lq{}Origin and Structure of the Universe\rq{} (\href{http://www.universe-cluster.de}{\texttt{www.universe-cluster.de}}).

This manuscript has been authored by Fermi Research Alliance, LLC under Contract No. DE-AC02-07CH11359 with the U.S. Department of Energy, Office of Science, Office of High Energy Physics. The United States Government retains and the publisher, by accepting the article for publication, acknowledges that the United States Government retains a non-exclusive, paid-up, irrevocable, world-wide license to publish or reproduce the published form of this manuscript, or allow others to do so, for United States Government purposes.

This paper has gone through internal review by the DES collaboration.

\addcontentsline{toc}{chapter}{Bibliography}
\bibliographystyle{mnras}
\bibliography{ref,des_y1kp_short}


\appendix

\section{Bernoulli masking of a count with Poissonian noise}
\label{app:bernoulli}
Assume that the true number of galaxies $N$ in a randomly selected volume follows a Poisson distribution around the expectation value $\bar{N}$ (which is conditional on the matter density inside the volume),
\begin{equation}
P(N)=\frac{e^{-\bar{N}}}{N!}\times \bar{N}^N \; .
\label{eqn:pn}
\end{equation}
If incompleteness and masking act on each galaxy with some detection probability $p$, then the observed number of galaxies in that volume $N_{\rm obs}$ is related to $N$ by a Bernoulli process,
\begin{equation}
P(N_{\rm obs}|N)=\frac{N!}{N_{\rm obs}!(N-N_{\rm obs})!}p^{N_{\rm obs}} (1-p)^{N-N_{\rm obs}} \; .
\label{eqn:pnobsn}
\end{equation}
The resulting probability distribution for $N_{\rm obs}$,
\begin{eqnarray}
P(N_{\rm obs})&=&\sum_N P(N_{\rm obs}|N) P(N) \\ \nonumber
&=& \frac{e^{-\bar{N}}}{N_{\rm obs}!}\left(p\bar{N}\right)^{N_{\rm obs}} \sum_{N=N_{\rm obs}}^{\infty}\frac{[(1-p)\bar{N}]^{N-N_{\rm obs}}}{(N-N_{\rm obs})!} \\ \nonumber
&=&\frac{e^{-p\bar{N}}}{N_{\rm obs}!}(p\bar{N})^{N_{\rm obs}} \; ,
\end{eqnarray}
is again Poisson-distributed around the expectation value $p\bar{N}$. Masking-induced scatter in $N_{\rm obs}$ is therefore correctly described by Poisson noise. This justifies the accounting for masking used in \autoref{sec:troughfinding}. Note that we have implicitly assumed intrinsic Poisson noise and independent random removal of galaxies due to masking, when in fact clustering implies a more complex form of Equations \ref{eqn:pn} and \ref{eqn:pnobsn}, yet at a level not relevant at first order.

\section{Choice of log-normal parameters for the simulated density and convergence fields}
\label{sec:lognormal_params}

In this Appendix, we describe how we chose the log-normal parameters (i.e., the minimum allowed values of the log-normal PDFs) and power spectra for generating simulated convergence and density fields that closely match the 2-point and 3-point auto- and cross-correlation statistics we expect from our fiducial model.

\subsection{Configuration of FLASK maps of simulated density contrast}

As we have shown in \citet{Oliver}, at a scale of $\theta_A=20'$ the PDF of the smoothed matter density contrast $\delta_{m,T}$ is well described by a zero-mean shifted log-normal distribution, when the parameters of the log-normal PDF are chosen such as to match the variance and skewness of $\delta_{m,T}$. For the cosmological parameters and redshift distributions of \textsc{redMaGiC} galaxies in the \emph{Buzzard} simulations, the variance and perturbation theory prediction for skewness result in a log-normal shift parameter of $\delta_0 = 0.669$. 

We input this parameter and an angular power spectrum computed by Limber's approximation from our fiducial matter power spectrum to the FLASK tool to generate maps of projected density contrast. FLASK will generate \textsc{healpix} maps of $\delta_{m,\mathrm{2D}}$ such that on the pixel scale of these maps $\delta_{m,\mathrm{2D}}$ is a log-normal random variable with $\delta_0 = 0.669$. In our case, the pixel scale is much smaller than the smoothing scale $\theta=20'$ for which we determined the log-normal parameter. Fortunately, a limit theorem derived by \citet{Szyszkowicz2009} ensures that a smoothed version of a log-normal random field, while not formally log-normal, is still well described by a log-normal field with the same shift parameter. We have verified this approximation to be accurate in our situation.

\subsection{Configuration of FLASK maps of simulated convergence fields}
\label{sec:flaskfields}
For the convergence field $\kappa$, defined in \autoref{eqn:kappa}, we again need to fix the parameters of the log-normal simulations. The expectation value of $\kappa_{\theta}$ around overdense or underdense lines of sight with given matter contrast $\delta_{m,T}$ is fully determined by the moments $\left\langle\delta_{m,T}^n\right\rangle$ and joint moments $\langle \delta_{m,T}^n\ \kappa_{\theta}\rangle$, $n\geq1$.

Hence, the expectation value of the density split lensing signal obtains contributions only from those redshifts where the distribution of tracers and the lensing efficiency kernel $W_{s}(w)$ overlap. The covariance of the signal, however, also has contributions from foreground and background structures at distances where only $W_{s}(w)$ is non-zero.

As pointed out by e.g.~\citet{2016MNRAS.459.3693X}, 2D projections of the 3D density contrast such as $\delta_{m,T}$ and $\kappa_{\theta}$ are not well described by a joint log-normal distribution if the kernels of their projection have a very different width along the line-of-sight. In our situation, the lensing kernel $W_{s}(w)$ is significantly broader than the distribution in comoving distance of our tracer galaxies. In order to still accurately match the higher-order statistics of $\delta_{m, \theta}$ and $\kappa_{\theta}$ with the FLASK log-normal simulations, we split $\kappa$ into a contribution from the overlap of tracers and $W_{s}(w)$ and a contribution from foreground and background structures, i.e.
\begin{equation}
\kappa(\hat n) = \kappa_{\mathrm{overlap}}(\hat n) + \kappa_{\mathrm{non-overlap}}(\hat n)
\end{equation}
with
\begin{equation}
\kappa_{\mathrm{overlap}}(\hat n) = \int_{w_{\min}}^{w_{\max}} \mathrm{d}w\ W_{s}(w)\ \delta_{m,\mathrm{3D}}(w\hat{\mathbf{n}}, w)
\label{eqn:koverlap}
\end{equation}
and
\begin{eqnarray}
\kappa_{\mathrm{non-overlap}}(\hat n) &=& \int_{0}^{w_{\min}} \mathrm{d}w\ W_{s}(w)\ \delta_{m,\mathrm{3D}}(w\hat{\mathbf{n}}, w) \nonumber \\
&& + \int_{w_{\max}}^{\infty} \mathrm{d}w\ W_{s}(w)\ \delta_{m,\mathrm{3D}}(w\hat{\mathbf{n}}, w)\ .
\end{eqnarray}
Here, $w_{\min}$ and $w_{\max}$ are the minimum and maximum comoving distances of our tracer population. We separately compute the power spectra of $\kappa_{\mathrm{overlap}}$ and $\kappa_{\mathrm{non-overlap}}$ using Limber's approximation and our fiducial matter power spectrum.

Instead of approximating the distribution of $\kappa_{\mathrm{overlap}}$ and $\delta_{m,T}$ with a joint log-normal distribution, we point out in \citet{Oliver} that it is better to further split $\kappa_{\mathrm{overlap}}$ into two contributions as
\begin{equation}
\kappa_{\mathrm{overlap}} = \kappa_{\mathrm{log-normal}} + \kappa_{\mathrm{uncorr.}}\ ,
\end{equation}
where only $\kappa_{\mathrm{log-normal}}$ is a log-normal variable and $\kappa_{\mathrm{uncorr.}}$ is assumed to be completely uncorrelated with $\delta_{m,T}$. The reason for this is the following: The density split lensing signal mainly depends on the moments $\left\langle \delta_{m,T}^2\right\rangle$ and $\left\langle \delta_{m,T}^3\right\rangle$ as well as
$\left\langle \delta_{m,T}\ \kappa_{\mathrm{overlap}} \right\rangle$ and $\left\langle \delta_{m,T}^2\ \kappa_{\mathrm{overlap}}\right\rangle$. Requiring our simulated convergence fields to obey our analytic predictions of these moments would already fix the log-normal PDF for $\kappa_{\mathrm{overlap}}$. Importantly, it would also fix the variance $\left\langle \kappa_{\mathrm{overlap}}^2\right\rangle$. And in general, this variance will disagree with the variance of $\kappa_{\mathrm{overlap}}$ as defined in \autoref{eqn:koverlap} that is predicted from our power spectrum. Splitting $\kappa_{\mathrm{overlap}}$ into $\kappa_{\mathrm{log-normal}}$ and $\kappa_{\mathrm{uncorr.}}$ solves this disagreement, since we can use the above moments to fix the log-normal PDF of $\kappa_{\mathrm{log-normal}}$ and attribute part of the variance of $\kappa_{\mathrm{overlap}}$ to $\kappa_{\mathrm{uncorr.}}$ to keep the total variance in agreement with our power spectrum. We then use FLASK to simulate the contributions to $\kappa_{\mathrm{overlap}}$ as two distinct random fields. We assume that $\kappa_{\mathrm{log-normal}}$ and $\kappa_{\mathrm{uncorr.}}$ are uncorrelated and that their power spectra are simply proportional to that of $\kappa_{\mathrm{overlap}}$. The proportionality factors are determined such that the two power spectra sum up to the total power spectrum of $\kappa_{\mathrm{overlap}}$ and also such that the variance of $\kappa_{\mathrm{log-normal}}$ is indeed the one predicted by its log-normal PDF.

Finally, while in \citet{Oliver} we allow the log-normal PDF to vary depending on the scale $\theta$ of the convergence field, we have to choose a fixed scale in order to generate log-normal random fields with FLASK. We consider $\theta = 20'$ as a reasonable choice. At larger scales the log-normal PDF of our formalism quickly transitions to a Gaussian PDF anyway, and we do not consider smaller scales in our analysis. The log-normal shift parameters we get this way are $\kappa_0=0.0088,0.0150,0.0181$ for our three DES source redshift bins and $0.0094$ for the SDSS sources. The remaining ingredient needed by FLASK to generate the field $\kappa_{\mathrm{log-normal}}$ is its cross power spectrum with $\delta_{m,\mathrm{2D}}$. Since we assume all other contributions to the convergence to be uncorrelated to the tracer density contrast, this is just the cross power spectrum of the total convergence and $\delta_{m,\mathrm{2D}}$. Also, the theorem by \citet{Szyszkowicz2009} ensures that generating log-normal fields with the above values of $\kappa_0$ at the pixel scale is sufficient to obtain the same log-normal properties also on larger smoothing scales.

The contributions $\kappa_{\mathrm{uncorr.}}$ and $\kappa_{\mathrm{non-overlap}}$ enter the covariance of our signal via 2-point statistics only.  Hence, we just use the sum of their power spectra to generate a single Gaussian random field, i.e. no log-normal shift parameters have to be determined for these components. Since we include 3 source bins in our analysis, we also have to include their cross-power spectra in our FLASK configuration. The cross-power spectra of the non-overlap contributions to the convergence can be computed straightforwardly using Limber's approximation. The cross-power spectra of $\kappa_{\mathrm{overlap}}$ between different source bins are split to cross-power spectra of the fields $\kappa_{\mathrm{log-normal}}$ and the fields $\kappa_{\mathrm{uncorr.}}$ in a similar way as the auto-power spectra for identical source bins: this time we assume that $\delta_{m,\mathrm{2D}}$ and the two convergence fields $\kappa_{\mathrm{log-normal}, i}$ and $\kappa_{\mathrm{log-normal}, j}$ corresponding to source bins $i$ and $j$ have a joint log-normal distribution. We then compute their combined third order moment with the method presented in \citet{Oliver}, which fixes this PDF and allows us to compute the covariance of $\kappa_{\mathrm{log-normal}, i}$ and $\kappa_{\mathrm{log-normal}, j}$. The cross power spectra are then split between the $\kappa_{\mathrm{log-normal}}$ and $\kappa_{\mathrm{uncorr.}}$ contributions such that they add up to the total cross power spectrum of $\kappa_{\mathrm{overlap}, i}$ and $\kappa_{\mathrm{overlap}, j}$ while also giving the correct covariance between the $\kappa_{\mathrm{log-normal}}$ contributions in each source bin. No cross-correlation is assumed for $\kappa_{\mathrm{log-normal}}$ and $\kappa_{\mathrm{uncorr.}}$ between any combination of source bins.

\section{Dependence of signal on \textsc{redMaGiC} variant}
\label{app:sysmaps}
As discussed in \autoref{sec:des_redmagic}, there are two variants of \textsc{redMaGiC} catalogs in DES Y1 (based on either \texttt{MAG\_AUTO} or \texttt{MOF} photometry), and either can optionally be corrected with a set of weights that removes the correlation of galaxy density with observational systematics. Our fiducial choice, as in \citet{wthetapaper} for the redshift range we use, is the \texttt{MAG\_AUTO} catalog with these weights applied. In this Appendix, we repeat our measurements with the remaining three variants of the \textsc{redMaGiC} catalogs to see whether there are appreciable differences in the recovered signals.

\begin{figure}
  \includegraphics[width=\linewidth]{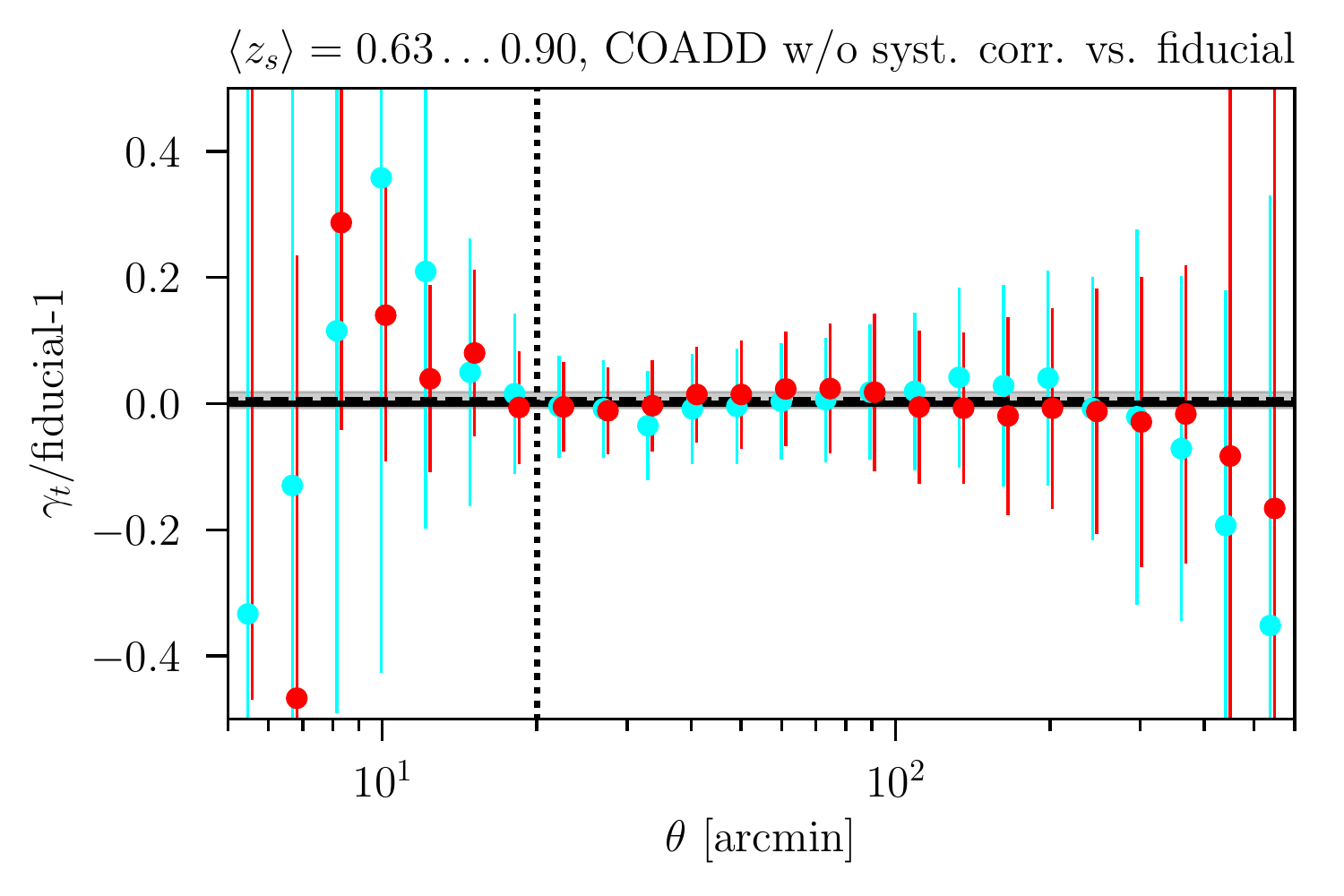}  
  \includegraphics[width=\linewidth]{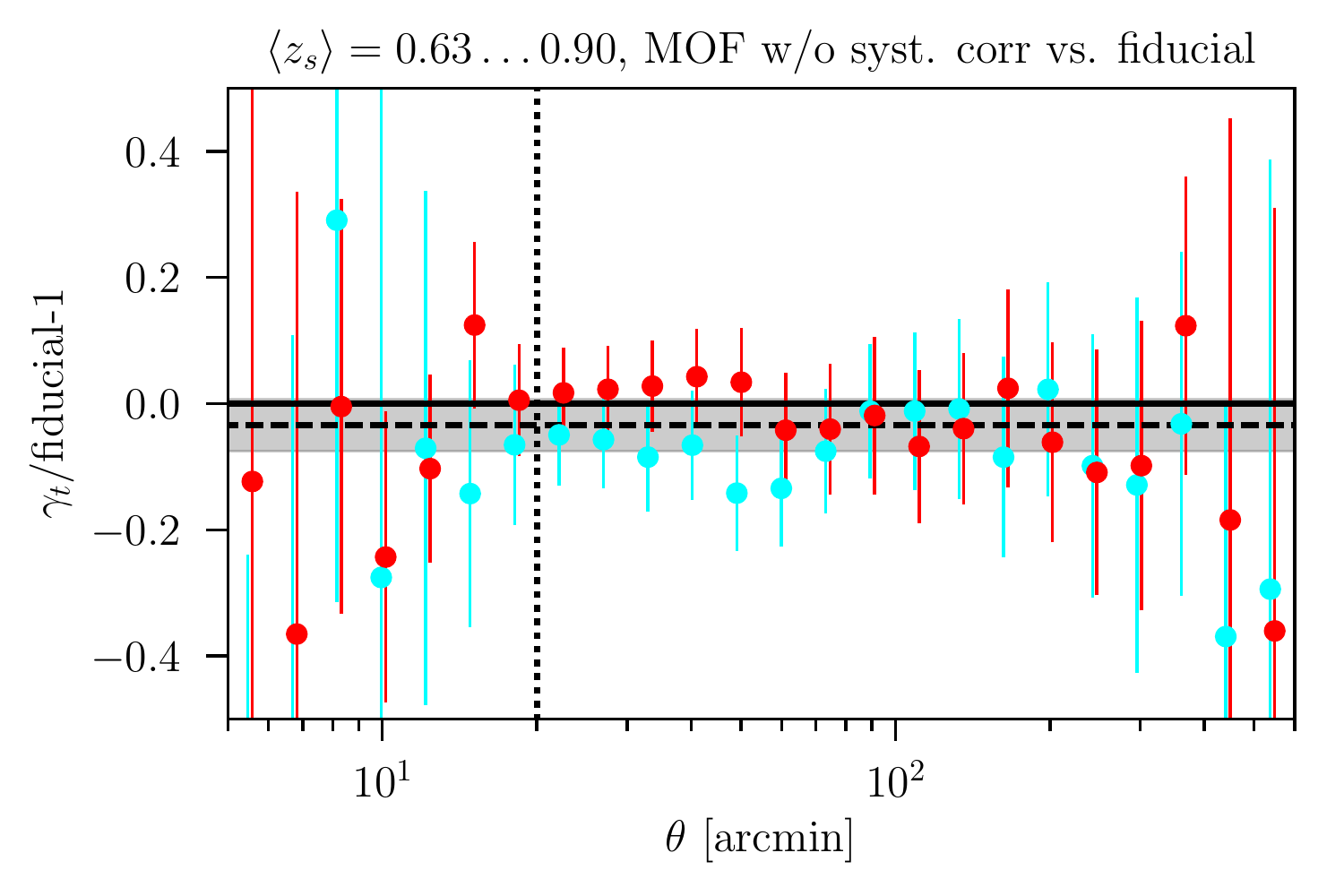}
  \caption{Ratio of tangential shear measured around troughs (cyan) and overdense lines of sight (red), identified using different variants of the \textsc{redMaGiC} catalog. The fiducial variant is based on coadd \texttt{MAG\_AUTO} photometry and corrects tracer density for its correlation with observational systematics. The top and bottom panel compare this to \textsc{redMaGiC} runs on the same photometry and multi-epoch MOF photometry, both without correcting for the correlation with systematics.}
\label{fig:redmagic_ratio}
\end{figure}

Results from this are shown in \autoref{tab:redmagic_ratio} and, for selected samples, in \autoref{fig:redmagic_ratio}. We find that the ratios of shear signals between the variants are consistent with unity, with a hint of lower signals in the \texttt{MOF} variant when not applying the weights to correct for systematics related density variations. This is consistent with the analysis of \citet{wthetapaper}, who found larger, significant correlations of galaxy density with observational systematics in this catalog.

\begin{table}
	\centering
	\caption{Shear ratios of \textsc{redMaGiC} variants}
	\label{tab:redmagic_ratio}
	\begin{tabular}{llrrr} 
		\hline
		$z_s$ bin & variant & $\gamma_t$ & $\gamma_{t,0}$ & $\gamma_{t,4}$ \\
		 &         &  ratio-1         &  ratio-1          &   ratio-1 \\
                  &         & $[10^{-2}]$& $[10^{-2}]$& $[10^{-2}]$ \\
        \hline
         $0.43-0.63$ & COADD, no corr. & $-0.2\pm2.7$ &  $0.7$ & $-1.1$ \\
         $0.63-0.90$ & COADD, no corr. &  $0.6\pm1.2$ &  $2.0$ &  $0.9$ \\
         $0.90-1.30$ & COADD, no corr  & $-1.2\pm1.3$ & $-0.2$ & $-2.2$ \\
        \hline
        $0.43-0.63$ & MOF, corr.      & $-4.1\pm7.1$ &  $7.5$ &$-15.7$ \\
        $0.63-0.90$ & MOF, corr.      & $-2.2\pm3.9$ & $-5.7$ & $-1.2$ \\
        $0.90-1.30$ & MOF, corr.      & $-1.2\pm4.3$ &  $5.5$ & $-7.8$ \\
        \hline
        $0.43-0.63$ & MOF, no corr.   & $-6.3\pm7.4$ &  $3.6$ &$-16.2$ \\
        $0.63-0.90$ & MOF, no corr.   & $-3.5\pm4.1$ & $-7.3$ &  $0.4$ \\
        $0.90-1.30$ & MOF, no corr.   & $-2.7\pm4.7$ &  $3.6$ & $-9.0$ \\
		\hline
	\end{tabular}
\end{table}

\section{Mock analysis of log-normal simulations}
\label{sec:lognormallikelihood}

We perform a set of tests of our model prediction code, our covariance estimation and likelihood pipeline using log-normal realizations of our data vector. 

We first check that the $\chi^2$ distributions of residuals of the individual log-normal realizations follow expectations. To this end, we add realizations of shape noise from random rotations of our actual source catalog to the log-normal data vectors. We have used $N_{\rm real}=960$ out of 1000 realizations of the sum of cosmic variance and shape noise to estimate the covariance matrix, and for these the mean $\chi^2$ when evaluated with the inverse of the estimated covariance matrix is equal, within uncertainties, to the number of degrees of freedom. 
For the additional 40 realizations that we have reserved for the purpose of testing, the mean $\chi^2$ is larger by a factor consistent with the inverse of \autoref{eqn:fah} \citep{Anderson,2007A&A...464..399H}. We apply this factor to our estimated covariance matrix for all following analyses. We also confirm that this correction is \emph{not} appropriate when estimating the cosmic variance and shape noise parts of our covariance independently and coadding them in the end, which is why we do not apply this (statistically desirable) procedure \citep[but cf.][for a possible way out]{2017arXiv170307786F}.

The mean data vector of all 1000 log-normal realizations is described well by the analytical prediction of \citet{Oliver}. The total $\chi^2$ of the residual of the mean vs. the model, at our fiducial scale cuts and for the DES analysis with three source redshift bins with 208 d.o.f., evaluated at the input cosmology and galaxy bias and the covariance matrix of \autoref{sec:covariance}, is 0.19, without the increase in the diagonal of the counts-in-cells covariance discussed in \autoref{sec:constructingcovariance}.
This is well below the statistical uncertainty $\left(\sqrt{\Var\left[\chi^2_{208\;\rm d.o.f}\right]}>20\right)$ and confirms that the analytical calculations used in \citet{Oliver} to get from input power spectra and skewness parameters for the matter and convergence fields to a prediction of our signals are derived and implemented correctly and to sufficient precision in our codes.

Next, we run a full likelihood analysis on the 40 reserved realizations to check the coverage, i.e.~whether the 68 percent confidence contour contains the input cosmology in a sufficient fraction of cases. Since nuisance parameters for measurement systematics are fixed to zero in these data vectors, we only vary $b, \alpha_0, \alpha_1, \Omega_{\rm m}, \sigma_8$, and in an additional run $\Delta S_3$ in these analyses.

\begin{figure}
  \includegraphics[width=\linewidth]{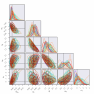}
  \caption{Realizations of likelihoods on 40 log-normal mocks, sampled with \textsc{emcee}, with the input parameters given by the dashed black lines. The coverage of these mock likelihoods is within expectations, i.e.~any input parameter lies within its marginalized 1$\sigma$ confidence interval $\approx68$ percent of the times.}
\label{fig:testlike_lognormal}
\end{figure}

\autoref{fig:testlike_lognormal} shows the resulting confidence contours. The fraction of times that these contain the true input cosmology (horizontal / vertical lines) is consistent with 68 percent. This statement is also true when adding the additional parameter $\Delta S_3$. The mean half-width of the marginalized confidence intervals for $(\Omega_{\rm m}, \sigma_8)$ are $(0.05, 0.07)$ and for $(\Omega_{\rm m}, \sigma_8,\Delta S_3)$ are $(0.07, 0.10, 0.28)$.

We repeat the test for the $(b,r)$ model, finding consistent coverage for $\sigma_8$ and marginally low coverage for $\Omega_{\rm m}$, potentially related to the asymmetry of the flat prior around $r$, which is correlated with $\Omega_{\rm m}$ (and which cannot physically take a value larger than 1, its true value in the log-normal simulations).

\begin{figure}
  \includegraphics[width=\linewidth]{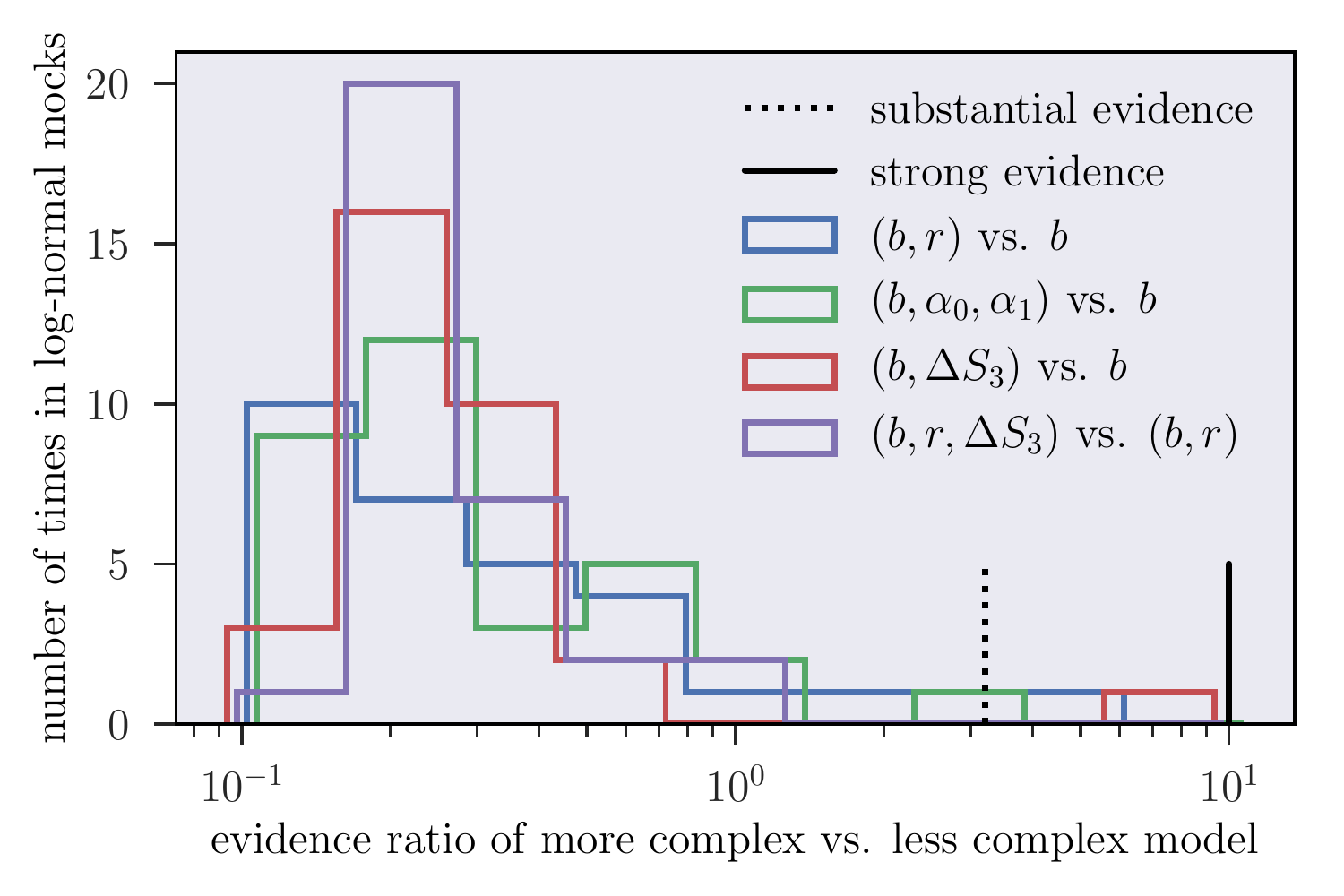}
  \caption{Model comparison based on the Bayes factor, i.e.~the ratio of evidences of the more complex over the less complex model, from likelihoods run on log-normal mocks. These mocks contain no stochasticity and no $\Delta S_3$ and hence should not favor the more complex models. Dotted and solid vertical lines indicate the Jeffreys scale for substantial and strong evidence, showing that this is indeed the case.}
\label{fig:modelcomp_lognormal}
\end{figure}

Finally, to test our methodology of model comparison with Bayes factors, we sample these likelihoods again with the \textsc{multinest} algorithm. We first confirm that parameter constraints derived from \textsc{multinest} and \textsc{emcee} closely match each other. We then determine ratio of Bayesian evidence between complex models and the most simple run (linear bias with no stochasticity, $\Delta S_3=0$). Since the log-normal simulations should be fully described by this simple model, there should be no evidence for any more complex model. We show in \autoref{fig:modelcomp_lognormal} that this is indeed the case.

\section{Mock analysis of \emph{Buzzard} simulations}
\label{app:mock}

We repeat the steps of \autoref{sec:lognormallikelihood}, but instead of log-normal realizations using density split lensing and counts-in-cells signals measured on the \emph{Buzzard} suite of $N$-body simulations.

The \emph{Buzzard} simulations are a suite of mock galaxy catalogs built on top of dark matter only $N$-body simulations. We describe only the essential details here and refer the reader to more complete descriptions in \citet{DeRose2017}. Each set of 6 \emph{Buzzard} DES Y1 catalogs is generated from a combination of 3 $N$-body lightcones runs using L-Gadget2, a version of Gadget2 \citep{Springel2005} optimized for memory efficiency. 2nd order Lagrangian perturbation theory initial conditions were generated using \textsc{2LPTIC} \citep{Crocce2006}. The lightcones were produced on the fly as the simulations ran from boxes of volumes $1050^3$, $2600^3$, and $4000^3\, (\Mpch)^{3}$ and mass resolutions of $2.7\times 10^{10}$, $1.3\times 10^{11}$, $4.8\times 10^{11}\/h^{-1}\Msol$ respectively. The lightcones are joined at redshifts 0.34 and 0.9, and arranged such that the highest resolution simulations are used at lower redshifts.

Galaxy catalogs are produced from the dark matter lightcones using the \textsc{ADDGALS} algorithm \citep{Wechsler2017}. \textsc{ADDGALS} uses the relation between large scale density and r-band absolute magnitude determined from a subhalo abundance matching \citep{Conroy2006, Reddick2013, Lehmann2017} algorithm run on a high resolution N-body simulation to place galaxies with magnitudes into a low resolution density field. Galaxies are then assigned SEDs from SDSS DR7 \citep{Cooper2011} based on the distance to their fifth nearest neighbor. DES $griz$ fluxes are generated from these SEDs and photometric noise is added to them using the DES Y1 depth map. These galaxies are then lensed using the multiple plane raytracing algorithm \textsc{CALCLENS} \citep{Becker2013} which uses a spherical harmonic transform Poisson solver allowing for curved sky boundary conditions. Two different versions of \textsc{ADDGALS} were used to create the catalogs referred to as \emph{Buzzard-v1.1} and \emph{Buzzard-v1.6} below. The main differences come from changes to the assumed luminosity function of galaxies, the evolution of the red fraction of galaxies with redshift and the use of different depth maps to produce photometric noise.

In order to obtain a simulated \textsc{redMaGiC} galaxy sample, we run the \textsc{redMaGiC} algorithm on the simulations with the same configuration as the data, yielding very similar photometric redshift and clustering properties to those found in DES Y1. A \metacal\ like sample is created by making signal to noise cuts in order to approximate the source density found in the data.

While the true input cosmology is known and identical between versions 1.1 and 1.6 of the \emph{Buzzard} mock catalogs, their model for early-type galaxy SEDs vary, and thus the \textsc{redMaGiC} selection is not the same. Consequently, we cannot use the values for bias and stochasticity parameters determined in \citet[][their sections 4.3.1 and 4.3.2]{Oliver} for version 1.1 as truth inputs for version 1.6. We repeat the analysis performed there to find that \textsc{redMaGiC} galaxies in $z=0.2-0.45$ have a somewhat larger bias and stochasticity in version 1.6 ($b,\alpha_0,\alpha_1=1.72, 1.36, 0.29$) than in version 1.1 ($b,\alpha_0,\alpha_1=1.54, 1.26, 0.29$). The same is true in the simpler parametrization with ($b,r=1.84,0.96$ vs. $1.62,0.96$). We note that the finding that these galaxies in the \emph{Buzzard-v1.6} simulation show large-scale stochasticity is in line with the analysis of \citet{Niall}.

The masks of the \emph{Buzzard-v1.1} simulations cover a somewhat smaller sky area (cf. \citeauthor{Oliver}, their fig.~1) than the full DES-SPT footprint (cf. \autoref{fig:skyplots}). We therefore generate a separate covariance matrix for \emph{Buzzard-v1.1} by cutting the mask of the \textsc{FLASK} realizations accordingly and repeating the procedure of \autoref{sec:cosmicvariance}. The larger galaxy bias in \emph{Buzzard-v1.6} causes a larger cosmic variance than the one derived in \autoref{sec:covariance} for \emph{Buzzard-v1.1} parameters. We account for this by simply re-scaling the \emph{Buzzard-v1.6} covariance matrix by a factor $1.05$ that brings the mean $\chi^2$ of the \emph{Buzzard-v.1.6} and \emph{v1.1} simulations to agreement.

We split the survey area by density and measure counts-in-cells and lensing signals in \emph{Buzzard-v1.1} and \emph{Buzzard-v1.6} mock catalogs as described in \autoref{sec:measurement}. As source galaxies, we use the approximated lensing source sample described above. For source redshift tomography, we define three bins based on BPZ redshift expectation values at $z_s=0.43-0.70,0.70-0.78,0.92-1.30$ in v1.1 and $z_s=0.43-0.66,0.66-0.76,0.76-1.00$. These limits were defined such that the mean true redshift matches that of the three highest redshift source bins in \citet{photoz}.  In measuring tangential shear profiles, we use true shear information from raytracing in\emph{Buzzard} \citep{calclens}. Shape noise is added at the level of the data vector using measurements of density split lensing made with randomly rotated source galaxy catalogs on DES Y1 data itself that were not used for estimating the covariance matrix (\autoref{sec:shapenoise}). We generate 62 data vectors in total, 28 (7 different shape noise realizations each) based on the 4 \emph{Buzzard-v1.1} $N$-body mocks and 34 (2 different shape noise realizations each) from 17 \emph{Buzzard-v1.6} mocks. Caution must be taken in interpreting results, especially on coverage, due to the fact that these realizations are not completely independent.

For a test of our model and the covariance matrix, we predict the signal, using the model of \citet{Oliver}, with the true input parameters for cosmology and the directly measured parameters for the $b,\alpha_0,\alpha_1$ model of galaxy biasing. In this, we use the mean true redshift distributions of \textsc{redMaGiC} and lensing source galaxies, and the mean \textsc{redMaGiC} counts in the \emph{v1.1} or \emph{v1.6} sets of \emph{Buzzard} simulations. For the source bias of the lowest redshift bin, we assume $b_{s1}=0.5$.

\begin{figure}
  \includegraphics[width=\linewidth]{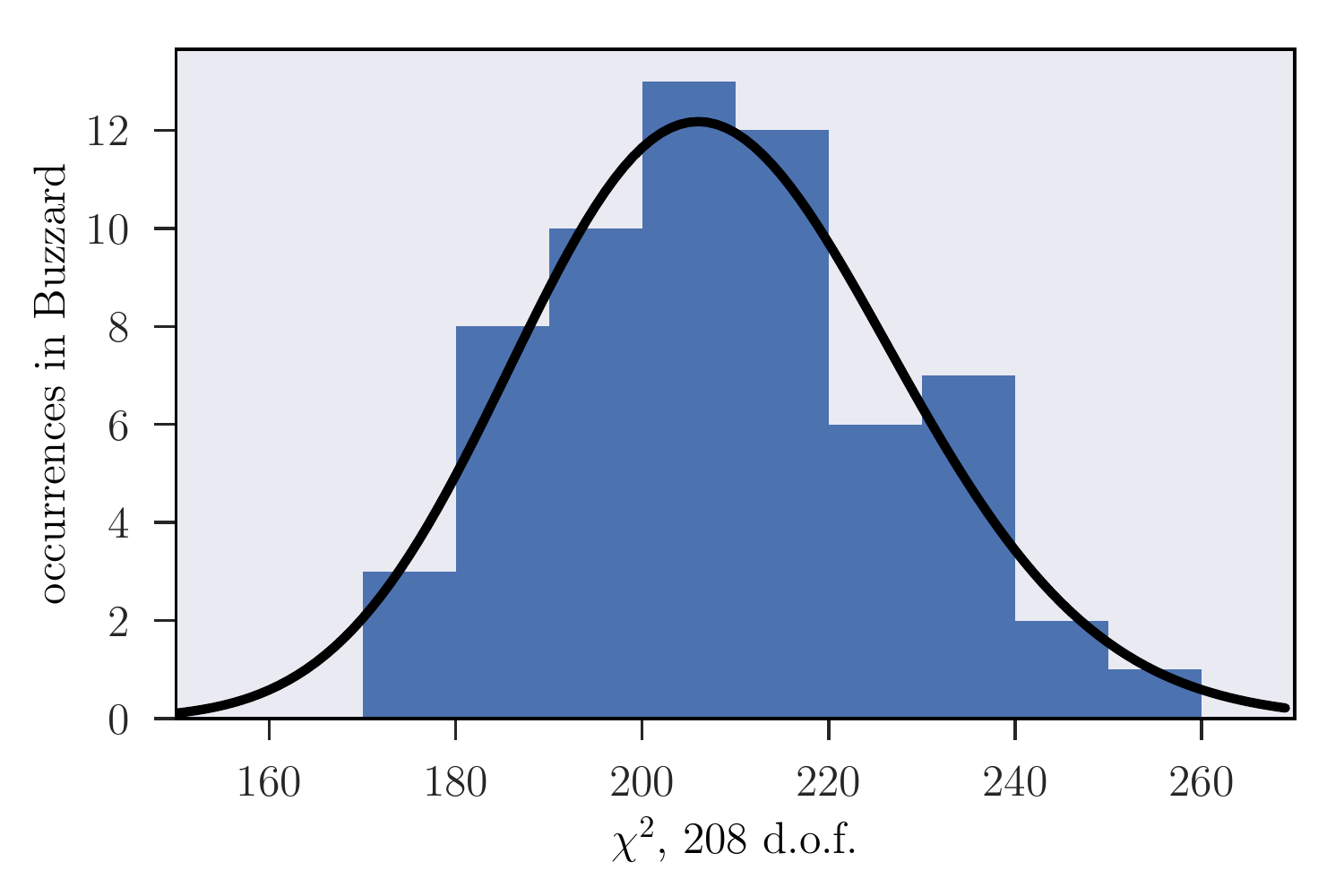}
  \includegraphics[width=\linewidth]{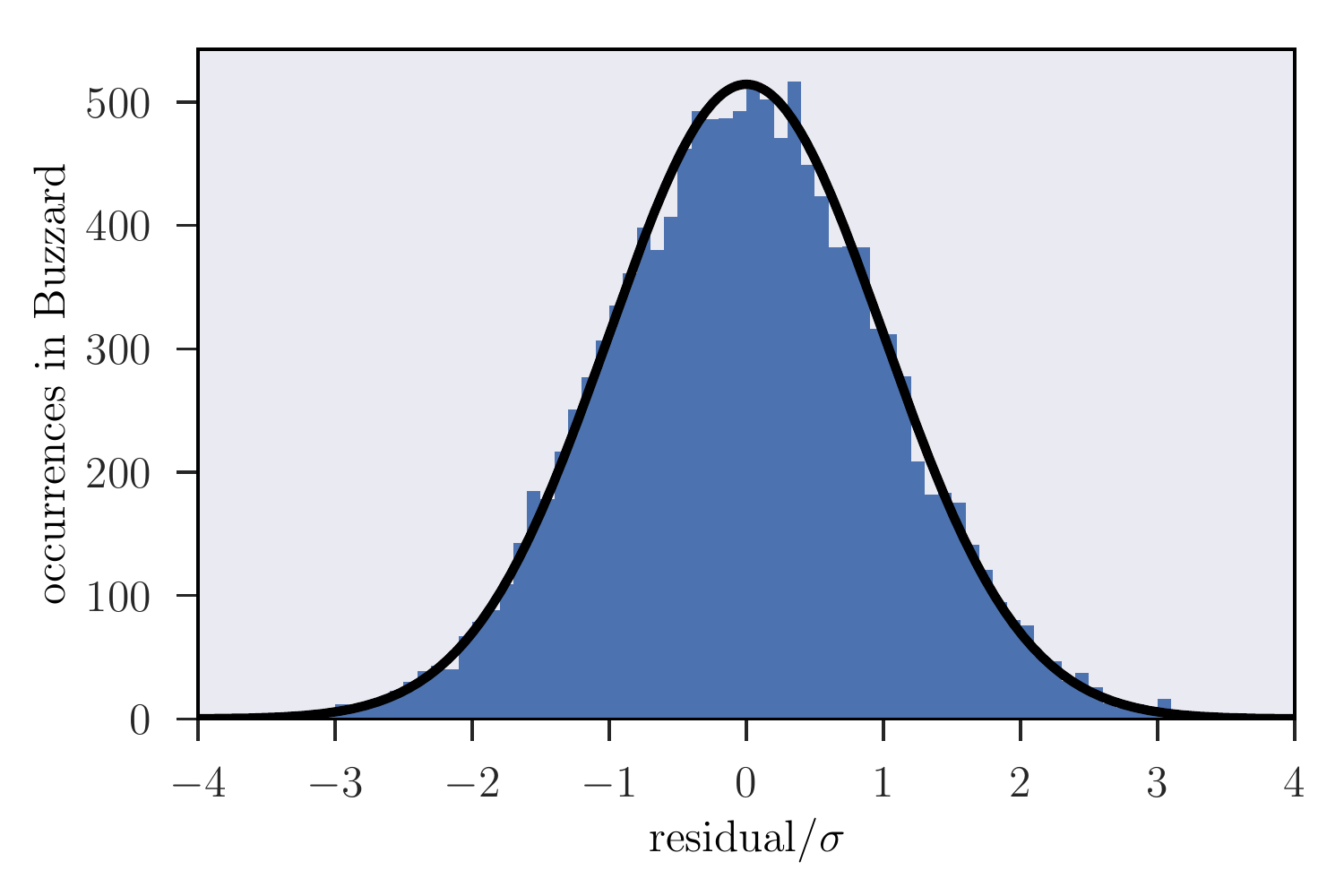}
  \caption{Residuals of measurements of lensing and counts-in-cells signal in \emph{Buzzard} relative to model evaluated at true parameters of the simulations. \emph{Top panel:} $\chi^2$ of each realization with 208 entries in the data vector. \emph{Bottom panel}: Residuals of individual data points in units of their expected standard deviation. Black lines indicate expected distributions. The figures use 62 realizations based on 21 independent N-body mocks from the \emph{Buzzard-v1.1} and \emph{v1.6} runs, to which different realizations of Y1 \textsc{metacalibration} shape noise were added.}
\label{fig:testmodel_buzzard}
\end{figure}

From the residuals of this model relative to the 62 realizations with 208 data points each, we generate the statistics shown in \autoref{fig:testmodel_buzzard}. The distribution of $\chi^2$ is consistent with expectations (left panel). The root-mean-square of individual data point residuals in units of the expected standard deviation according to the covariance matrix is slightly larger than unity, but by less than 5 percent (right panel). We have confirmed that in the case of the \emph{Buzzard-v1.1} simulations it is indeed consistent with unity, and suspect that the small increase in \emph{Buzzard-v1.6} might be due to the larger cosmic variance caused by the increased \textsc{redMaGiC} galaxy bias, which we have not fully accounted for in the covariance matrix (see discussion above). There is thus no apparent non-Gaussianity in the distribution of residuals \citep[but cf.][]{2017arXiv170704488S}.  These statistics look very similar when using the $b,r$ instead of the $b,\alpha_0,\alpha_1$ model, i.e.~the prediction evaluated at the true cosmology and directly measured stochasticity is a good fit to the simulations.

For a test of our inference methodology, we run mock likelihood chains on these \emph{Buzzard} realizations, with $\Omega_m,\sigma_8,b$, stochasticity parameters, and galaxy bias of sources in the lowest redshift bin $b_{s1}$ as free parameters. All other parameters are fixed to their input values, and we use true redshift distributions for the predictions.

\autoref{fig:testmodel_buzzard} shows the most relevant run, of the fiducial $b,\alpha_0,\alpha_1$ biasing model on the 17 independent \emph{Buzzard-v1.6} realizations with 2 independent versions of shape noise each. Coverage of cosmological parameters is within expectations for 17 independent realizations (58 percent for $\Omega_m$, 62 percent for $\sigma_8$). We find a mean source bias of $b_{s1}=0.62$. 

For the $b,r$ model, in contrast, coverage of $\sigma_8$ is low (only 18 percent) because $\sigma_8$ is biased high (with the best fit being 0.93 on average, significantly above the true input $0.82$). We confirm that this is not only caused by the asymmetry of the allowed parameter range around the directly measured $r=0.96$: allowing linear extrapolation of the model to $r>1$ still yields a coverage of only 33 percent in $\sigma_8$. We also confirm that the same low coverage is found regardless of whether we use \textsc{emcee} or \textsc{MultiNest} as a sampler.

\begin{figure}
  \includegraphics[width=\linewidth]{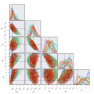}
  \caption{Realizations of likelihoods on 34 mock realizations based on the \emph{Buzzard-v1.6} catalogs with the true parameters given by the dashed black lines. The coverage of these mock likelihoods is within expectations, i.e.~any input parameter lies within its marginalized 1$\sigma$ confidence interval $\approx68$ percent of the times. Likelihoods are sampled with \textsc{multinest}, which is a significant improvement in speed. While the appearance is more patchy, results are consistent with long \textsc{emcee} chains, as we have checked from a subset of dual runs on identical simulations.}
\label{fig:converage_buzzard}
\end{figure}

\begin{figure}
  \includegraphics[width=\linewidth]{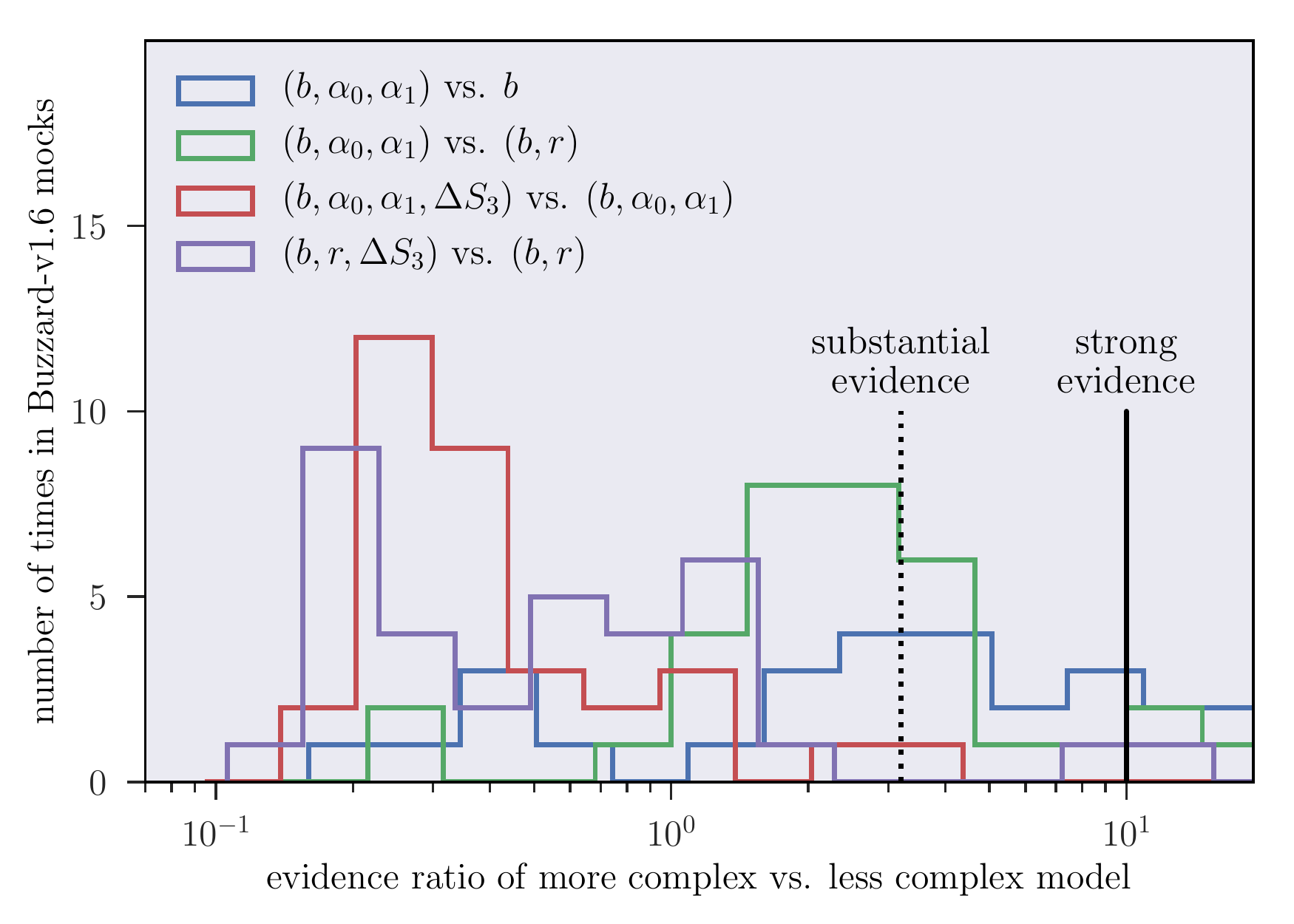}
  \caption{Model comparison based on the Bayes factor, i.e.~the ratio of evidences of the more complex over the less complex model, from likelihoods run on \emph{Buzzard-v1.6} mocks. These mocks contain no $\Delta S_3$ but do show a complex relation of galaxy count and matter density and hence may favor the more complex models. Dotted and solid vertical lines indicate the Jeffreys scale for substantial and strong evidence, showing that this is indeed the case.}
\label{fig:modelcomp_buzzard}
\end{figure}

Some understanding might be gained from the Bayesian model comparison we perform in \autoref{fig:modelcomp_buzzard}. While the \emph{Buzzard} simulations should not prefer a model with free $\Delta S_3$, it is up to tests like this to determine what level of complexity is needed to describe the stochasticity in the galaxy count distribution. Notably, in some cases the $b,\alpha_0,\alpha_1$ model is strongly preferred to the $b,r$ model, and in one case the $b,r$ model prefers the introduction of $\Delta S_3$ as a free parameter, potentially to compensate for the excess skewness in a galaxy count distribution with density-dependent stochasticity. Both could be an indication of density dependent stochasticity actually being present in \emph{Buzzard-v1.6}. In this case, the high bias of $\sigma_8$ in the $b,r$ model seen in the coverage tests could be a partial compensation of the missing variance.

\begin{figure}
  \includegraphics[width=\linewidth]{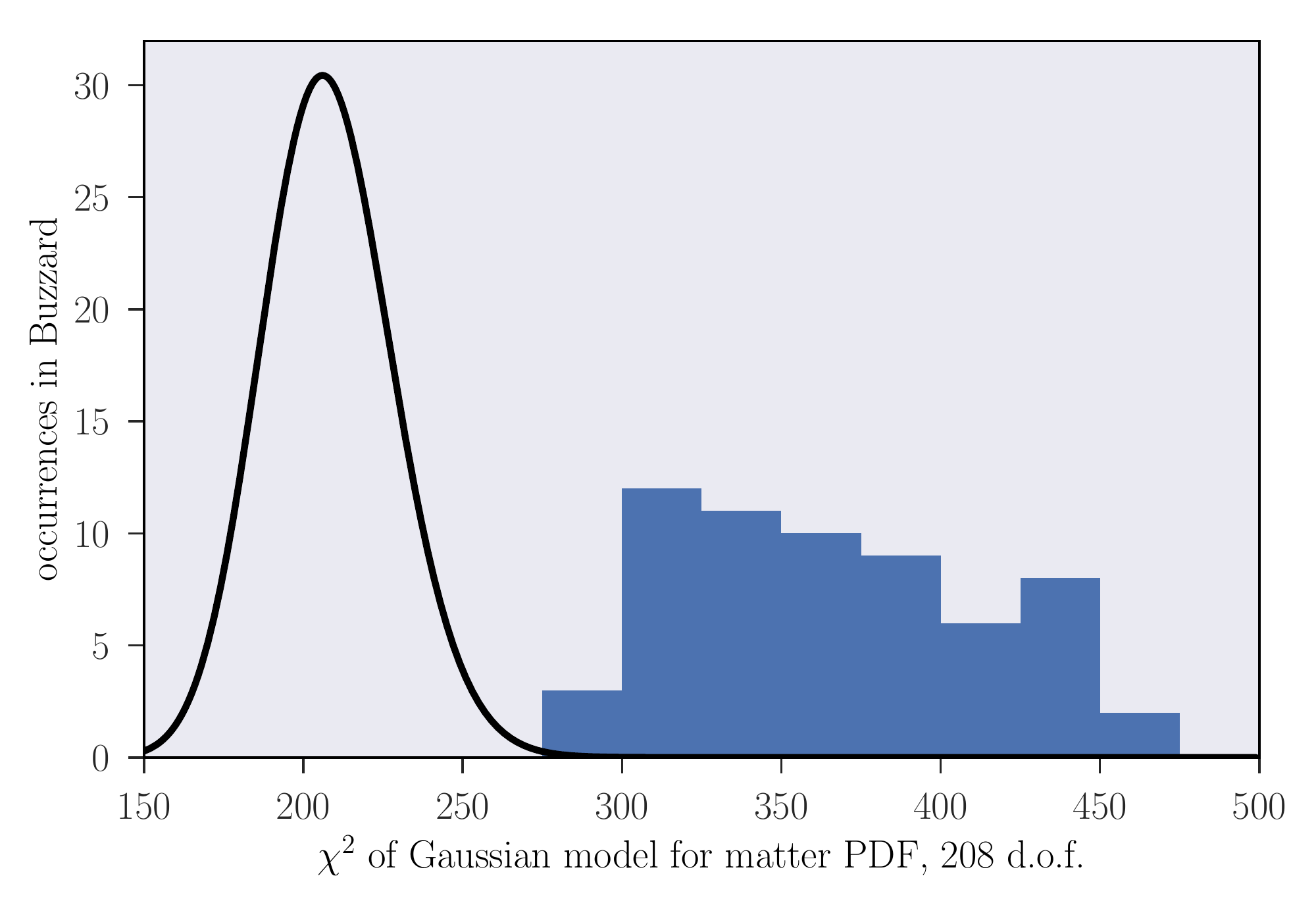}
  \caption{$\chi^2$ of comparing \emph{Buzzard} realizations of the data vector to a model of the signal in which the skewness of the matter density field was set to 0, i.e.~the smoothed matter density field was assumed to be Gaussian. In all realizations, the Gaussian model is excluded at more than $3\sigma$.}
\label{fig:testmodel_buzzard_gaussian}
\end{figure}

We conclude that the $b,\alpha_0,\alpha_1$ model is required for describing the galaxy distribution in \emph{Buzzard-v1.6} sufficiently well. In this framework, the coverage test and model comparison test for adding $\Delta S_3$ as a free parameter have results that are within expectations for a reliable inference scheme. Whether or not the complexity in \emph{Buzzard} is realistic or caused, in part, by peculiarities of the simulations (such as the stitching of simulation boxes or the placement of galaxies) is not clear at this point \citep[see also][]{Niall}.

Finally, we test how well the \emph{Buzzard} realizations of our data vector are fit in a model in which the smoothed matter field $\delta_{m,T}$ is assumed to be Gaussian with the variance predicted by the power spectrum, i.e.~to have no skewness. This is a model that provided a reasonable fit to the data in \citet{2016MNRAS.455.3367G}. \autoref{fig:testmodel_buzzard_gaussian} shows the $\chi^2$ of the residuals of data and Gaussian model, where we have used true cosmology and directly measured galaxy bias $(b,\alpha_0,\alpha_1)$ parameters for the latter. Even for the least $\chi^2$ among the 62 realizations, the large $\chi^2$ allows us to exclude the Gaussian model at $p<0.001$.

\section{Clustering constraints on source and lens redshift distributions} 

For both our \textsc{redMaGiC} tracer population and our lensing source galaxies, we calibrate the mean value of redshift distributions using clustering redshifts. In these, excess angular correlations with thinly-sliced spectroscopic or spectroscopic-like samples are used to determine the redshift distributions \citep{2008ApJ...684...88N, 2013arXiv1303.4722M, 2017arXiv170708256D, xcorrtechnique, redmagicpz, xcorr}

For the case of DES Y1, this calibration and its systematic uncertainties are described in \citet{xcorrtechnique,redmagicpz,xcorr}. Here, we give the missing details relevant to the specific tracer galaxy selection in DES and the tracer and source samples in SDSS not described in those papers.

\subsection{\textsc{redMaGiC} galaxies}
\label{sec:redmagiczbias}

\begin{figure*}
  \includegraphics[width=0.48\linewidth]{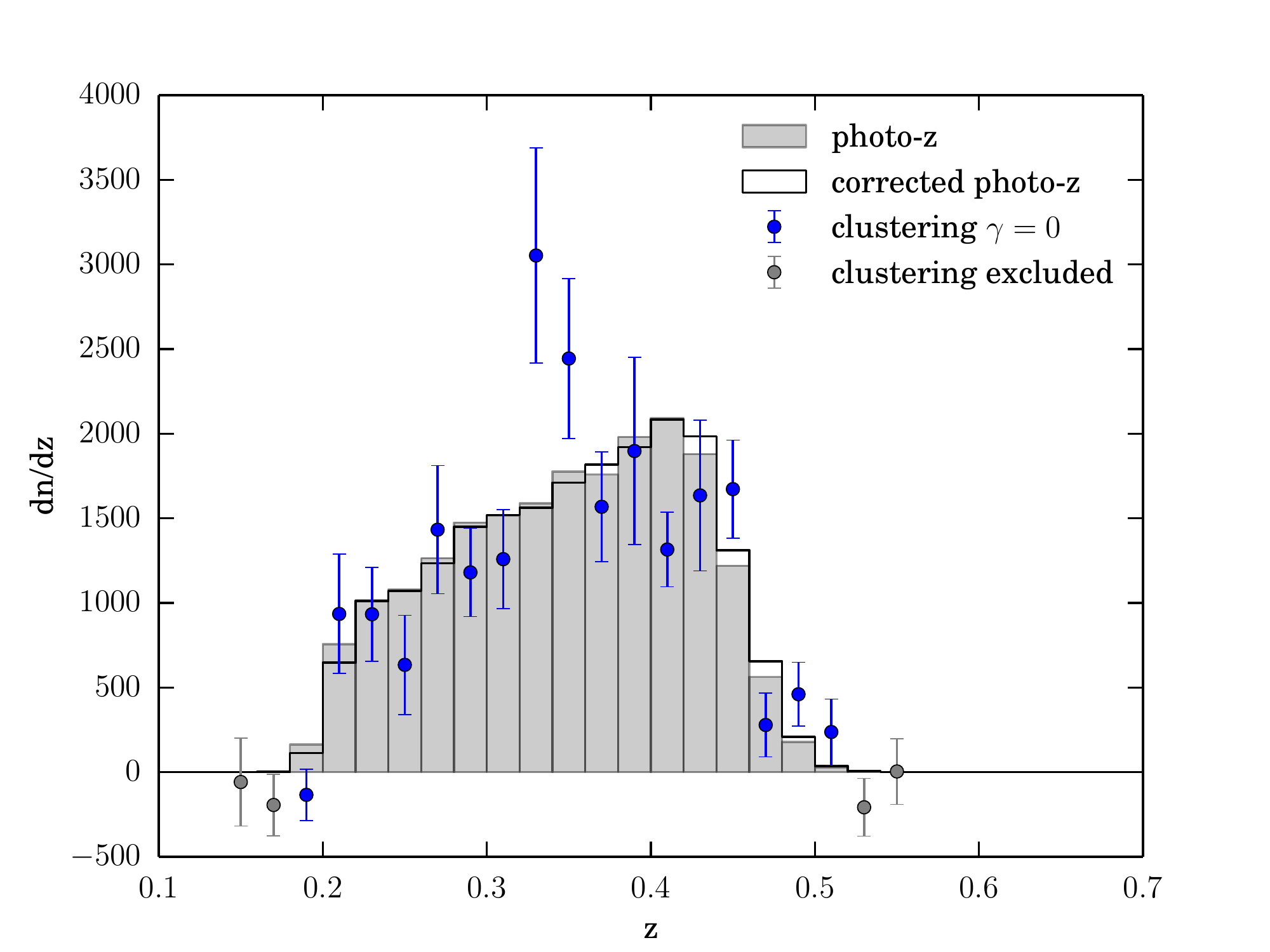}
  \includegraphics[width=0.48\linewidth]{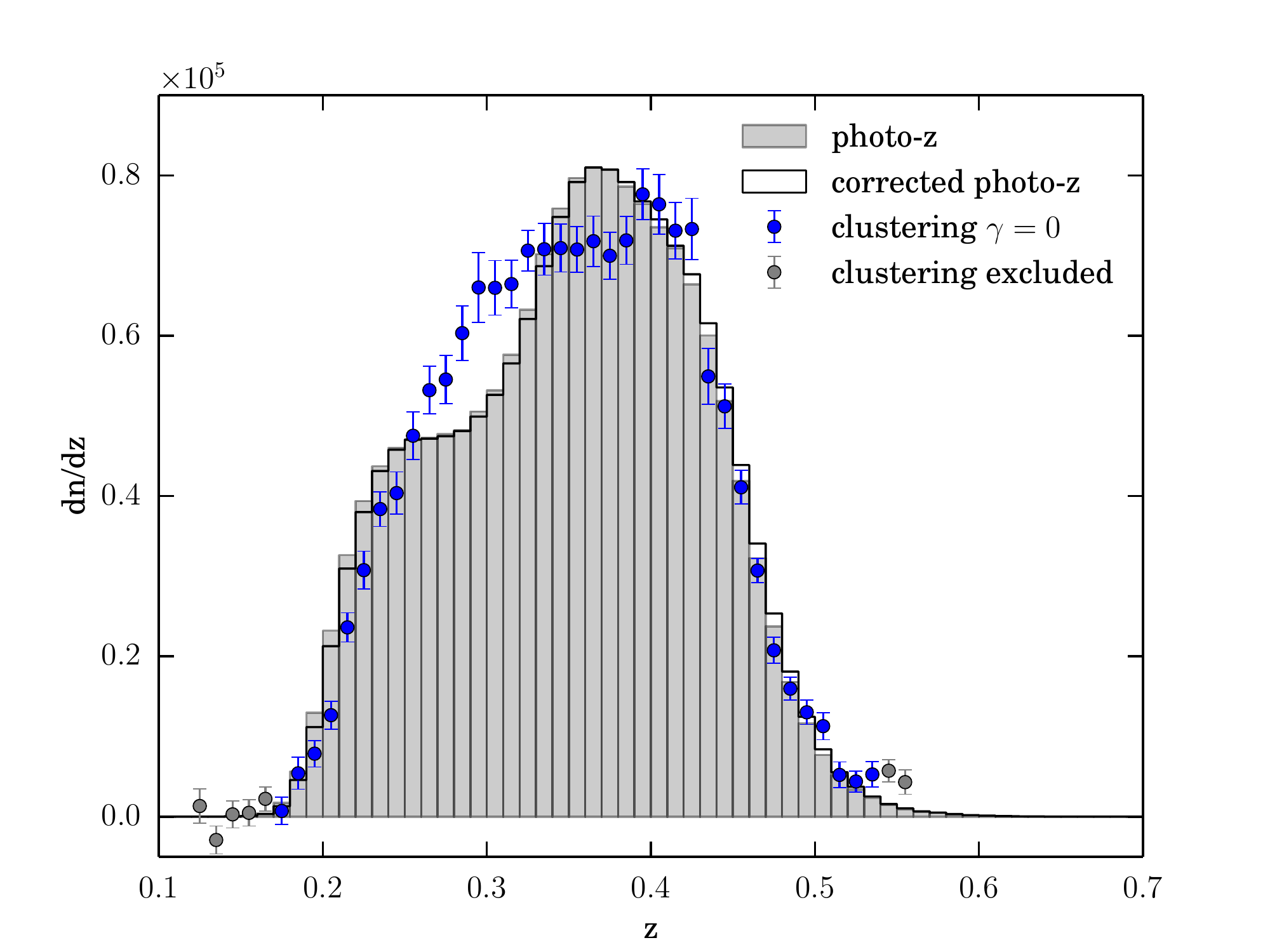}
  \caption{Comparison of the distribution of DES (\emph{left panel}) and SDSS (\emph{right}) redMaGiC photometric redshifts with the estimated redshift distribution from clustering using SDSS DR12 LOWZ and CMASS as a reference sample. By shifting the photo-z distribution to fit the mean of the clustering estimate, we estimate the photometric bias to be $\Delta z=0.003 \pm 0.008$ (DES) and $\Delta z=0.002 \pm 0.006$ (SDSS).}
\label{fig:redmagic_clustering}
\end{figure*}

\begin{figure}
  \includegraphics[width=0.9\linewidth]{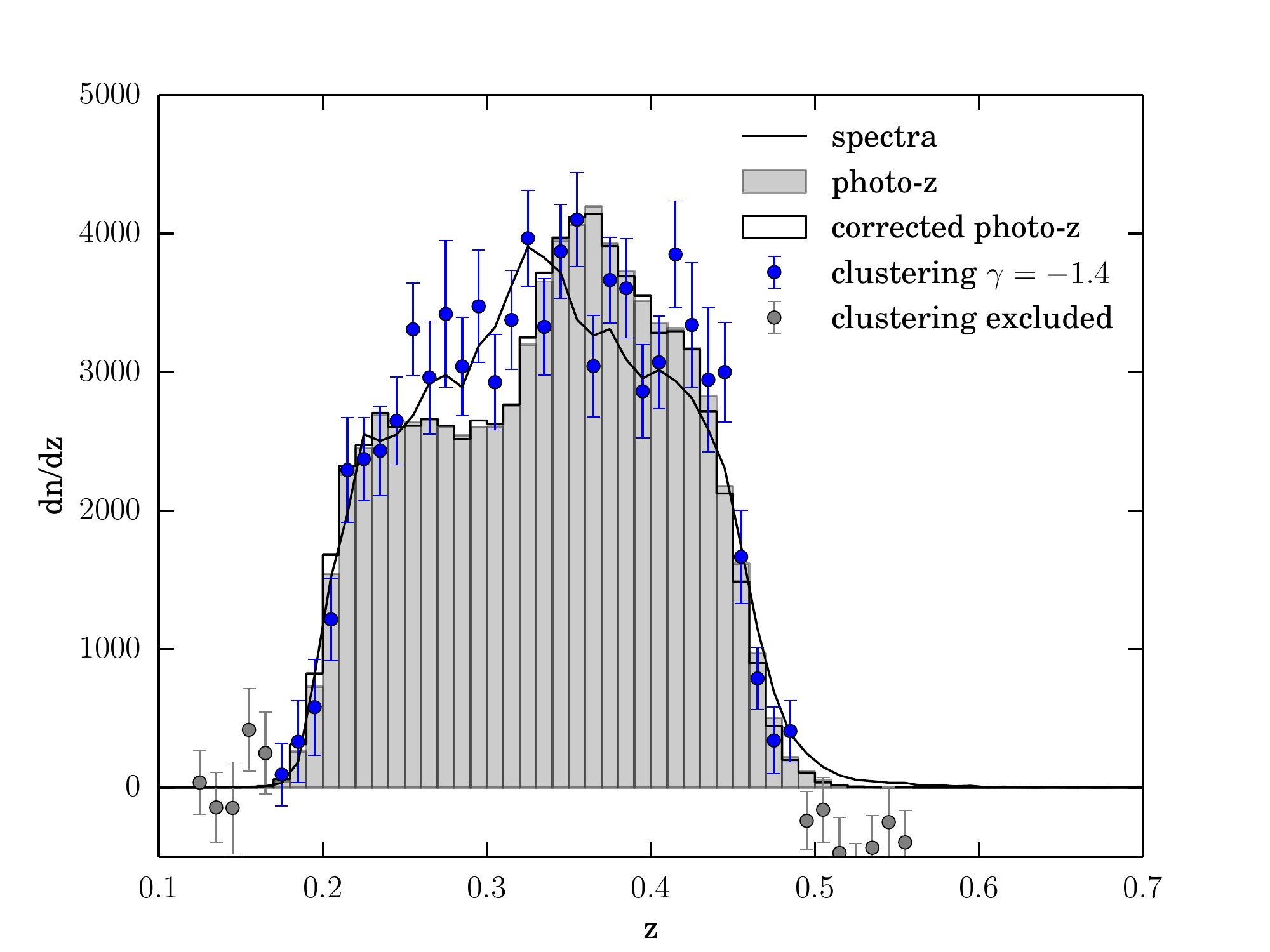}
  \caption{Clustering measurements on a subset of the SDSS lens sample that has spectroscopic measurements. ($79,583, 4.9 \%$ of the full sample.) The clustering measurement contains a correction for the evolution of galaxy bias, $\gamma=-1.4$ which best fits the true redshift distribution mean. This subsample gives a photo-z bias of -0.0015.}
\label{fig:redmagic_clustering_spectra}
\end{figure}

We perform a clustering redshift analysis to assess the accuracy of the \textsc{redMaGiC} photometric redshift algorithm. In this, we follow the same techniques as described in \citet{redmagicpz}, yet with a different redshift cut on the redMaGiC sample.

We cross-correlate the SDSS and DES redMaGiC high density sample, selected by the photometric estimate $z_{\rm red}=0.2-0.45$, with the BOSS spectroscopic samples LOWZ and CMASS \citep{2013AJ....145...10D}. The clustering redshift measurement follows the method described in \citet{2013MNRAS.431.3307S}, using physical scales of 0.5 to 1.5 Mpc. Statistical error estimates are from jackknife resampling.

As described in \citet{redmagicpz}, the main systematic to this measurement is the redshift evolution of the galaxy bias of the two samples across the redshift bin. We can measure the evolution of the spectroscopic samples using autocorrelations. The amplitude of the autocorrelations of \textsc{redMaGiC} though will be impacted by photo-z errors. Results in \citet{redmagicpz} suggest that the quantity $b_{l} \sqrt{w_{\rm mm}}$, where $b_{l}$ is the galaxy bias of \textsc{redMaGiC} and $w_{\rm mm}$ is the autocorrelation of the matter density on the scales we measure, shows very little evolution with redshift for \textsc{redMaGiC}. We parameterize the galaxy bias evolution with $b_{l} \sqrt{w_{\rm mm}} \propto (1+z)^\gamma$. In \citet{redmagicpz}, we assumed that $\gamma$ is in the range $0.0\pm2.0$, but found that is shows less spread around 0 for wider redshift bins. Since the redshift range $z_T=0.2-0.45$ used in this work is significantly wider than that used in the bins of \citet{redmagicpz}, we assume $\gamma=0 \pm 1.5$ here. These choices of $\gamma$ range broadly fit the various estimates of $\gamma$ from the auto-correlations of \textsc{redMaGiC} on the full DES sample, the Stripe 82 sample which contains the galaxies that overlap with BOSS, and simulations as measured in \citet{redmagicpz}.

After the galaxy bias correction is selected, the clustering measurement is also narrowed to $\pm 2.5 \sigma$ around the mean of the clustering redshift distribution estimate, with $\sigma$ being the standard deviation of that estimate. This cut is indicated in Figures \ref{fig:redmagic_clustering} and \ref{fig:redmagic_clustering_spectra}. This is done since the clustering measurement can be noisy and biased where the signal is low. We then fit for a single photometric bias parameter, $\Delta z=z-z_{\rm phot}$, where $z$ is the clustering redshift mean and $z_{\rm phot}$ is the photometric redshift mean over the redshift range selected by the $\pm 2.5 \sigma$ cut.

The results of this analysis are shown in Table \ref{tab:redmagic_clustering_bias} and Figures \ref{fig:redmagic_clustering}. The measurement on DES \textsc{redMaGiC} can only be done on a subsample of 20,347 galaxies in Stripe 82, which has overlap with the BOSS spectroscopic samples. For this reason, the statistical uncertainty on the DES measurement is significantly larger than on SDSS \textsc{redMaGiC}. For both samples, the dominant systematic error is the uncertainty in the \textsc{redMaGiC} galaxy bias evolution parameterized by $\gamma$. The uncertainty in $\gamma$ of  $\pm 1.5$ leads to approximately an uncertainty of $\Delta z$ of $\pm 0.006$. Figure \ref{fig:redmagic_clustering_spectra} shows the cross correlation redshift estimate using just a subsample of SDSS \textsc{redMaGiC} that has spectra. This is a biased sample that is brighter and likely has a different photo-$z$ bias and galaxy bias, though it appears to confirm some similar trends seen in Figure \ref{fig:redmagic_clustering}, such as a lower galaxy density around $z=0.4$ and higher density around $z=0.3$ compared to the photo-z code.

\begin{table}
	\centering
	\caption{Cross-correlation estimates of photo-$z$ bias on \textsc{redMaGiC}}
	\label{tab:redmagic_clustering_bias}
	\begin{tabular}{llrr}
		\hline
		Sample &$\Delta z$ & $\delta \Delta z$ (syst) & $\delta \Delta z$ (stat) \\
        \hline
         DES & $0.003\pm0.008$ &  $0.006$ & $0.005$ \\
        \hline
         SDSS & $0.002\pm0.006$ &  $0.006$ &  $0.001$ \\
		\hline
	\end{tabular}
\end{table}

\subsection{SDSS source galaxies}

\label{sec:sourcezbias}
\begin{figure}
  \includegraphics[width=0.9\linewidth]{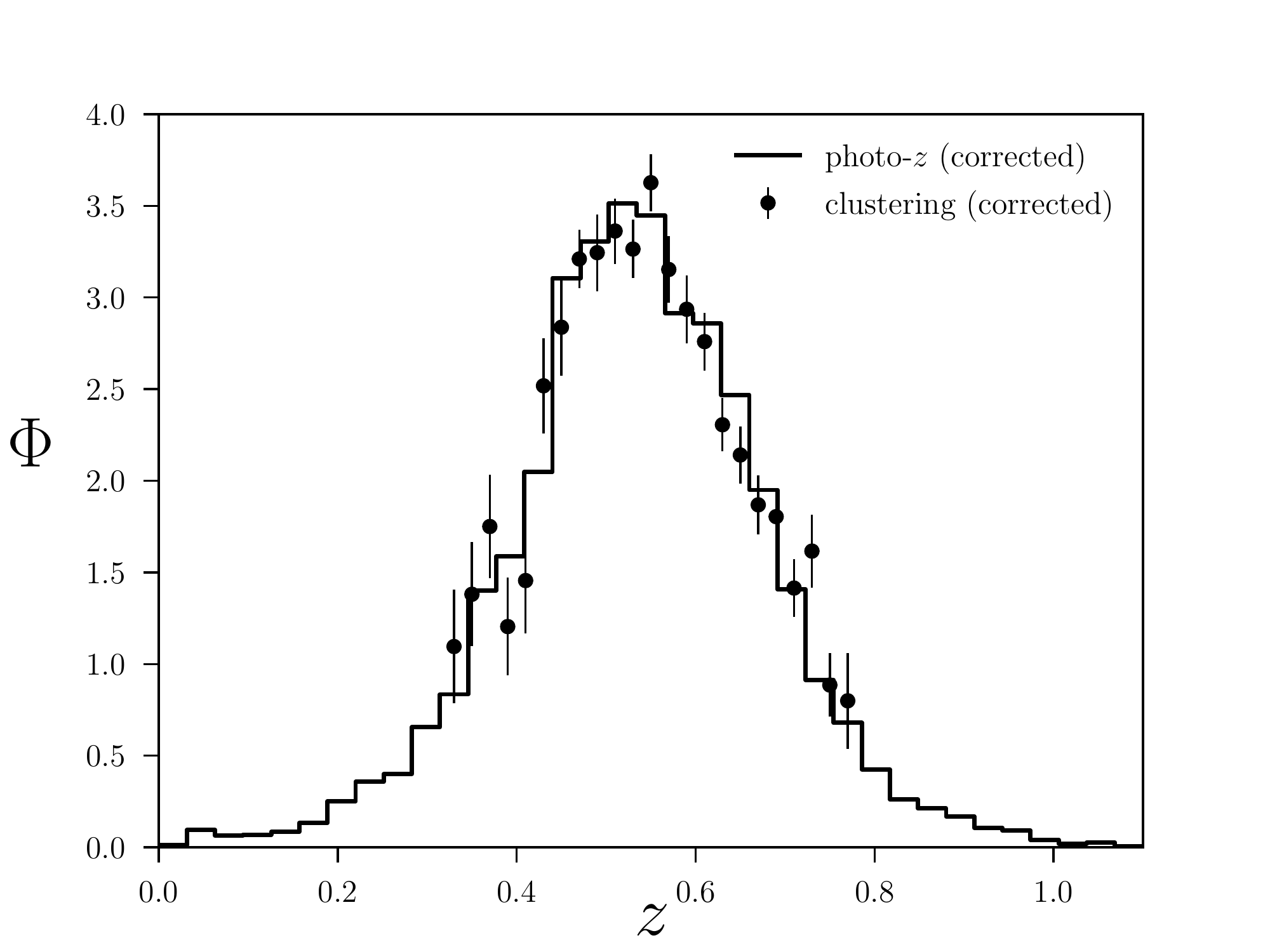}
  \caption{
Redshift distribution of combined SDSS samples as measured by photo-$z$ (solid line) and clustering (points) with SDSS DR12 spectra \citet{2015ApJS..219...12A}. The photo-$z$ is calibrated with a mean redshift offset of $\Delta_z = -0.014 \pm 0.011$, while the clustering estimates are calibrated with a bias evolution model of $b_{\mathrm{PZ}} b_{\mathrm{spectra}} \propto (1 + z)^{\gamma}$ with $\gamma = -2.0 \pm 0.6$.}
\label{fig:pzsdss}
\end{figure}

We measure the excess angular clustering from 500 to 1500 kpc between SDSS source galaxies selected and weighted as in \autoref{sec:sdss_shapes} and the SDSS DR12 spectroscopic galaxy sample. This method is similar to that detailed in \citet{2013arXiv1303.4722M}. Statistical errors in clustering-$z$ are estimated by jackknife resampling. We find that reliable clustering signal can only be recovered over a subset of the sample, from $0.32 < z < 0.80$, but this is sufficient to calibrate the photo-$z$ distribution with our method, where we assume the \emph{shape} of the source $n(z)$ is correctly estimated by photo-$z$ and only determine its \emph{shift}. More details for the calibration procedure may be found in \citet{xcorr} as well as in \citet{xcorrtechnique} and \citet{redmagicpz}. 

We fit a relative redshift bias, $z \rightarrow z - \Delta_z$ in the photo-$z$ distribution, and parameterize the clustering bias evolution in the clustering-$z$ as a power-law with free exponent, $b_{\mathrm{PZ}} b_{\mathrm{spectra}} \propto (1 + z)^{\gamma}$. As we found in \citet{xcorrtechnique} we expect that systematic uncertainties in modeling the underlying bias evolution to dominate over our statistical uncertainties. Systematic uncertainties in \citet{xcorrtechnique} are of the order $\sigma_{\Delta z} \approx 0.01$, while our statistical uncertainty is $\sigma_{\Delta z} = 0.002$. In quadrature, these combine to be an uncertainty of $\sigma_{\Delta z} = 0.011$. We find $\Delta_z = -0.014 \pm 0.011$ and $\gamma = -2.0 \pm 0.6$, indicating only a marginal preference for a mean offset in redshift, and moderate combined bias evolution. We note that the bias evolution measured is the \textit{product} of the bias evolutions of the lens and source samples, and so the bias evolution measured here will likely differ from the calibration of the lens bin even though both use the same reference galaxies and scales.

\end{document}